\documentclass[numberedappendix,twocolumn,iop,apj]{emulateapj} 
\usepackage{times}

\usepackage[section]{placeins}

\usepackage{amssymb,amsmath}
\usepackage{graphicx,subfigure}
\usepackage{color}
\usepackage{rotating}
\usepackage{ulem} 
\usepackage{float}
\DeclareMathOperator{\sech}{sech}

\newcommand{\be}{\begin{equation}}
\newcommand{\ee}{\end{equation}}

\newcommand{\gray}{$\gamma$-ray}

\newcommand{\fermilat}{{\it Fermi}--LAT}
\newcommand{\fermi}{{\it Fermi}}
\newcommand{\gardian}{{\it GaRDiAn}}

\newcommand{\GP}{{\it GALPROP}}

\newcommand{\frankie}{{\it FRaNKIE}}

\newcommand{\galtoolslib}{{\it GALTOOLSLIB}}
\newcommand{\galgas}{{\it GALGAS}}
\shorttitle{Galactic High-Energy Interstellar Emission Models in 3D}
\shortauthors{Porter~et~al.}

\begin{document}

%\linenumbers


\title{High-Energy Gamma Rays from the Milky Way: Three-Dimensional Spatial Models for the Cosmic-Ray and Radiation Field Densities in the Interstellar Medium}

\author{
T.~A.~Porter\altaffilmark{1,$\dagger$},
G.~J\'ohannesson\altaffilmark{2,3}, and 
I.~V.~Moskalenko\altaffilmark{1}
}
\altaffiltext{$\dagger$}{email: tporter@stanford.edu}
\altaffiltext{1}{W. W. Hansen Experimental Physics Laboratory and Kavli Institute for Particle Astrophysics and Cosmology, Stanford University, Stanford, CA 94305, USA}
\altaffiltext{2}{Science Institute, University of Iceland, IS-107 Reykjavik, Iceland}
\altaffiltext{3}{AlbaNova Univ. Center Nordita, Roslagstullsbacken 23, SE-106 91 Stockholm, Sweden }

\begin{abstract}
  High-energy \gray{s} of interstellar origin 
  are produced by the interaction of cosmic-ray (CR) particles with the diffuse gas and radiation fields in the Galaxy.
  The main features of this emission are well understood and are reproduced
  by existing CR propagation models employing 2D Galactocentric cylindrically
  symmetrical geometry.
  However, the high-quality data from instruments like the \fermi\ Large Area
  Telescope reveal significant deviations from the model predictions on
  few to tens of degree scales indicating the need to include the details of the Galactic spiral structure and thus require 3D spatial modelling.
  In this paper the high-energy interstellar emissions from the Galaxy are
  calculated using the new release of the \GP\ code employing 3D
  spatial models for the CR source and interstellar radiation field (ISRF)
  densities.
  Three models for the spatial distribution of CR sources are used that are 
  differentiated by their relative proportion of input luminosity attributed to the smooth disc or spiral arms.
  Two ISRF models are developed based on stellar and dust spatial density distributions taken from the literature that reproduce local near- to far-infrared observations.
  The interstellar emission models that include arms and bulges for the CR source and ISRF densities provide plausible physical interpretations for features found in the residual maps from high-energy \gray{} data analysis.
  The 3D models for CR and ISRF densities provide a more realistic basis that can be used for the interpretation of the non-thermal interstellar emissions
  from the Galaxy. 
  \end{abstract}


%%%%%%%%%%%%%%%%%%%%%%%%%%%%%%%%%%%%%%%%%%%%%%%%%%%%%%%%%%%%%%%%%%%%

\section{Introduction}
\label{sec:intro}
\setcounter{footnote}{0}

The high-energy \gray{} sky is dominated by the emissions 
produced by cosmic-ray (CR) particles interacting with matter and radiation
fields in the interstellar medium (ISM) of the Milky Way.
Observations of these interstellar emissions began with the OSO-III satellite in the late 1960s \citep{1972ApJ...177..341K}, and were followed by the
space-borne experiments SAS-2 and COS-B in the early- and mid-1970s, COMPTEL and EGRET 
on the {\it Compton Gamma-Ray Observatory (CGRO)} (1990s), and the present-day 
\fermi-LAT \citep{2009ApJ...697.1071A}.
Each of these instruments has represented a significant advance over its
predecessor, with the \fermi-LAT providing the highest-sensitivity
data to date for $\gtrsim30$~MeV energies.

Because \gray{s} are not deflected by magnetic fields, and their absorption
in the ISM is negligible over Galactic distances up to
energies of $\sim 30$~TeV \citep{2006ApJ...640L.155M},
they directly probe CR intensities and spectra in distant locations, far beyond the
comparatively small region accessible by direct CR measurements.
Describing the interstellar emissions using models for the CR propagation 
and interactions in the ISM has been very successful in explaining 
many features of the multi-wavelength diffuse spectrum of the 
Galaxy \citep[for a review see][]{2007ARNPS..57..285S}.

Physical modelling codes such as 
\GP{} \citep{1998ApJ...493..694M,1998ApJ...509..212S,2000Ap&SS.272..247M,2000ApJ...537..763S,2011CoPhC.182.1156V}, 
can reproduce the general features of the interstellar \gray{} emission 
over the whole sky, showing that the CR physics and interactions producing it 
are well-understood.
However, it is the residuals from when interstellar emission models (IEMs) are subtracted from the data 
that provide the potential for identifying new phenomena in high-energy 
\gray{s}.
Their understanding requires a careful assessment of the 
the modelling inputs, in particular those related to the CR source and ISM
densities.

To date the most extensive study of high-energy IEMs 
has been made by the \fermilat\ team \citep{2012ApJ...750....3A} using a 
grid of 128 {\it a-priori} \GP\ models normalised to reproduce 
local CR data.
The grid entries are categorised by 4 CR source spatial density models from the 
literature (supernova remnants, 2 pulsar, and OB-stars), multiple CR 
propagation halo sizes, and other parameters related to the interstellar gas.
These IEMs employ a 2D\footnote{Here and elsewhere throughout this paper when the dimensionality is mentioned it refers to the purely {\it spatial} dimensions. The other axes, such as energy/frequency, are present independent of the spatial dimensionality.} Galactocentric cylindrical symmetry, which
has been the norm since the {\it CGRO}-era due to the limited quality of
the \gray{} data, information on ISM distributions, and computing resources
available.
Examination of the residual maps developed in that work shows $\sim$few to tens of degrees scale 
features that are asymmetric about the meridian defined by Galactic longitude $l = 0^\circ$ and the Galactic plane.
Some of them are likely related to large-scale structure in the 
CR and ISM distributions that are not properly accounted for by the 2D-based IEMs.
Another notable analysis is the 
\fermilat\ team investigation of the \gray{} emission toward the inner 
Galaxy \citep[][ hereafter IG16]{2016ApJ...819...44A}, which
uses a subset of the 128--model grid as baseline IEMs to develop estimates for the fore/background from the Galaxy, and enable 
extraction of the \gray{} emission from within $\sim1$~kpc about the Galactic
centre (GC).
An attempt to compensate for the modelling limitations of the
2D IEMs is made in that work by introducing new degrees of freedom for
the inverse Compton (IC) component and fitting the \gray{} emission outside
of a $15^\circ \times 15^\circ$ region about the GC to estimate the 
fore/background.
An interesting result found from the application of the procedure developed by IG16 is that for Galactocentric radii $R\sim3-8$~kpc the baseline IEM predictions need to be scaled upward by $\sim 20-30$\% to account for the positive residuals for Galactic longitudes $15^\circ \lesssim |l| \lesssim 80^\circ$.
In addition, they found that the IC emission within $R\sim 1-2$~kpc of the GC is strongly dominant compared to that produced by CRs interacting with the gas there.
These somewhat puzzling results are difficult to further interpret using the 2D-based IEMs.
The larger parameter space afforded by 3D models is a logical next step
in the evolution of physics-based modelling of the high-energy interstellar
emissions from the Galaxy.

The \GP{} code has been capable of 3D CR propagation calculations since the beginning \citep{1998ApJ...509..212S,2001AIPC..587..533S,2001ICRC....5.1964S} but this mode has had
limited usage because of the available data quality and computing resources necessary.
Even the current data and theory does not provide for a complete 3D model of the ISM to be built.
Thus, studying the effects of 3D structures on various observables is still at its early stages.

It was recognised early on that the 3D spatial density distribution of the CR sources within $\sim 1$~kpc
of the Solar system is particularly important for modelling the CR electron/positron data \citep{1970ApJ...162L.181S,1971ApL.....9..169S}, which has been
subsequently expanded on with time-dependent calculations
\citep[e.g.,][]{1995PhRvD..52.3265A,2003ICRC....4.1989S,2004ApJ...601..340K}.
At larger scales, spiral arm models for the CR source density have been considered using analytic methods \citep[e.g.,][]{2003NewA....8...39S} and with 3D numerical calculations employing toy-model set ups \citep[e.g.,][]{2009PhRvL.103k1302S,2013PhRvL.111b1102G,2014ApJ...782...34B,2015APh....70...39K}, but their focus has been on the CR particle spatial distributions in the ISM while consequences for the associated diffuse emissions were omitted.

Only relatively recently has 3D numerical modelling been made of the high-energy interstellar emissions from the Galaxy: \citet{Johannesson:2013qqi,Johannesson:2015qqi} using \GP{}, \citet{2017arXiv170107285K} and \citet{2017arXiv170107288N} with the {\it PICARD} code, and by \citet{2017MNRAS.466.3674N} with a Monte Carlo code.
The \GP{} calculations use both 3D CR source and gas density models, while the {\it PICARD}-based and \citet{2017MNRAS.466.3674N} works employ a spiral arm model for the CR source densities but 2D models for the ISM densities.
So far 3D interstellar radiation field (ISRF) models have not been 
used, which are necessary for comprehensive treatment of the CR
electrons and positrons and their high-energy emissions.

In this paper, a study is made of the high-energy interstellar 
emissions calculated using 3D models for the CR source and ISRF densities 
in the Galaxy.
The new release of the \GP\ code is employed for the CR propagation and 
high-energy interstellar emission calculations\footnote{The data products and configuration files for reproducing the calculations made in this paper will be made available at the \GP\ website (http://galprop.stanford.edu).}. 
Three CR source density models are considered that have the injected CR power
distributed as 
an axisymmetric smooth disc, 50/50\% smooth disc/spiral arms, or 100\% spiral
arms.
This work also uses the
{\it F}ast {\it Ra}diation {\it N}umerical {\it K}alculation of 
{\it I}nterstellar {\it E}mission ({\it FRaNKIE}) code \citep[][]{2008ApJ...682..400P,Porter:2015fyg} to calculate two ISRF Galaxy-wide spectral intensity distributions based on self-consistently derived 
stellar and dust density models taken from the literature that 
reproduce local near- to far-infrared observations.
It will be shown that the combination of 3D CR source and ISRF density models
produces an interstellar emission intensity at Earth that is more structured than the 2D case.
Features related to the non-axisymmetric structures included in the CR source and ISRF density models can be identified in the interstellar emission residual maps, which can provide useful information on the CR and ISRF densities far beyond where local CR data are able to probe.
It will also be shown that, with the addition of a population of CR sources distributed according to the stellar bulge/bar spatial densities from either of the ISRF models, the models with spiral arms provide a plausible physical explanation for the puzzling results from the analysis of high-energy \gray{} toward the inner Galaxy obtained by the \fermilat\ team (IG16).

\section{Models}
\label{sec:models}

\subsection{Cosmic Ray Propagation (\GP{} code)} 
\label{sec:galprop}

Theoretical understanding of CR propagation in the ISM is the 
framework that the \GP{} code is built around.
The key idea is that all CR-related data, including direct measurements, \gray{s}, sychrotron radiation, etc., are subject to the same physics and must therefore be modelled simultaneously.
The \GP{} project now has nearly 20 years of development behind it.
The original FORTRAN90 code has been public since 1998, and a rewritten C++ version was produced in 2001.
The last major public source code release was v54 \citep{2011CoPhC.182.1156V}.
A new version of the \GP{} code is made available with this paper following improvements made over a number of years \citep{Moskalenko:2015ptr,GalBayesII}
with the new features summarised in Appendix~\ref{app:gpv55}.
The latest releases are always available at the dedicated website, which also provides the facility to run \GP{} via a web browser interface\footnote{http://galprop.stanford.edu/webrun.}. 
The website also contains links to all \GP{}--team publications together with detailed information on CR
propagation and the different versions of the code and supporting data sets.
Below a brief review of the CR production and propagation relevant to the
present paper is made. 
Further details are given by \citet{2011ApJ...729..106T} and
more information can be found in the review by \citet{2007ARNPS..57..285S}.

\GP{} numerically solves the system of partial 
differential equations describing the particle transport 
with a given source distribution and boundary conditions for all species of CRs.
Propagation is described using the diffusion-reacceleration equation, which has proven to be remarkably successful at modelling transport processes in the ISM. 
The processes involved include diffusive reacceleration and, for nuclei, nuclear spallation, secondary particle production, radioactive decay, electron capture and stripping, electron knock-on, and electron K-capture, in addition to energy loss from ionisation and Coulomb interactions. 
For CR electrons and positrons, important processes are the energy losses due to ionisation, 
Coulomb scattering, Bremsstrahlung (with
the neutral and ionised gas), IC scattering, and 
synchrotron emission.

Galactic properties on large scales, including the diffusion coefficient, halo size,
Alfv\'en  velocity and/or advection velocity, as well as
the mechanisms and sites of CR acceleration, can be probed by measuring stable and radioactive secondary CR nuclei. 
The ratio of the halo size to the diffusion coefficient can be constrained by measuring the abundance of stable secondaries such as B. 
Radioactive isotopes ($^{10}$Be, $^{26}$Al, $^{36}$Cl, $^{54}$Mn) then allow the resulting degeneracy to be lifted \citep[e.g.,][]{1998A&A...337..859P,1998ApJ...509..212S,1998ApJ...506..335W,2001ICRC....5.1836M}.
However, the interpretation of the peaks observed in the 
secondary-to-primary  ratios (e.g., B/C, [Sc+Ti+V]/Fe) 
around energies of a few GeV/nucleon, remains model-dependent.

CR propagation in the heliosphere is described by the \citet{1965P&SS...13....9P} equation.
The modulated fluxes significantly differ from the interstellar spectra below energies of $\sim$20-50 GeV/nucleon, but correspond to the ones actually measured by balloon-borne and spacecraft instruments.
Spatial diffusion, convection with the solar wind, drifts, 
and adiabatic cooling are the main mechanisms that determine transport of CRs to the inner heliosphere. 
These effects have been incorporated into realistic (time-dependent, three-dimensional) models \citep[e.g.,][]{2003JGRA..108.1228F,2006ApJ...640.1119L,2004AnGeo..22.3729P,2017arXiv170406337B}.
The ``force-field'' approximation that is ordinarily used \citep{1968ApJ...154.1011G}, instead characterises the modulation effect as it varies over the solar cycle using a single 
parameter -- the ``modulation potential''.
Despite having no predictive power, the force-field approximation is a useful low-energy parameterisation
of the modulated spectrum for a given interstellar spectrum. 

The focus in this work is on the CR induced \gray{} emission from
different realisations of the CR source and ISRF distributions.
The propagation model is therefore limited to diffusive re-acceleration with an isotropic and
homogeneous diffusion coefficient that has a power-law dependence with rigidity.
The CR injection spectra are modelled as rigidity-dependent 
broken power laws with parameters derived as by a recent study \citep{GalBayesII}.
Electrons, protons, and He each have two breaks 
while elements with $Z>2$ are modelled with a single break.
The extra break
for the low-mass elements is to model the spectral change observed at
rigidities above 100~GV \citep{2011Sci...332...69A,PRL:1141103, PRL:1131102}.
The CR source density distribution is a mixture of a disc and 
4 spiral arms that have the same exponential scale-height (200~pc) perpendicular to the Galactic plane.
The smooth disc spatial density follows the radial distribution of pulsars
as given by \citet{2004A&A...422..545Y}.
The spiral arm spatial density is that of the 4 major arms in the R12 model (Sec.~\ref{sec:modelr12}) and assumes an identical injection of CR power by each arm.
Three such models are considered with a 100\% (2D) disc, 
50\% disc/50\% arm, and 100\% spiral arm contribution.
These are termed SA0, SA50, and SA100, following the proportion of injected CR luminosity by the spiral arms.
The normalisation for the injected CR power of each source density model is obtained by 
requiring the propagated CR intensities agree with the local CR observations, where all calculations made in this
paper use
the IAU recommended Sun-GC distance of $R_S = 8.5$~kpc~\citep{1986MNRAS.221.1023K}.

\subsection{Interstellar Radiation Field}
\label{sec:isrfmodels}

The ISRF is the result of emission
by stars, and the scattering, absorption, and re-emission of absorbed
starlight by dust in the ISM.
Early models for the ISRF were
motivated to enable predictions of the Galactic IC emission at MeV \gray{}
energies \citep{1974csgr.conf..229C,1976ApJ...208..893S,1976A&A....52...69P,1977A&A....59..233B}.
\citet{1983A&A...128..212M} (MMP83) presented calculations of the ISRF from UV to
far-infrared wavelengths at a few locations with a model that has been widely used.
\citet{1985A&A...145..391B} extended the MMP83 work to study the IC emission in the context
of COS-B data, but noted the model shortcomings toward the inner Galaxy
because of the artificial cut-off for $R \lesssim 3$~kpc in the stellar spatial
distribution.
\citet{1991JPhG...17..987C} augmented the MMP83 stellar
model with the dust emission results described by \citet{1986A&A...155..380C}
and re-calculated the spatial distribution of the ISRF energy density.
This work was used by the EGRET team as input to their
modelling of the high-energy interstellar \gray{} emission
\citep{1993ApJ...416..587B,1997ApJ...481..205H}.
\citet{2000ApJ...537..763S} employed stellar population models \citep{1992ApJS...83..111W} and dust emissivities derived
from {\it COBE}/DIRBE data \citep{1997ApJ...480..173S} to re-evaluate
the ISRF energy density distribution.

Despite the increasing sophistication of the inputs these early calculations
were not self-consistent because they did not
couple the starlight absorption by dust with re-emission.
\citet{2005ICRC....4...77P} describe the first work
to combine the stellar emission with a self-consistent dust 
scattering/absorption/re-emission radiation transfer 
calculation using the
\frankie\ code.
\citet{2008ApJ...682..400P} extended the calculations to 
produce also the spatially varying anisotropic intensity of the ISRF using
a 2D stellar and dust density distribution.
The \fermi-LAT team~\citep{2012ApJ...750....3A} used a revision of the \citet{2008ApJ...682..400P} work for the study of high-energy \gray{} interstellar emission models calculated using the \GP\ CR propagation code.
The spectral intensity distribution used for the \citet{2012ApJ...750....3A} study represents the latest publicly available model for the ISRF that is distributed from the \GP\ website\footnote{As this paper was being finalised \citet{2017arXiv170506652P} described the results of a calculation for the Galaxy-wide ISRF using an alternative 2D axisymmetric model and radiation transfer code.}.

The Galactic structure is spatially complex with spiral arms \citep[e.g.,][and references therein]{2014ApJS..215....1V,2016AJ....151...55V}, a central region that is dominated by a bulge/bar complex \citep[e.g.,][]{2000MNRAS.317L..45H,2005ApJ...630L.149B,2007A&A...465..825C}, and warped stellar/dust discs \citep[e.g.,][hereafter F98]{1998ApJ...492..495F}.
Sophisticated 3D models have been developed for the spatial densities of stellar populations
\citep[e.g.,][]{1986A&A...157...71R,1992ApJS...83..111W}.
But their emphasis is on calculating counts of stars as observed
from Earth.
Three-dimensional spatial models that include asymmetric elements, such
as a central bulge/bar, spiral arms, and warped stellar/dust discs, have been fit to infrared data \citep[e.g., F98, ][]{2001ApJ...556..181D}
but they obtain only emissivities and do not make radiation transfer
calculations.
The most comprehensive 3D modelling to date \citep[][hereafter R12]{2012A&A...545A..39R} uses the SKY model \citep{1992ApJS...83..111W,1993AJ....105.1860C,1994AJ....107..582C} as input to the
HYPERION code \citep{2011A&A...536A..79R} to self-consistently calculate
predictions for local observed intensities from near- to far-infrared
wavelengths.
R12 produced an optimised version of the SKY model that is 
consistent with {\it Spitzer} and {\it IRAS} data in a narrow band of latitude about the Galactic plane for longitudes 
$-60^\circ \leq l \leq 60^\circ$.
However, the spectral intensity was only evaluated locally for comparison with
the data.

There is no unique description for the Galactic structure even with the
diversity of models so far employed to fit for the different spatial
distributions of stars and dust in the Galaxy.
A hybrid combining different spatial elements and their parameters from
the collection studied in the literature without a full fitting procedure
can lead to inconsistent models because of convariance between the presence
of individual elements and their derived parameters,
particularly for the major components such as the stellar disc, arms,
and bulge.
Because the purpose of this paper is to make the first steps to
incorporating 3D spatial structure into the description of the ISRF
employed by CR propagation and interstellar emissions codes like \GP\ such
an optimisation is not made.
Instead two models for the stellar and dust distributions are chosen from the literature: the R12 and F98 models, whose references are given above.
They employ different spatial densities for both the stars and dust but
produce very similar intensities to the data for near- to far-infrared
wavelengths at the Solar system location (see below).
The spatial differences can be summarised as the former model has an
axisymmetric bulge and smooth disc, and major spiral arms and local arm
segments near the Solar system,
while the latter has a non-axisymmetric bulge and smooth disc, where the details
are given in Sec.~\ref{sec:modelr12} and~\ref{sec:modelf98}.

The \frankie\ code is used to calculate the Galaxy-wide spectral intensity distribution for the R12 and F98 models, which is necessary for the lepton IC energy losses and \gray{} production.
Its solution method for the radiation transport has evolved from the ray-tracing method initially employed by \citet{2005ICRC....4...77P} to using simulation techniques similar to other Monte Carlo radiation transfer codes \citep[e.g.,][]{2001ApJ...551..269G,2008A&A...490..461B,2012A&A...545A..39R,2015A&C.....9...20C} where luminosity
packets are injected according to user-specified
spatial and spectral distributions, and the propagation history of each
packet is traced to produce observables that are recorded at observer locations.
The propagation includes self-consistent scattering, absorption, and re-emission by dust, the optical properties and spatial distributions of which are also user-specified.
A description of the code is provided by \citet{Porter:2015fyg} and recent
developments will be covered elsewhere (Porter~et~al.,~in~prep.).
The essential components for a model `run' (stellar and dust distributions)
are described directly below, and the simulation geometry and other details
are given in Sec.~\ref{sec:rtcalcs}.

\subsubsection{Stellar Luminosity Density}
\label{sec:lummodel}

The input spectral luminosity density follows the SKY model approach
of representing the stellar content of the Galaxy using a table of spectral
types \citep[also similar to earlier work, e.g.,][]{1987PASP...99..453G}.
The table includes early-type stars,
main-sequence stars, asymptotic branch giants and other giants/super-giants,
and exotics.
There are 87 types \citep[see Table~2 of][]{1992ApJS...83..111W}
characterised by a 
local disc number density, fractional contribution across the spatial 
components of the
SKY model (bulge, disc, ring, arms, halo), scale-height perpendicular to
the Galactic plane, and spectral information for photometric filters
from UV to mid-infrared wavelengths.
Number densities of individual types were determined by \citet{1992ApJS...83..111W} by adjusting them to obtain a luminosity function agreeing with data at visual and infrared wavelengths.
The relative contribution of the stellar types across the spatial components
was assigned by \citet{1992ApJS...83..111W} based on physical considerations (see their Sec.~2.2.4 for 
details): the arms and star-forming ring contain
predominantly young stars (mainly spectral types O through A), with the
smooth disc containing all types, and the bulge absenting the young
and hot stars and star-forming regions.

R12 noted that the spectral characteristics of the early-type stars in the
SKY model did not adequately represent the UV portion of the spectrum, which is
important for the heating of the smallest dust particles.
They improved the spectral representation at these wavelengths by 
using photospheric models from \citet{2004astro.ph..5087C} that were 
rescaled to the \citet{1992ApJS...83..111W} absolute magnitudes.
In addition, R12 modified the early-type stellar content of
the spiral arms to further optimise the model, in particular depleting
arms 1 and 1$^\prime$ of stellar types other than O and B, and increasing the
number densities of O- and B-types for arms 2 and 2$^\prime$
(see Sec.~\ref{sec:modelr12} for the meaning of the arm designations).
Both of these modifications are employed in this paper.

\begin{figure}[ht]
  \includegraphics[scale=0.85]{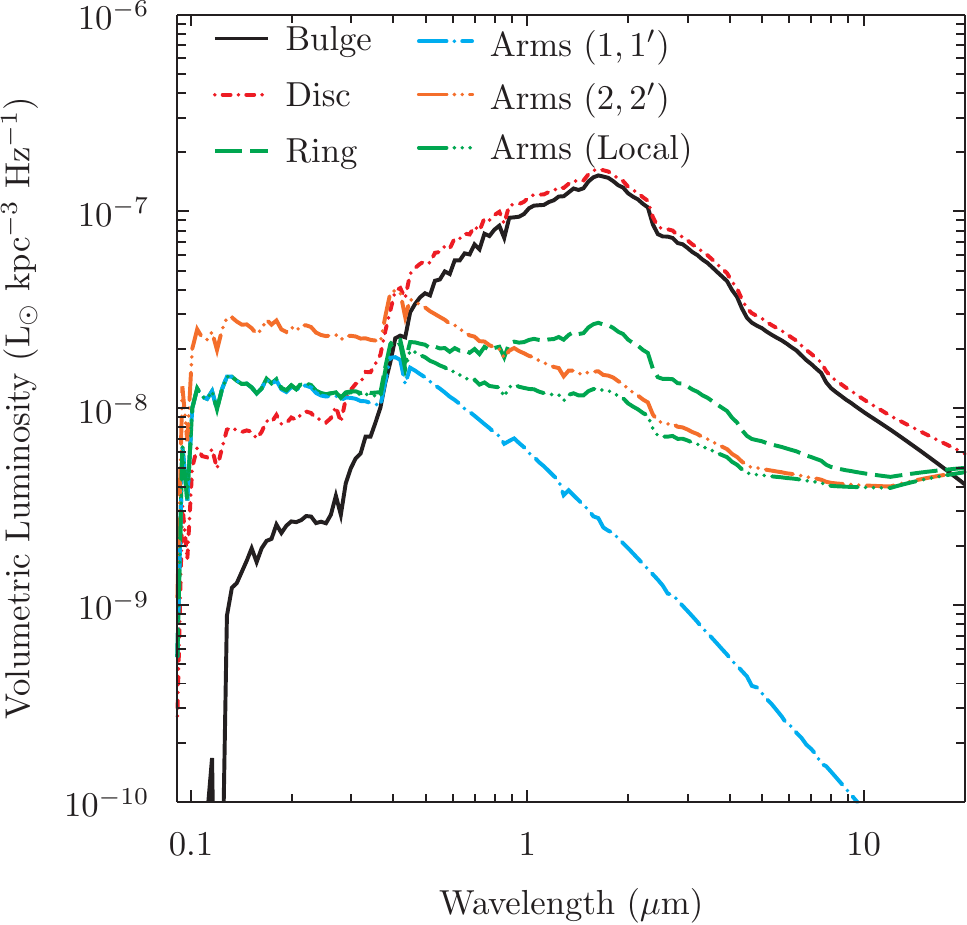}
  \caption{Spectral luminosity density for the major R12 model 
    components.
    Line styles: solid, bulge; short-dash-dot, disc; long-dash, ring; long-dash-dot, arms 1 and $1^\prime$; long-dash-dot-dot, arms 2 and $2^\prime$; long-dash-triple-dot, local arms.
    See Sec.~\ref{sec:modelr12} for the meaning of the arm designations.
    \label{fig:compluminosity}}
\end{figure}

Figure~\ref{fig:compluminosity} shows the spectral luminosity density for the
major components of the R12 model at their normalisation position in the 
plane (GC for the bulge, $R_S$ for the disc and arms, $R_R = 0.45R_S$ for the ring -- see Sec.~\ref{sec:modelr12}).
Each curve is obtained by summing over the 87 type spectra weighted by
their spatial densities for the respective geometrical component.
For the F98 model, the bulge, disc, and local arm luminosity densities shown
in Fig.~\ref{fig:compluminosity} are used for the bar, old disc, and `average'
arm components (see Sec.~\ref{sec:modelf98}).

\subsubsection{Dust Properties}
\label{sec:dustproperties}

The dust composition model of \citet{2007ApJ...657..810D} is used in this
paper.
The model describes a mixture of neutral and ionised polycyclic aromatic hydrocarbon (PAH)
molecules, carbonaceous, and amorphous silicate dust grains based on earlier
work \citep{2001ApJ...554..778L} with adjustments \citep{2003ARA&A..41..241D}
and optimisations to obtain better agreement with {\it Spitzer} data.
Each species is characterised by a size distribution function \citep{2001ApJ...548..296W}.
The dust grains in the model are assumed spherical, with the absorption and scattering cross sections computed for PAH and carbonaceous grains \citep{2001ApJ...554..778L} and 
`smoothed' astronomical silicate \citep{1984ApJ...285...89D,1993ApJ...402..441L,2001ApJ...548..296W} using a Mie code with data taken from B.~T.~Draine's website
\footnote{https://www.astro.princeton.edu/$\sim$draine/dust/dust.diel.html.}.

\subsubsection{R12 Spatial Model}
\label{sec:modelr12}

The R12 spatial model has the same 5 stellar spatial components as the SKY
model: axisymmetric bulge,
exponential disc, star-forming ring and spiral arms (4 major and 2 minor local `spurs'), and halo.
Each stellar spatial component is characterised with a normalising factor,
spatial
density distribution, and spectral luminosity density (Sec.~\ref{sec:lummodel}).
R12 retain the same global normalising factors for the bulge, ring, arms, and halo ($\rho_B,\rho_R,\rho_A,\rho_H$) as determined by \citet{1992ApJS...83..111W}.
The R12 optimisations made compared to the SKY model are for the
geometric parameters and stellar content of the spiral arms and dust density distribution to better describe {\it Spitzer} and {\it IRAS} data.

The bulge stellar density has the form

\begin{equation}
  \rho_{\rm bulge} (R,Z) = \rho_B \sum_i \rho_{\rm bulge,i} S^{-1.8} e^{-S^3}
  \label{r12:bulgedensity}
\end{equation}

\noindent
with $\rho_B = 3.6$ and

\begin{equation}
  S = \frac{\sqrt{R^2 + K_1 ^2 Z^2}}{R_1}
  \label{r12:bulgecoord}
\end{equation}

\noindent
where the bulge axis ratio and scale-radius are $K_1 = 1.6$
and $R_1 = 2$~kpc, respectively.
The sum is over the individual space densities for the $i=1,$~...~$,87$ stellar types used for the spectral luminosity model (Sec.~\ref{sec:lummodel}).
  
The stellar disc has the functional form

\begin{equation}
  \rho_{\rm disc} (R,Z) = \sum_i \rho_{{\rm disc},i} e^{-(R-R_S)/D_R} e^{-|Z|/H_{Z,i}}
  \label{r12:discdensity}
\end{equation}

\noindent
where the radial scale-length is $D_R = 3.5$~kpc, $\rho_{{\rm disc},i}$ and
$H_{Z,i}$ are the space density and scale-height perpendicular to the
Galactic plane, respectively, of the $i$-th stellar type.

The star-forming ring stellar density is Gaussian in $R$ centred on $R_R = 0.45 R_S$ with
half-width $\sigma_R = 0.064 R_S$

\begin{equation}
  \rho_{\rm ring} (R,Z) = \rho_R \sum_i \rho_{{\rm ring},i} e^{-(R-R_R)^2/2\sigma_R^2} e^{-|Z|/H_{Z,i}} 
  \label{r12:ringdensity}
\end{equation}

\noindent
where $\rho_R = 25$ and the sum is again over stellar types
with the scale-heights $H_{Z,i}$ as for Eq.~\ref{r12:discdensity}.

The spiral arm component has 4 main arms and 2 local (to the Solar system) `spurs'. 
In the original SKY model the 4 main logarithmic
arms have equal weight with a single local arm segment.
\citet{1994AJ....107..582C} split the local segment into two smaller spurs.
The R12 arms are optimised by adjusting the start positions and changing
the density profile perpendicular to the arm axis (from a step function for
the SKY model to a Gaussian), and with twice larger
in-arm densities for the O and B spectral types for arms 2 and $2^\prime$ (Table~\ref{table:r12armparams}), and eliminating all but the O and B spectral types
from arms 1 and $1^\prime$.
R12 identify arms 1 and 1$^\prime$ as the Norma and Sagittarius arms, and
arms 2 and $2^\prime$ as the Scutum-Centaurus and Perseus arms, respectively.

The functional form for the arm shape is

\begin{equation}
  \phi_j(R) = \alpha_j \log \left( \frac{R}{R_{{\rm min},j}} \right) + \phi_{{\rm min},j}
  \label{r12:armwinding}
\end{equation}

\noindent
where $\alpha_j$, $R_{{\rm min},j}$, and $\phi_{{\rm min},j}$ are parameters that are given in Table~\ref{table:r12armparams}.
The major arms have angular extent 6 radians starting from their minimum radii, while the local spurs are considerably shorter (0.55 radians). 
The arms follow the radial and vertical profiles as the stellar disc (Eq.~\ref{r12:discdensity}) with the sum over stellar types including the fractional
densities as the other components, and the global
normalisation pre-factor $\rho_A = 5$.
For the main arms ($j = 1, 1^\prime, 2, 2^\prime$) the Gaussian profile
perpendicular to the arm axis has $\sigma_j = 0.55$~kpc.
The local spurs are modelled as a step function following the original
SKY model with density given by Eq.~\ref{r12:discdensity} inside and zero elsewhere.

\begin{deluxetable}{llccccc}[tb]
\tablecolumns{7}
\tablecaption{R12 spiral arm parameters.
\label{table:r12armparams}}
\tablehead{
\colhead{Arm} &
\colhead{$\alpha$} &  
\colhead{R$_{\rm min}$} &
\colhead{$\phi_{\rm min}$} &
\colhead{Extent} &
\colhead{$\sigma$} &
\colhead{Width}\\
& & (kpc) & (rad) & (rad) & (kpc) & (kpc)
}
\startdata
1 & 4.18 & 3.800 & 0.234 & 6.00 & 0.55 & \\
1' & 4.18 & 3.800 & 3.376 & 6.00 & 0.55 & \\
2 & 4.19 & 4.500 & 5.425 & 6.00 & 0.55 & \\
2' & 4.19 & 4.500 & 2.283 & 6.00 & 0.55 & \\
L & 4.57 & 8.100 & 5.847 & 0.55 & & 0.30 \\
L' & 4.57 & 7.591 & 5.847 & 0.55 & & 0.30
\enddata
\end{deluxetable}

The fifth component is the ellipsoidal stellar halo, but its contribution 
is minor and so is omitted from the calculations.

The dust spatial density for the R12 model is an exponential disc with a central hole:

\begin{equation}
  \rho_{\rm dust} (R,Z) = \rho_{\rm dust,0} e^{-\frac{|Z|}{H_{\rm dust}}}
  \times
  \begin{cases}
    f_0 e^{-\frac{(R - \mu_0)^2}{2\sigma_0^2}}, \,\,\,\mbox{$R < R_{\rm smooth}$}\\
    e^{-\frac{R}{D_R}}, \,\,\,\mbox{$R \geq R_{\rm smooth}$}
  \end{cases}
  \label{r12:dustdensity}
\end{equation}

\noindent
where $\rho_{\rm dust} = 10^{-25}$~g~cm$^{-3}$, $H_{\rm dust} = 100$~pc, and
$R_{\rm smooth}$, $f_0$ ensure a smooth transition between the
exponential radial profile and the inner hole:

\begin{equation}
  R_{\rm smooth} = \sigma_0^2/D_R + \mu_0,
  \label{r12:dustrsmooth}
\end{equation}

\begin{equation}
  f_0 = \frac{e^{-R_{\rm smooth}/D_R}}{e^{-(R_{\rm smooth} - \mu_0)^2/2\sigma_0^2}}.
  \label{r12:dustf0}
\end{equation}

\noindent
R12 tried various $\mu_0$ and $\sigma_0$, finding that
$\mu_0 = 4.5$~kpc and $\sigma_0 = 1$~kpc provide
adequate agreement for the far-infrared data.
These values are used in this paper.

\subsubsection{F98 Spatial Model}
\label{sec:modelf98}

The F98 model has a non-axisymmetric stellar bulge
and exponential discs for the stellar and dust distributions.
F98 determined the parameters of the spatial components 
by fitting to {\it COBE}/DIRBE near-infrared ($1.25-4.9$~$\mu$m) data.
A separate emissivity normalisation was obtained from the fits for the bar and disc stellar and dust distributions for each of the 4 DIRBE wavebands used in the study, but the spatial
parameters (e.g., stellar or dust disc radial scale-length) remain the same 
across all wavebands.
F98 evaluated different models based on the bar radial spatial profile: Model S ($\sech^2$), Model E (exponential-to-power), and Model P (power-law-with-core).
The best-fit `Model S' for $R_S = 8.5$~kpc is used in this paper.

The central region of the Galaxy in this model is dominated by an ellipsoidal bar.
Its spatial distribution is given by

\begin{equation}
  \rho_{\rm bar} (R,\phi,Z) = \rho_{\rm bar,0} \sech^2 (R_b)  \\
  \times
  \begin{cases}
    1, \,\,\,\mbox{$R \leq R_{\rm end}$}\\
    e^{-[(R_b - R_{\rm end})/H_{\rm end}]^2}, \,\,\,\mbox{$R > R_{\rm end}$}
  \end{cases}
  \label{f98:baremissivity}
\end{equation}

\noindent
where $\rho_{\rm bar,0}$ is the normalisation, $R_b$ is the bar radial coordinate:

\begin{eqnarray}
  R_b^{C_{||}} & = & R_\perp ^{C_{||}} + \left( |Z^\prime|/A_Z \right)^{C_{||}}\\
  R_\perp ^{C_\perp} & = & \left(|X^\prime|/A_X\right)^{C_\perp} + \left(|Y^\prime|/A_Y\right)^{C_\perp}
    \label{f98:barradius}
\end{eqnarray}

\noindent
with $X^\prime$, $Y^\prime$, and $Z^\prime$ evaluated in the bar frame,
which is described by body-centred axes with a rotation\footnote{For the F98 coordinate system this rotation is clockwise. In this paper the coordinate system employed makes this an anti-clockwise rotation.} with
respect to the Sun-GC line of $\phi_{\rm bar}$ and pitch angle $\beta_{\rm bar}$
where the latter is the angle between the bar major axis and the Galactic
plane, and $A_X, A_Y, $ and $A_Z$ are the bar axis scale-lengths.
  
The stellar and dust discs are exponential in $R$ with an inner and outer
truncation, with a $\sech^2$ distribution in $Z$ and
$R$-dependent warping.
Their spatial distribution is given by the expression

\begin{equation}
  \rho_{\rm disc} (R,Z) = \rho_{\rm disc,0} H(R,\phi) e^{-R/H_R} \sech^2 \left( \frac{Z - \overline{Z}}{H_Z} \right)
  \label{f98:discemissivity}
\end{equation}

\noindent
where $\rho_{\rm disc,0}$ is the disc normalisation, $H(R,\phi)$ is the hole function describing the inner radial truncation:

\begin{equation}
  H(R,\phi) = 1 - e^{-(R_H/D_H)^{\alpha_H}}
  \label{f98:holefn}
\end{equation}

\noindent
with $D_H$ the hole scale radius,
$R_H ^2 = {X^\prime}^2 + (\epsilon Y^\prime)^2$ evaluated in bar centred
coordinates with $X^\prime$ the bar major axis and $\epsilon$ the hole
eccentricity, $H_R$ is the disc radial scale-length, $H_Z$ is the disc
scale-height, and $\overline{Z} \equiv \overline{Z}(R,\phi)$ models the warp
so that $\overline{Z} = 0$ for $u = R - R_w  \leq 0$ while for $u > 0$

\begin{equation}
  \overline{Z}(R,\phi) = A (c_1 u + c_2 u^2 + c_3 u^3) \sin(\phi - \phi_w)
  \label{f98:zwarpscale}
\end{equation}

\noindent
with a straight line of nodes at azimuth $\phi_w$.
For the `old' stellar disc $A = 1$ while it is a fit parameter for the dust
disc.
The stellar and dust discs are smoothly truncated for $R > R_{\rm max}$ by
setting $H_R = 0.5$~kpc in Eq.~\ref{f98:discemissivity}.

Table~4 (column~3) from F98 and Fig.~12 from the same work (for the disc warp polynomial coefficients)
give the values for the various spatial parameters described above, and they are used in this paper.
The emissivities for the old stellar disc and bar from F98 are given only for the 4 DIRBE wavebands used in that analysis.
For the \frankie\ calculations the spectral luminosity density from UV to longer wavelengths is needed. 
To set these for the old stellar disc and bar the disc and bulge spectral luminosity densities from Fig.~\ref{fig:compluminosity} are folded with the DIRBE
spectral responses\footnote{See https://lambda.gsfc.nasa.gov/product/cobe/c\_spectral\_res.cfm} and scaled to the F98 emissivities to provide the correct normalisation.
On the other hand the emissivities for the dust disc over the DIRBE wavebands
determined by F98 represent a mixture contributions by stars, scattering, and stochastic heating of very small dust grains by early-type stars.
However, the emissivity of the latter is not linearly related to the input luminosity of the young stars that are acting as heating sources and so it is not possible to obtain a simple rescaling to determine the normalisation for the spectral luminosity density as done for the old stellar disc and bar.
To account for the contributions by young stars the spatial distribution given by F98 for the heating sources for the dust disc is used: it is a warped exponential disc with the same
radial scale-length as the old disc, but with a scale-height/warp that follows the dust disc.
The original SKY model spiral arm spectral luminosity density
(`local' arm in Fig.~\ref{fig:compluminosity}) is taken for its input luminosity spectrum, because this represents an average spectrum that does not have any of the major arm specific optimisations introduced  by R12.
The young star input luminosity for the F98 model is then obtained using the \frankie\ calculations (Sec.~\ref{sec:rtcalcs}) by adjusting its normalisation so that the predicted 3.5/4.9~$\mu$m profiles, which also include the old disc and bar contributions, agree with the data.
The dust spatial distribution employs the F98 dust disc assuming the fitted parameters for the scale-lengths/heights, hole, and warp from that paper.
Its normalisation is taken to be the same as the R12 dust density at $R_S = 8.5$~kpc, corresponding to a central normalisation of $\rho_{\rm dust} = 1.5 \times 10^{-25}$~g~cm$^{-3}$ using the dust properties model described in Sec.~\ref{sec:dustproperties}\footnote{The convention of F98 is followed here: the central normalisations are given assuming the absence of a hole in the spatial distribution.}.

\section{Calculations}

\subsection{Cosmic Rays}
\label{sec:crcalc}

\begin{deluxetable}{lcl}
\tablecolumns{3}
\tablewidth{0pc}
\tablecaption{Data used for determining the propagation parameters
   \label{tab:CRdata}}
\tablehead{
\colhead{Instrument} &
\colhead{Isotopes} &
\colhead{Reference\tablenotemark{a}}
}
\startdata
AMS-02 (2011-2016) & B/C & I \\
AMS-02 (2011-2013) & e$^-$ & II \\
AMS-02 (2011-2013) & H & III \\
AMS-02 (2011-2013) & He & IV \\
HEAO3-C2 (1979-1980) & B, C, O, Ne, Mg, Si & V \\
Voyager-1 (2012-2015) & H, He, B, C, O, Ne, Mg, Si & VI \\
PAMELA (2006-2008) & B, C & VII \\
\enddata
\tablenotetext{a}{I: \citet{PRL:1171102}, II: \citet{PRL:1131102}, III:
\citet{PRL:1141103}, IV: \citet{PRL:1151101}, V: \citet{Engelmann:1990}, VI:
\citet{Cummings:2016}, VII: \citet{Adriani:2014}}
\end{deluxetable}

The generation of secondary CRs depends on the amount of matter
traversed as their primary counterparts propagate through the ISM from their sources.
Thus, the secondary intensities and spectra in the Galaxy depend on the details of the
assumed propagation model, as well as the spatial distributions of the
CR source and interstellar gas densities.
The parameters of the CR propagation therefore need to be tuned dependently on the CR source spatial density and gas density\footnote{The 2D gas model that has been standard in \GP\ for many years is used in this paper \citep{1998ApJ...509..212S,2002ApJ...565..280M}. The effect of 3D gas models will be considered separately (J\'{o}hannesson~et~al.,~in~prep.).} models for a self-consistent
description when reproducing observations of CR secondary species,
such as B.

\begin{deluxetable}{lccc}[htb]
\tablecolumns{4}
\tablewidth{0pc}
\tablecaption{Final model parameters. \label{tab:CRparameters} }
\tablehead{
\colhead{Parameter} & 
\colhead{SA0} & 
\colhead{SA50} &
\colhead{SA100}  
}
\startdata
$D_{0,xx}$ [$10^{28}$ cm$^2$s$^{-1}$]\tablenotemark{a} & $4.37$            & $4.47$            & $4.71$            \\
$\delta$\tablenotemark{a}                              & $0.494$           & $0.508$           & $0.483$           \\
$v_{A}$ [km s$^{-1}$]                                  & $7.64$            & $9.19$            & $7.34$            \\
$\gamma_0$\tablenotemark{b}                            & $1.47$            & $1.61$            & $1.66$            \\
$\gamma_1$\tablenotemark{b}                            & $2.366$           & $2.350$           & $2.381$           \\
%$\gamma_2$\tablenotemark{b}                            & $2.66  \pm 0.03$  & $2.64  \pm 0.03$  & \nodata           \\
$R_1$ [GV]\tablenotemark{b}                            & $3.64$            & $3.92$            & $4.12$            \\
%$R_2$ [GV]\tablenotemark{b}                            & $18.7 \pm 0.6$    & $18.9 \pm 0.1$    & $\infty$          \\
$\gamma_{0,H}$\tablenotemark{b}                        & $1.74$            & $1.78$            & $1.74$            \\
$\gamma_{1,H}$\tablenotemark{b}                        & $2.350$           & $2.342$           & $2.351$           \\
$\gamma_{2,H}$\tablenotemark{b}                        & $2.178$           & $2.206$           & $2.207$           \\
$R_{1,H}$ [GV]\tablenotemark{b}                        & $5.78$            & $6.18$            & $5.62$            \\
$R_{2,H}$ [GV]\tablenotemark{b}                        & $304$             & $226$             & $332$             \\
$\Delta_{He}$                                          & $0.026$           & $0.018$           & $0.017$           \\
$\gamma_{0,e}$\tablenotemark{b}                        & $1.63$            & $1.66$            & $1.61$            \\
$\gamma_{1,e}$\tablenotemark{b}                        & $2.765$           & $2.756$           & $2.756$           \\
$\gamma_{2,e}$\tablenotemark{b}                        & $2.378$           & $2.332$           & $2.329$           \\
$R_{1,e}$ [GV]\tablenotemark{b}                        & $5.95$            & $6.18$            & $6.06$            \\
$R_{2,e}$ [GV]\tablenotemark{b}                        & $103$             & $120$             & $100$             \\
$J_p$ [$10^{-9}$ cm$^{-2}$ s$^{-1}$ sr$^{-1}$ MeV$^{-1}$]\tablenotemark{c} & $4.598$ & $4.562$ & $4.599$           \\
$J_e$ [$10^{-11}$ cm$^{-2}$ s$^{-1}$ sr$^{-1}$ MeV$^{-1}$]\tablenotemark{c} & $1.221$ & $1.209$ & $1.250$           \\
$q_{0,^{4}He}/q_{0,p}\times10^{-6}$\tablenotemark{d}           & $93892$           & $94157$           & $93416$           \\
$q_{0,^{12}C}/q_{0,p}\times10^{-6}$\tablenotemark{d}           & $ 2882$           & $ 2867$           & $ 2746$           \\
$q_{0,^{16}O}/q_{0,p}\times10^{-6}$\tablenotemark{d}           & $ 3780$           & $ 3873$           & $ 3645$           \\
$q_{0,^{20}Ne}/q_{0,p}\times10^{-6}$\tablenotemark{d}          & $  356$           & $  358$           & $  333$           \\
$q_{0,^{24}Mg}/q_{0,p}\times10^{-6}$\tablenotemark{d}          & $  644$           & $  654$           & $  609$           \\
$q_{0,^{28}Si}/q_{0,p}\times10^{-6}$\tablenotemark{d}          & $  742$           & $  762$           & $  718$           \\
$\Phi_{HEAO3-C2}$ [MV]\tablenotemark{e}                & $ 857$            & $ 849$            & $ 827$            \\
$\Phi_{PAMELA}$ [MV]\tablenotemark{e}                  & $ 578$            & $ 578$            & $ 572$            \\
$\Phi_{AMS}$ [MV]\tablenotemark{e}                     & $ 610$            & $ 640$            & $ 594$            \\
\enddata
\tablenotetext{a}{$D(R) \propto \beta R^{\delta}$.  $D(R)$ is normalised to $D_0$ at 4~GV.}
\tablenotetext{b}{The injection spectrum is parameterised as $q(R) \propto R^{\gamma_0}$ for $R < R_1$, $q(R) \propto R^{\gamma_1}$ for $R_1 < R < R_2$, and $q(r) \propto R^{\gamma_2}$ for $R > R_2$.  The spectral shape of the injection spectrum is the same for all species except H.}
\tablenotetext{c}{The proton and e$^-$ flux is normalised at the Solar
location at a kinetic energy of 100~GeV.}
\tablenotetext{d}{The injection spectra for isotopes are adjusted as a ratio
of the proton injection spectrum at 100~GeV.  The isotopes not listed here
have the same value as found in \citet{GalBayesII}.}
\tablenotetext{e}{The force-field approximation is used for the solar modulation of CRs. The modulation potential is assumed to be dependent only on the time and location of the observations.}
\end{deluxetable}

\begin{figure*}[tb!]
\center{
\includegraphics[width=0.49\textwidth]{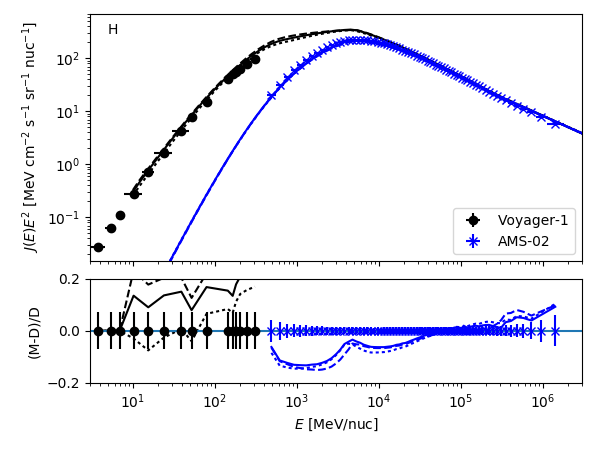}
\includegraphics[width=0.49\textwidth]{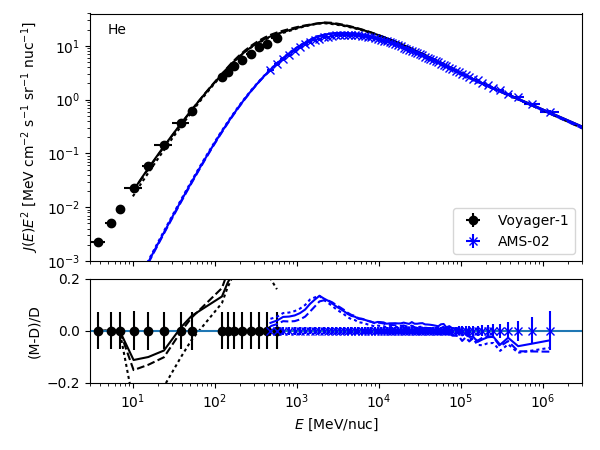}\\
\includegraphics[width=0.49\textwidth]{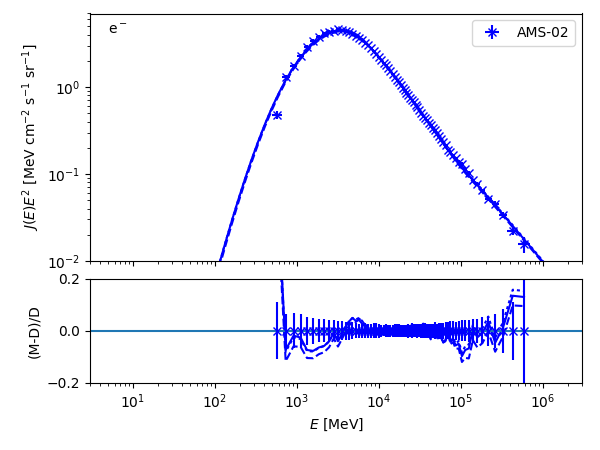}
\includegraphics[width=0.49\textwidth]{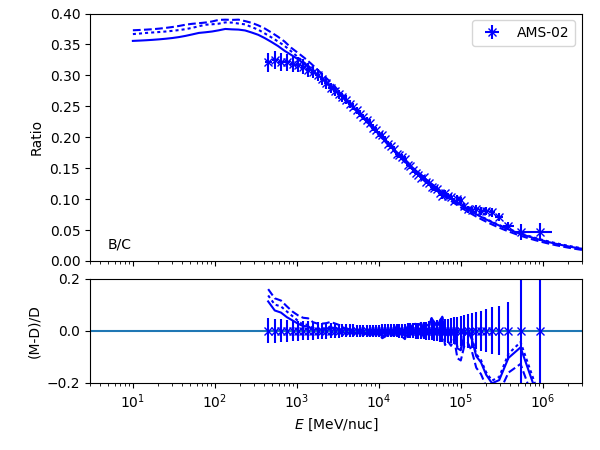}\\
\includegraphics[width=0.49\textwidth]{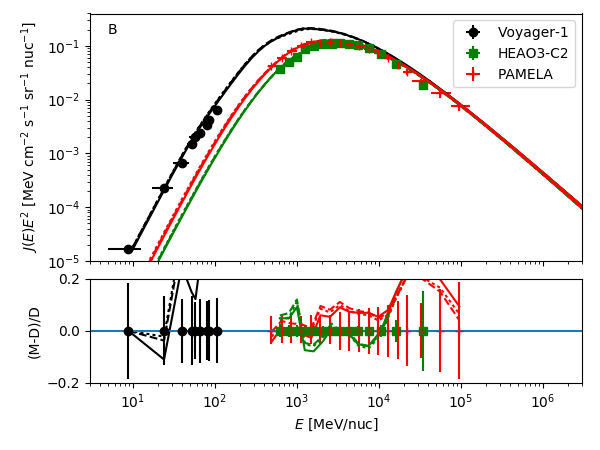}
\includegraphics[width=0.49\textwidth]{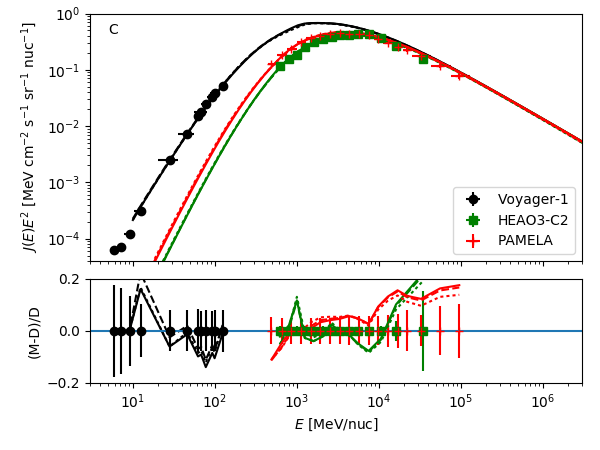}
}
\caption{CR data with CR source density model SA0 (solid curve), SA50 
(dotted curve), and SA100 (dashed curve)
overlaid.  Protons (top left), He (top right), e$^-$ (centre left), 
B/C (centre right), B (bottom left), and C (bottom right).}
\label{fig:Spectra1}
\end{figure*}

\begin{figure*}[tb!]
  \subfigure{
    \includegraphics[scale=0.55]{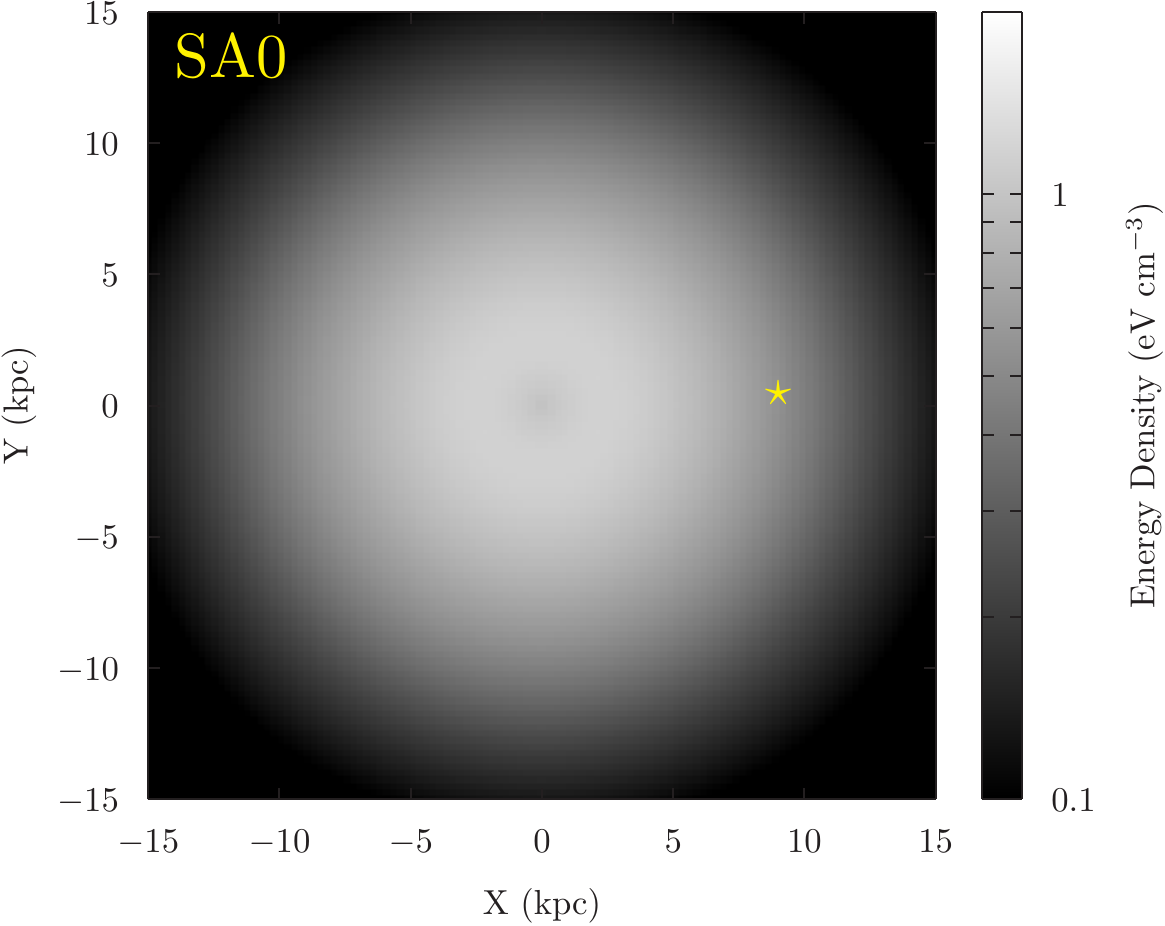}
    \includegraphics[scale=0.55]{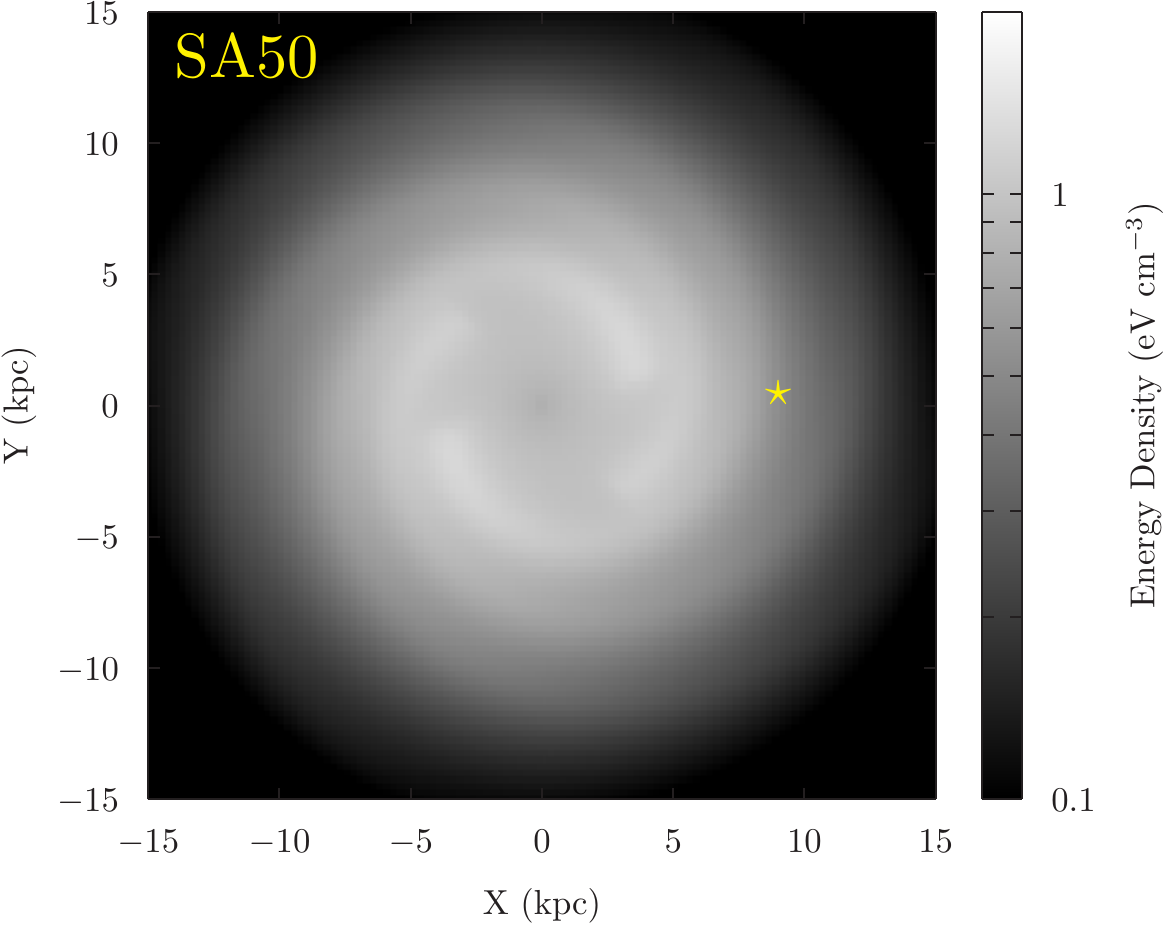}
    \includegraphics[scale=0.55]{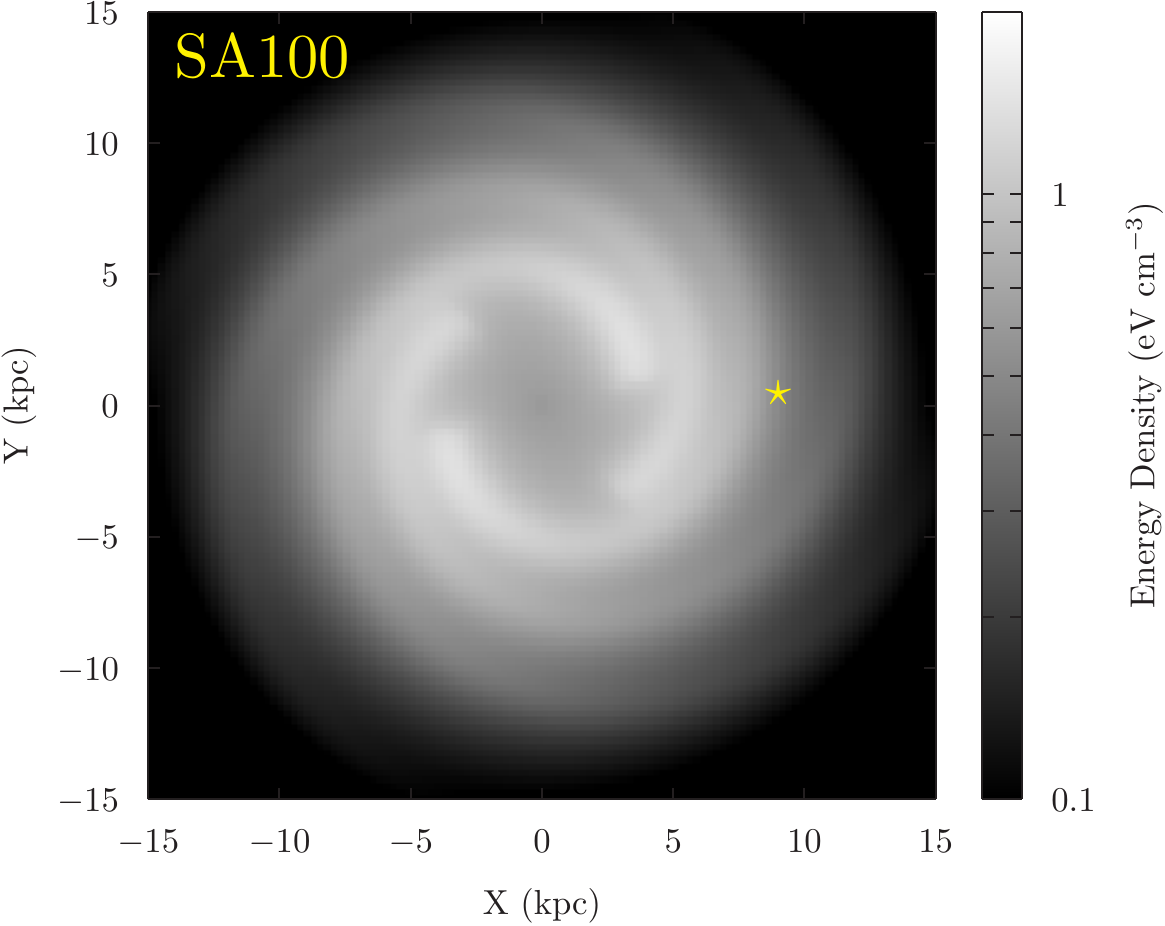}
  }
  \caption{Total CR energy densities at the plane for the
    SA0 (left), SA50 (centre), and SA100 (right) source density
    models used in this paper.
    The yellow star marks the location of the Solar system for each.
    The maximum of the energy density is $\sim 1.2-1.5$~eV~cm$^{-3}$ depending on density model.
  \label{fig:crenergydensity}}
\end{figure*}

For each of the SA0, SA50, and SA100 CR source densities the propagation model parameter tuning is
made using a maximum-likelihood fit employing the data listed in Table~\ref{tab:CRdata}.
To reduce the number of parameters in each fit the procedure is split into two stages, similar to the analysis described by \citet{Cummings:2016}.
The propagation model parameters that are fit for are listed in Table~\ref{tab:CRparameters}.
There is a strong degeneracy between the halo height and the normalisation of the diffusion coefficient.
Even though using radio-nuclei ($^{10}$Be, $^{26}$Al, $^{36}$Cl, $^{54}$Mn) constrains the halo size significantly the range of possible values remains quite wide.
Instead of fitting for both simultaneously, the halo height is set to 6~kpc,
in good agreement with previous analyses \citep[e.g.,][]{2005AIPC..769.1612M,2013MNRAS.436.2127O,GalBayesII}.
The first stage of the procedure fits for the other
propagation model parameters
together with the injection spectra and abundances of $Z>2$ elements.
With the propagation parameters and the injection spectra for $Z>2$
determined, they are held constant.
The injection
spectra for electrons, protons, and He are then obtained in the second
stage of the procedure.
To reduce the number of parameters the injection spectrum of He is coupled
with that of the
protons such that the location of the breaks are the same, and all the indices
are smaller by a value $\Delta_{He}$ that is a parameter in the fit procedure.
This is similar to the linking of the proton and He spectra for the analysis described by \citet{GalBayesII}.
Fourteen parameters are determined by the first stage of the procedure, while the second stage fits for fifteen parameter values.

The results of the fitting procedure
are given for the SA0, SA50, and SA100 CR source density models in Table~\ref{tab:CRparameters}.
The calculations are made for a Cartesian spatial 
grid with dimensions $\pm 20$~kpc for the $X$ and $Y$ coordinates, with $\Delta X,Y,Z = 0.125$~kpc and CR kinetic energy grid covering 10~MeV/nucleon to 1~TeV/nucleon
with logarithmic spacing at 10 bins/decade.
The span and sampling of the spatial and energy grids is chosen to enable realistic and efficient computations given the available resources\footnote{Increasing the energy grid sampling by a factor of 2 only produces a change in the propagated CR intensities at maximum of $\sim 2$\%. The runtime and memory consumption is increased by a proportional factor for the finer energy grid but would not substantially alter the results or conclusions.}.
The spatial grid sub-division size allows adequate sampling of the CR and ISM density distributions.
The $X,Y$ size of the grid is sufficient to ensure that CR leakage from the
Galaxy is determined by the size of the confinement region perpendicular to the
Galactic plane at the Solar system location where the propagation model parameters are fitted; it has been shown that there is only a weak effect on parameters determined for 2D models using 20~kpc and 30~kpc maximum radial boundaries \citep{2012ApJ...750....3A}.

The change in
parameters between the models is small but statistically significant. 
However, there is no obvious trend for most of the parameters.
That is, the
values for those of SA50 are not always between
the values for SA0 and SA100.
Note that the propagation parameters $v_A$ and $\delta$ determined here differ from those obtained by \citet{GalBayesII} and \citet{Cummings:2016} because of the datasets employed.
 The larger value of the delta parameter comes from the reduced Alfv\'{e}n speed obtained by the fits: higher Alfv\'{e}n speeds create a larger bump around $\sim1$ GeV for the B/C than exhibited by the AMS-02 data that are used for this paper.

The CR model spectra and data are shown in Fig.~\ref{fig:Spectra1}.
The calculated spectra agree well with the data and are generally well within
the data uncertainties, where the experiment systematic and statistical are added in quadrature.
The model predictions are very similar being within $\sim 5$\% of each other.
But some deviations from the data occur, in particular the He spectrum is
over-predicted between 1 and 10~GeV while the proton spectrum is
under-predicted for the same energy range.
This is an indication that more freedom is needed for the injection spectrum
of He than is allowed by the fitting procedure.
However, the discrepancies are within $\sim 20$\%, which is a sufficient level
of agreement for the calculations in this paper.

For the $Z>2$ elements the last data point
in the spectrum from HEAO3-C2 seems to be over-predicted.
Without further observations at higher
energies it is difficult to say if this is due to unknown systematic uncertainties in the data or incorrect model predictions.
The PAMELA C spectrum indicates that the modelled spectrum may be slightly too hard, but this small discrepancy for the high-energy CR spectra will not affect the interstellar emission
calculations (Sec.~\ref{sec:gammacalc}). 
The modelled spectra agree well with the B/C observations from AMS-02, even
though there is some under-prediction at the highest energies while
over-predicting at the lower energies.
The low-energy AMS-02 data is in tension with the B and C spectra
of HEAO3-C2 and PAMELA for the same energy ranges.
This may be due to the force-field approximation that is used for treating
the solar modulation here, which is likely providing an inadequate description
during the high solar activity period of the AMS-02 data taking.
There are also hints of this disagreement in the proton and He spectra, where the Voyager data is not well matched by the models.

Figure~\ref{fig:crenergydensity} shows the spatial distribution for the
integrated CR energy densities at 
the plane for each of the source density models used in this paper.
The bulk of the CR energy density (and hence CR pressure) comes from protons with momenta $\sim 1$--few~GeV/c.
The long residence times at these energies produce a smoothing of the CRs in the ISM compared to the initial source density distributions.
This is particularly evident for the SA100 model (right panel) where the bulk of the injection
along each of the arms is narrowly concentrated, but inside the Solar circle the subsequent propagation smoothes the CRs into a quasi-ring of high energy density for $R\sim3-7$~kpc.
Outside the plane the decrease of the total CR energy density is
approximately logarithmic with increasing $|Z|$ near the GC, and drops to zero
at the halo edge according to the boundary condition there.

\begin{figure*}[ht!]
  \subfigure{
    \includegraphics[scale=0.85]{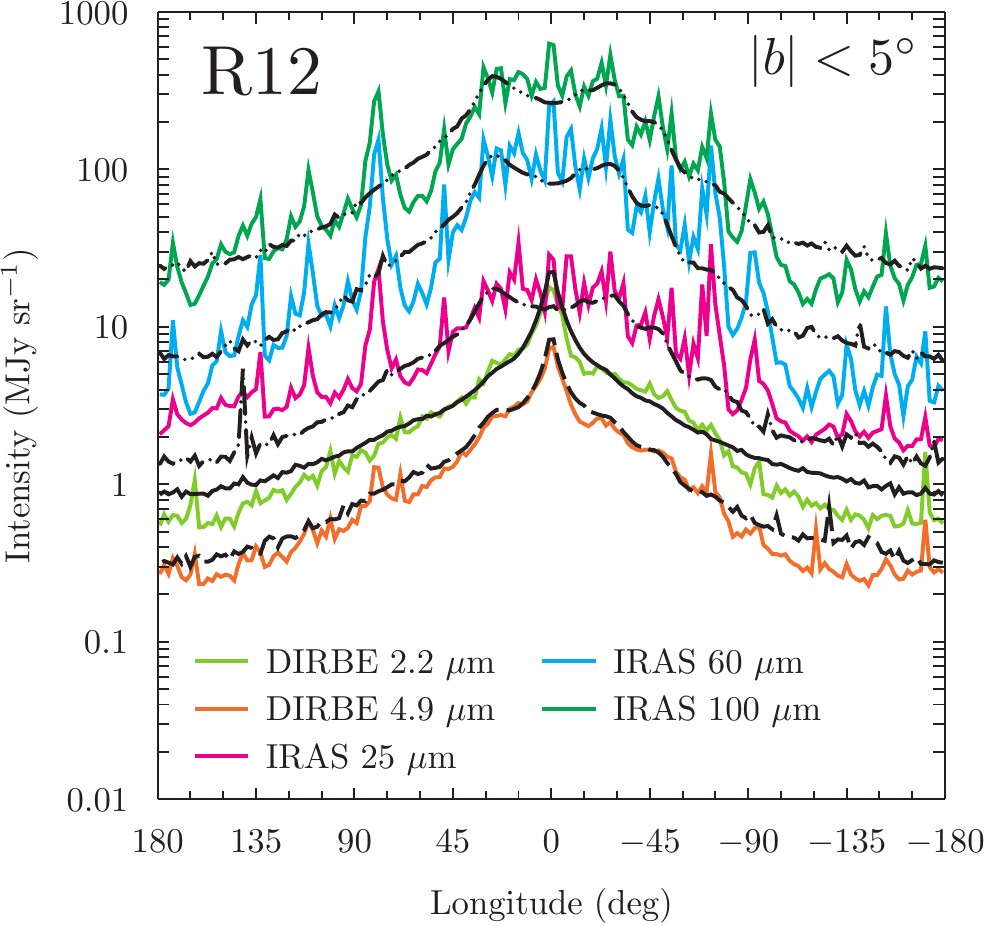}
    \includegraphics[scale=0.85]{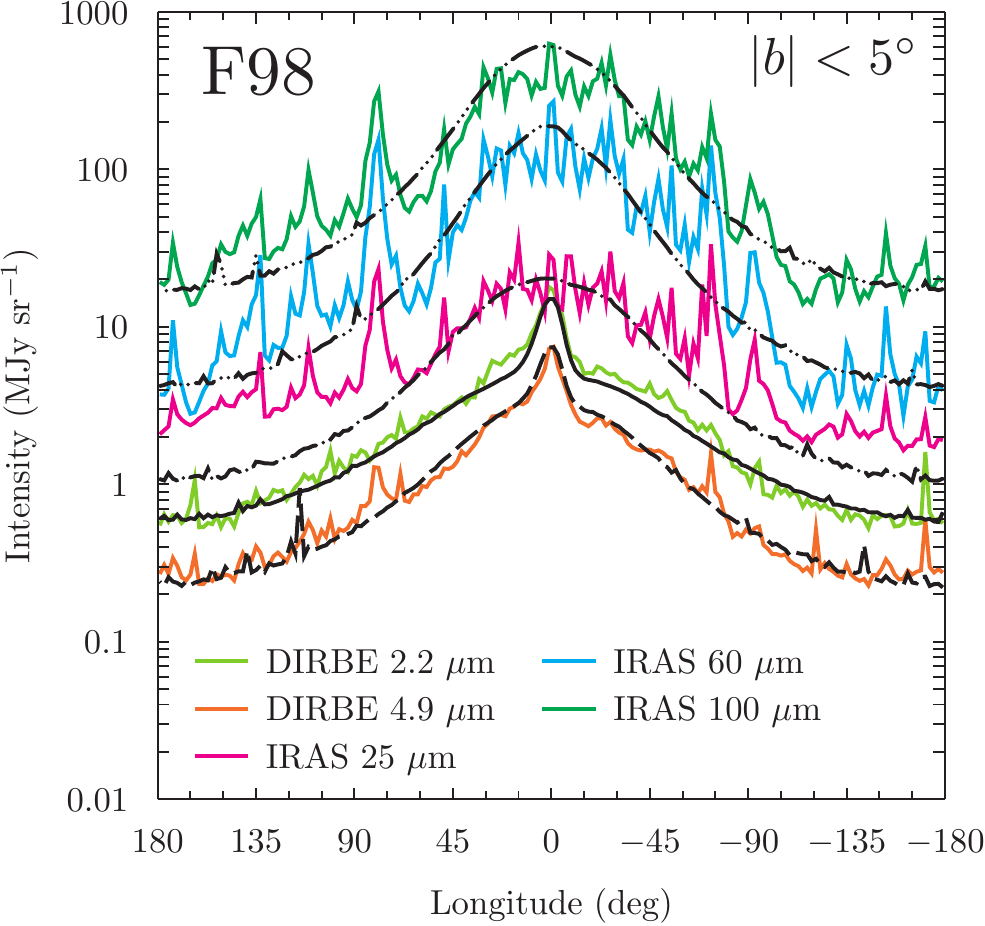}
  }
  \caption{Longitude profile averaged over $-5^\circ < b < b^\circ$ for
    the modified SKY (left, R12) and {\it COBE}/DIRBE bar/disc (right, F98) 
    models, respectively.
    Coloured curves show the data for {\it COBE}/DIRBE and reprocessed
    {\it IRAS}, while black lines show the model predictions
    convolved with the instrumental bandpass overlaid on the data for 
    {\it COBE}/DIRBE 2.2~$\mu$m (solid), 4.9~$\mu$m (long-dashed), and 
    {\it IRAS} 25~$\mu$m (long-dash-dot), 60~$\mu$m (long-dash-dot-dot), and
    100~$\mu$m (long-dash-triple-dot) bands, respectively.
  \label{fig:isrflongprofiles}}
\end{figure*}

The corresponding total injected CR powers for the SA0, SA50, and SA100 models are $5.32\times10^{40}$, $5.14\times10^{40}$, and $4.93\times10^{40}$ erg~s$^{-1}$, respectively.
For the SA0 model the injected CR proton power is $5.00\times10^{40}$, primary electrons $1.77\times10^{39}$, and He and heavier nuclei $1.41\times10^{39}$ erg~s$^{-1}$.
Likewise, the injected CR powers for the SA50 model are $4.83\times10^{40}$, $1.77\times10^{39}$, and $1.33\times10^{39}$ erg~s$^{-1}$, and for
the SA100 model are $4.62\times10^{40}$, $1.77\times10^{39}$, and $1.31\times10^{39}$ erg~s$^{-1}$, for protons, primary electrons, and He and heavier nuclei, respectively.
These injected powers are within $\sim 20$\% of those found by \citet{2010ApJ...722L..58S} using 2D \GP\ models and within a factor $\sim 2$ of the canonical estimate for the total CR injected power of $10^{41}$~erg~s$^{-1}$ \citep[e.g.,][ and references therein]{2017A&A...597A.117D}.

The major difference in injected powers by the different CR source density distributions is for the nuclei.
Because the normalisation for the CR spectra is to the data, which are collected at the Solar system location, more power is
required for the SA0 model because the region of highest source density is further away than for the other density distributions.
\citet{2010ApJ...722L..58S} noted also that the required injected CR power changes with varying halo height, but this is not a consideration in this paper because the size of the CR confinement region is constant.

\subsection{Interstellar Radiation Field}
\label{sec:rtcalcs}

The simulation volume for the radiation transfer calculations is a
box with dimensions $\pm 15$~kpc for the $X,Y$ coordinates and $\pm3$~kpc
for the $Z$ coordinate because this effectively encapsulates all of the
input stellar luminosity and dust for the R12 and F98 models,
and hence ensures that computation resources are not wasted for
regions that contribute negligibly to the spectral intensity distribution.
A Cartesian grid is used to segment the simulation volume.
The $X,Y$ coordinates have regular spacing $\Delta X,Y = 0.125$~kpc, while 
the $Z$ coordinate uses a logarithmic spacing with 25 bins 
covering 0.001 to 3~kpc plus an additional linear bin for that 
closest to the Galactic midplane (52 in total).
The wavelength grid spans 0.0912--10000~$\mu$m with 256\footnote{The \frankie\ code uses both CPUs and accelerators \citep[see][]{2013arXiv1311.4627P} with specific optimisations that require the wavelength/frequency gridding to be a power of 2 and multiple of the largest machine vector size.} logarithmically spaced bins.

To record the spectral intensity distribution a cylindrical grid is used because of the radial and angular dependence of the spatial densities for the R12 and F98 models.
Any choice of grid spacing is a compromise balancing computational resources
and overall accuracy.
The following grid is found to provide adequate sampling for the typical 
sizes of simulation volumes used for \GP{} CR propagation and interstellar 
emission calculations (halo sizes $\sim 4-10$~kpc perpendicular to the 
Galactic plane and maximum $X,Y$ boundaries $\sim 20$~kpc).
The spacing is variable in Galactocentric radius with $\Delta R = 0.1$~kpc near the GC, then $\Delta R = 0.25$~kpc to the Solar circle, $\Delta R = 1$~kpc beyond to 15~kpc, and $\Delta R = 5$~kpc beyond that to $R = 30$~kpc.
Azimuthally the spacing is at $\Delta \phi = 10^\circ$.
The $Z$-coordinates for the camera locations are $\pm20, \pm10, \pm5, \pm2, \pm1, \pm0.5, \pm 0.25, \pm0.1,$ and 0~kpc.
Note that the ISRF sampling grid entirely encloses the spatial grid for the \GP{} calculations made in this paper.

Each ISRF model calculation (R12, F98) uses $5\times10^{8}$ luminosity packets with
HEALPix Nside = 8 maps\footnote{\GP{} anisotropic IC \gray{} calculations 
have been tested using the R12 model calculated with Nside = 4, 8, and 16 
spectral intensity maps to evaluate the best compromise between accuracy, 
run-time, and disc storage for the ISRF data files. With the Nside = 8 and 16 
resolutions the position-dependent intensity traces asymmetries 
(by arms, etc.) sufficiently for the \gray{} calculations. There is no 
discernable difference for the IC calculations using the Nside = 8 and 16 
maps, but the run-time and storage on-disc for the latter is $\sim 4\times$ 
higher.}
to determine the Galaxy-wide spectral intensity distributions that are
employed for the calculations with \GP{}.
These statistics are sufficient because integrating
over the ISRF spectral intensity when determining the $e^\pm$ energy losses and IC emissivities smooths any artifacts due to Monte Carlo noise.
However, for the data/model comparison for the line profiles higher statistic runs are made using $5\times10^{10}$ luminosity packets with HEALPix Nside = 1024 maps.
Even with the higher statistics runs there is unavoidable Monte Carlo noise typically where the intensities are lowest.

Figure~\ref{fig:isrflongprofiles} shows the predicted 
longitude profiles for latitudes $-5^\circ < b < 5^\circ$ for the 
R12 (left) and F98 (right) models overlaid
with the data from {\it COBE}/DIRBE
\footnote{{\it COBE}/DIRBE ZSMA maps retrieved from https://lambda.gsfc.nasa.gov/.} and the reprocessed {\it IRAS} 
data\footnote{IRIS -- Improved Reprocessing of the IRAS Survery \citep{2005ApJS..157..302M}: https://www.cita.utoronto.ca/~mamd/IRIS/IrisOverview.html.}.
The all-sky model intensity maps are convolved with the instrumental 
band passes and point spread functions, then averaged over the latitude band
to construct the profiles.
The models use density distributions for the stars and dust that do not distinguish individual sources.
Therefore the comparison with data is made without point-source subtraction.

The bulk features of the intensity profiles are generally successfully
reproduced by each model.
The R12 model traces the structure of the data more closely for $-60^\circ \leq l \leq 60^\circ$ because this is the region that it has been optimised over.
Outside of this longitude range the R12 intensities tend to be higher
than the data at near-infrared wavelengths.
This is due to the scale-length of the stellar disc, which produces a shallower
profile than the data indicate toward the outer Galaxy, and the presence of
the local arms that also contribute in this region.
The F98 model agrees well with the near-infrared data over all longitudes
due to the smaller radial scale-length for the stellar discs.
However, the structure in the data for $-60^\circ \leq l \leq 60^\circ$
is not reproduced because the spiral arms are treated as an averaged disc
component for this model.
For the mid- and far-infrared wavebands the better correspondence of the
profile structure by the R12 model is again no surprise because it was
optimised to reproduce the features in the data tracing the spiral
arm tangents.
The F98 model adequately traces the intensity profile given its comparative
simplicity to the R12 model. 
Both models under-predict the mid-infrared in the outer Galaxy.
For the R12 model this is most likely due to mismodelling of the stellar
luminosity content of the local arms.
For the F98 model it is their absence that causes the lower intensity in these regions.

\begin{figure*}[htb!]
  \subfigure{
    \includegraphics[scale=0.85]{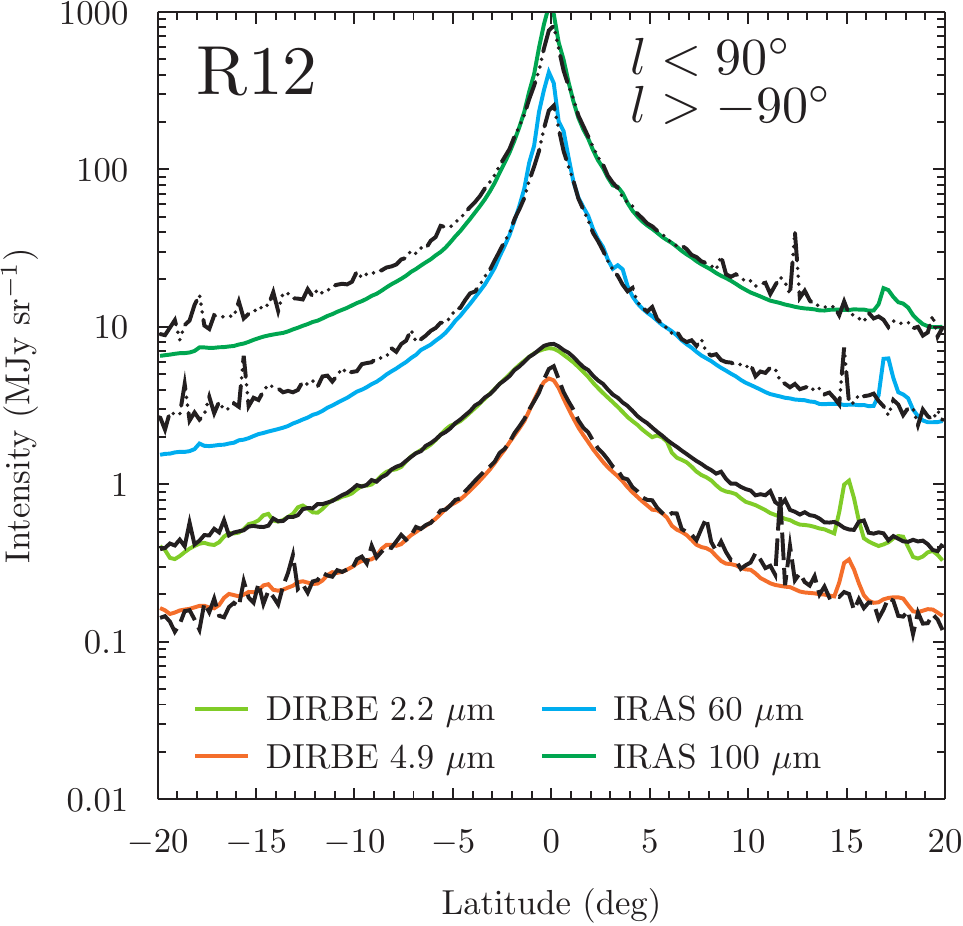}
    \includegraphics[scale=0.85]{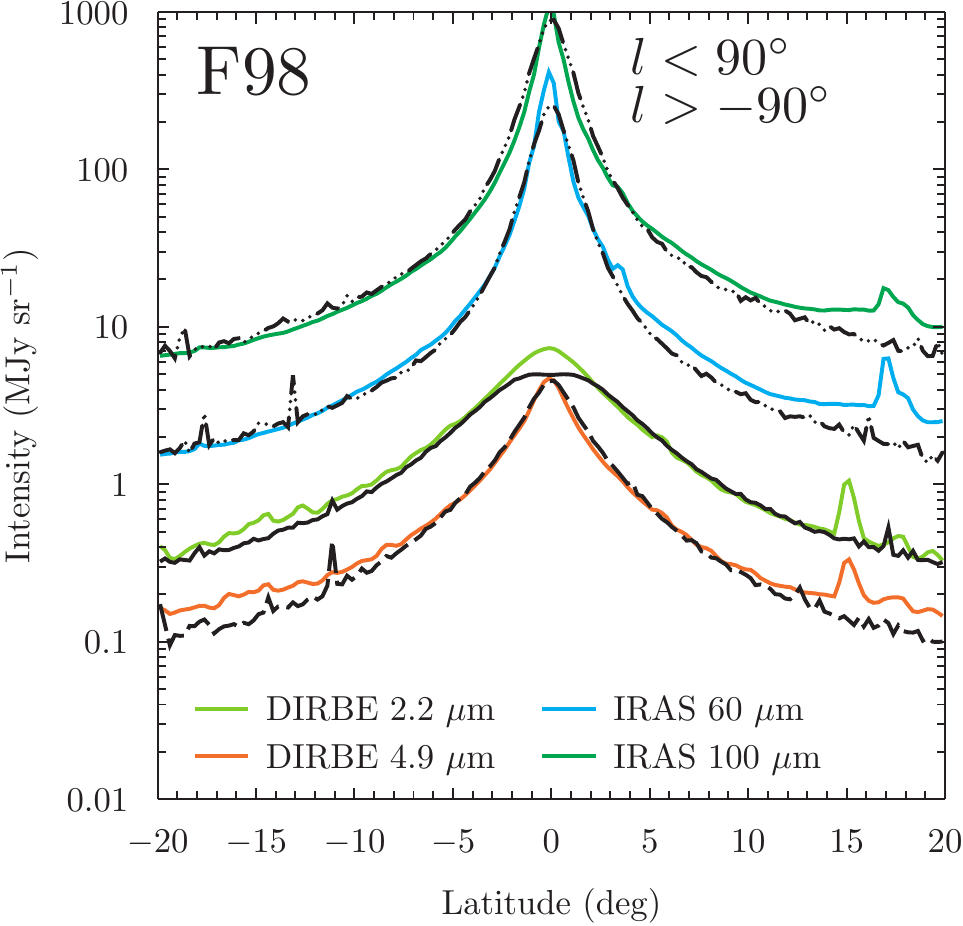}
  }
  \subfigure{
    \includegraphics[scale=0.85]{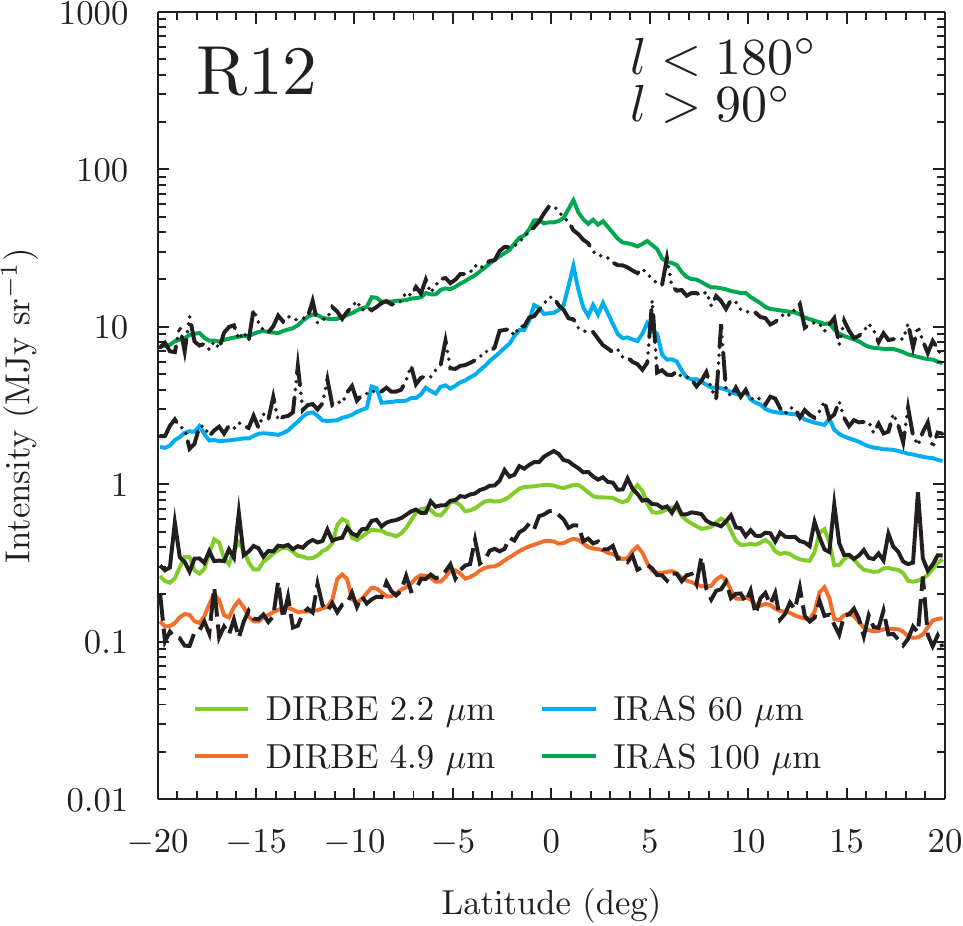}
    \includegraphics[scale=0.85]{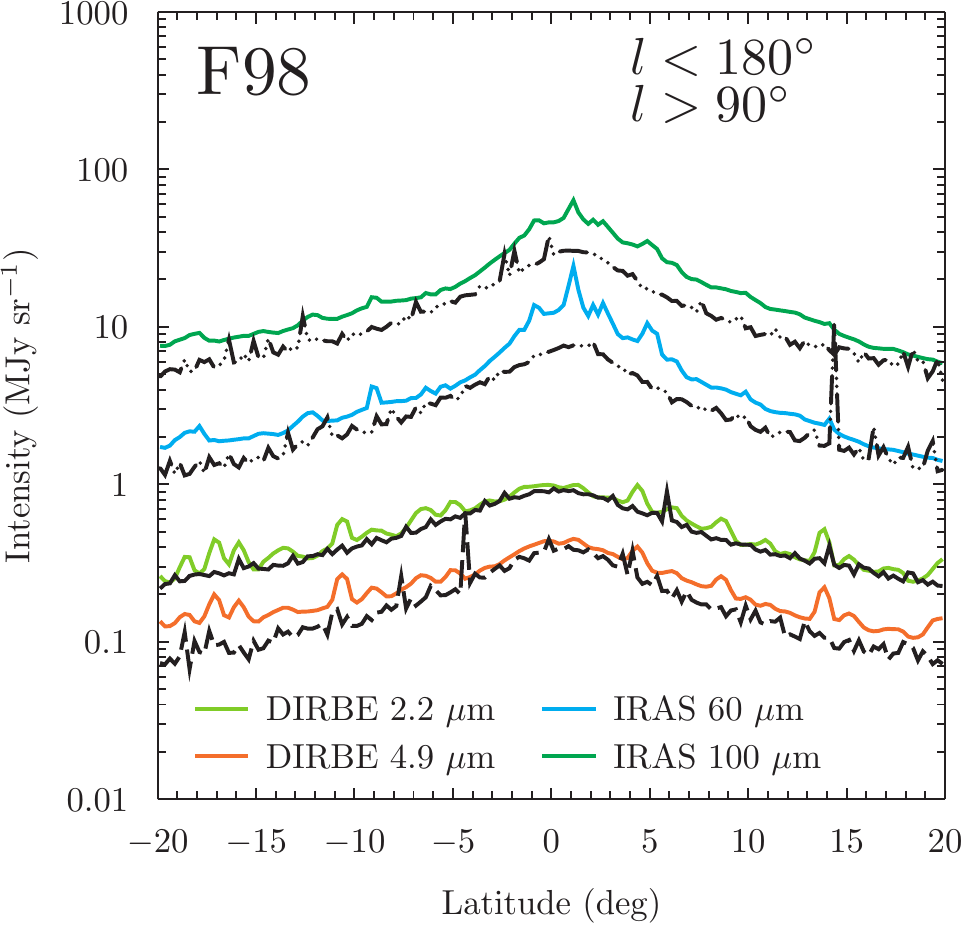}
  }
  \subfigure{
    \includegraphics[scale=0.85]{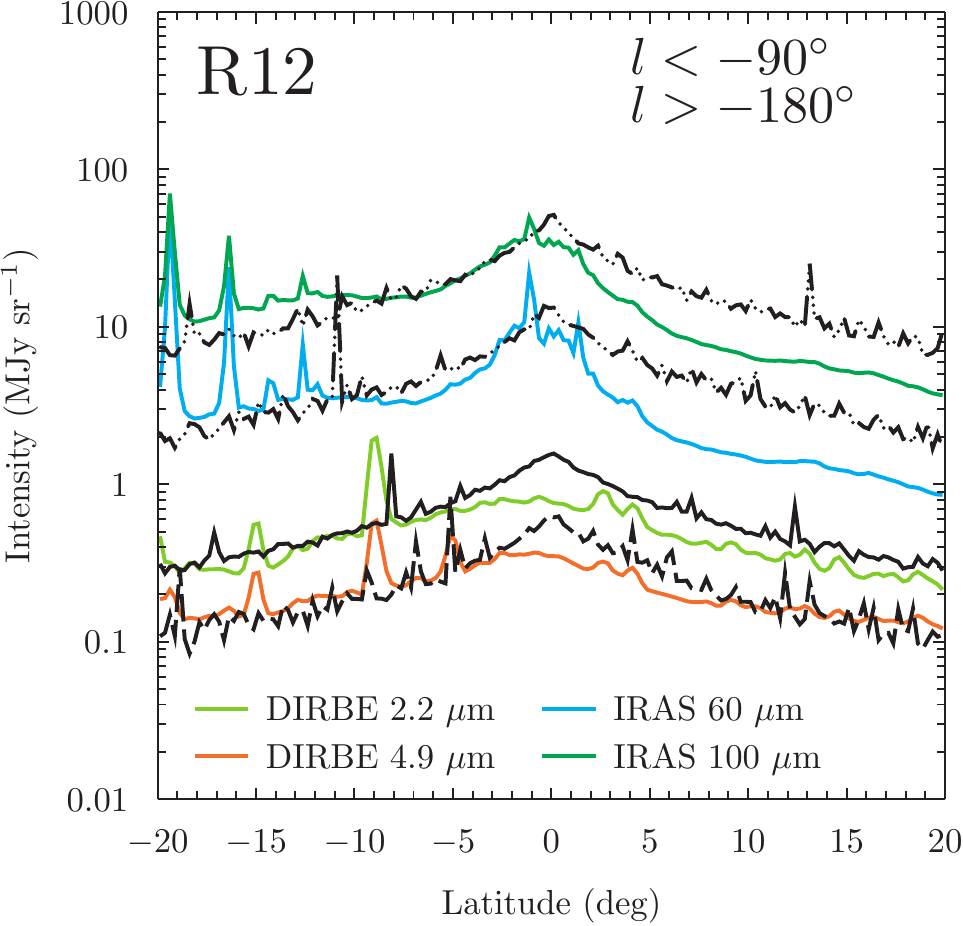}
    \includegraphics[scale=0.85]{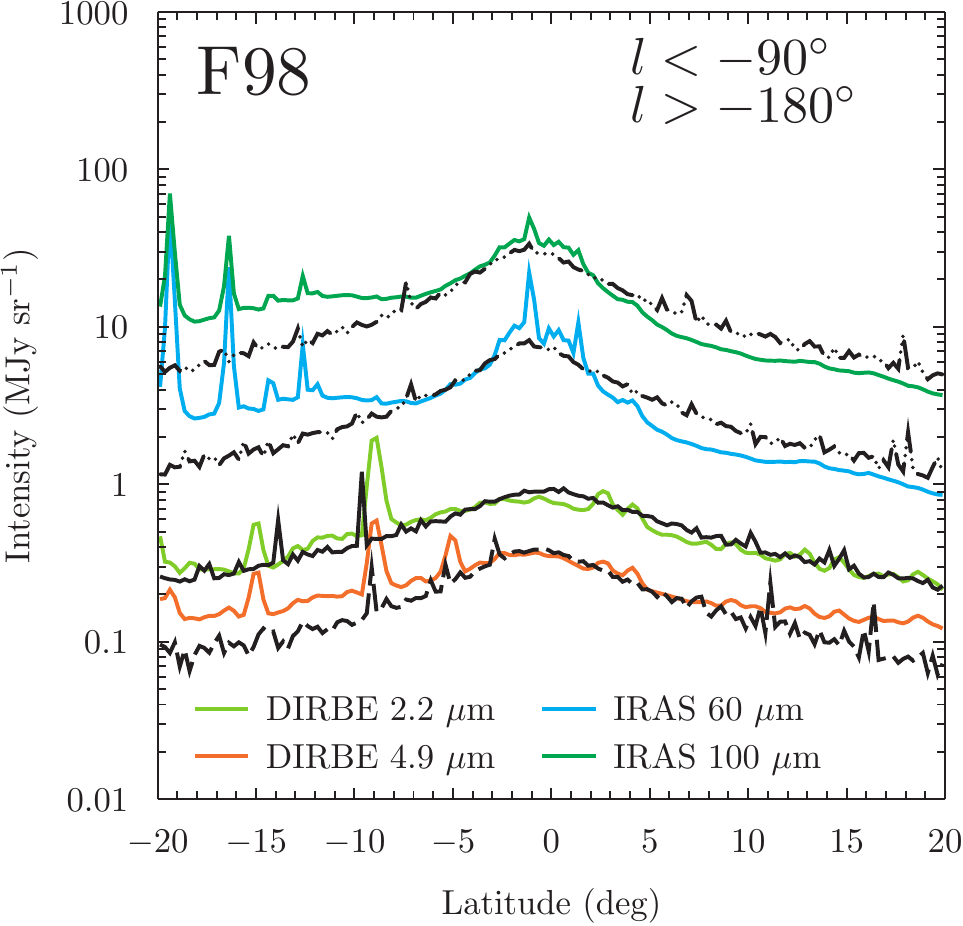}
  }
  \caption{Latitude profiles for
    the R12 (left) and F98 (right) models, respectively.
    Line colours and styles as in Fig.~\ref{fig:isrflongprofiles}, except 
    the {\it IRAS} 25~$\mu$m band is not shown to allow for straightforward
    separation of the profiles.
  \label{fig:latprofiles}}
\end{figure*}

Figure~\ref{fig:latprofiles} shows the predicted latitude profiles and data 
for the Galactic quadrants 1 and 4 (top), 2 (centre), and 3 (bottom).
For the inner quadrants both R12 and F98 perform fairly well at reproducing the profiles.
The major difference is the lower intensity by the F98 model within a couple of degrees either side of the Galactic plane for the $2.2$~$\mu$m band.
It comes from the larger dust column due to the shorter radial
scale-length and smaller inner `hole' region for the dust disc in the F98 model.
For the outer Galactic quadrants (2 and 3 -- middle and bottom rows) the R12 model intensities are too high and display insufficient asymmetry about the plane
compared to the data.
These mismatches are a result of the stellar disc radial scale-length and local
arm populations discussed earlier, and the lack of warping for both the
stellar and dust discs.
The F98 intensities match the asymmetries in the data for the outer quadrants
much better and mostly reproduce the intensities over all wavebands well.
There are under-predictions for the second quadrant and out of the plane that
are most likely from the absence of the local arms, which contribute
over all wavebands via the direct stellar emission along with the associated
dust heating.

\begin{figure*}[htb!]
  \subfigure{
    \includegraphics[scale=0.85]{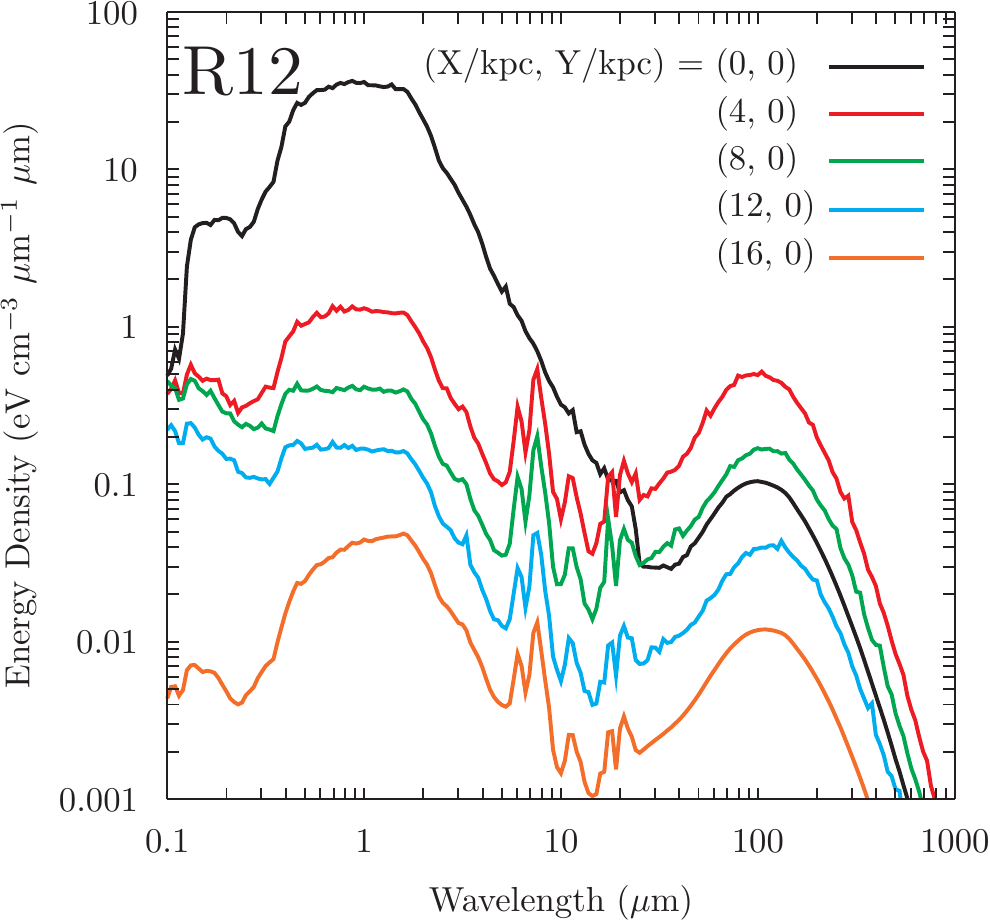}
    \includegraphics[scale=0.85]{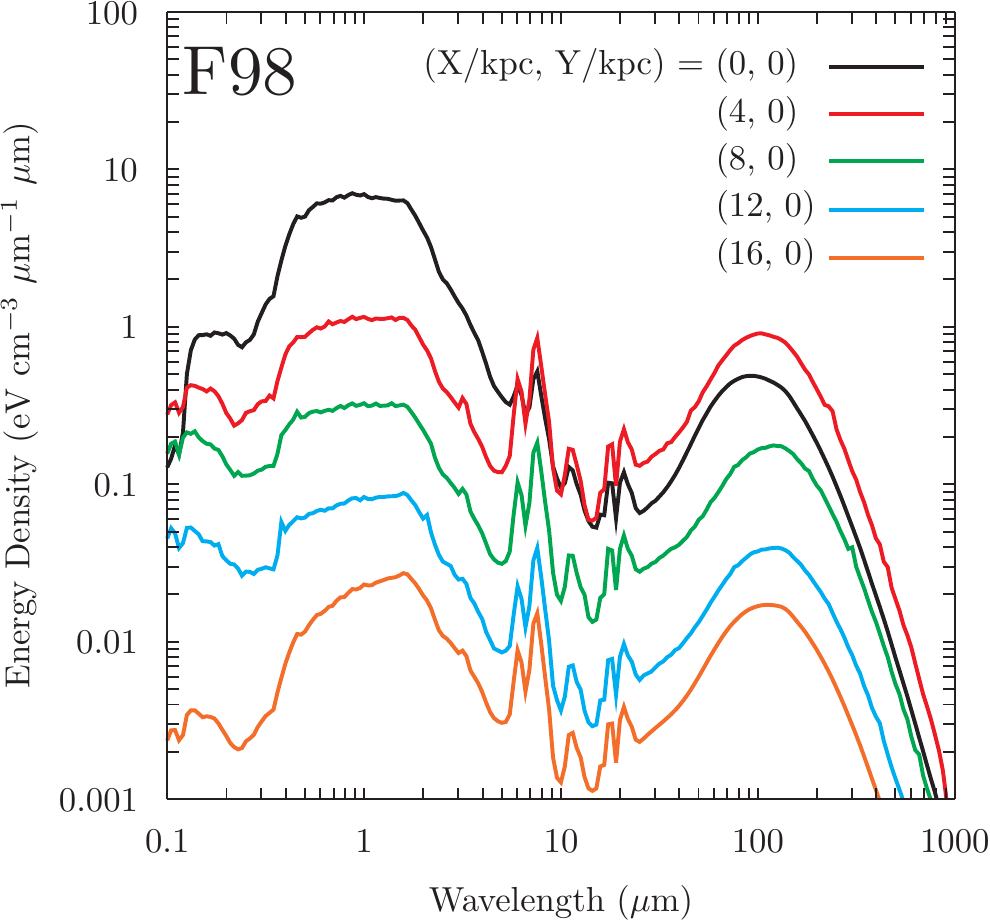}
  }
  \caption{ISRF spectral energy density for the 
    R12 (left) and F98 (right) models, respectively, showing
    the variation with positive $X$ coordinate in the Galactic plane.
  \label{fig:isrfsed}}
\end{figure*}

\begin{figure*}[htb!]
  \subfigure{
    \includegraphics[scale=0.8]{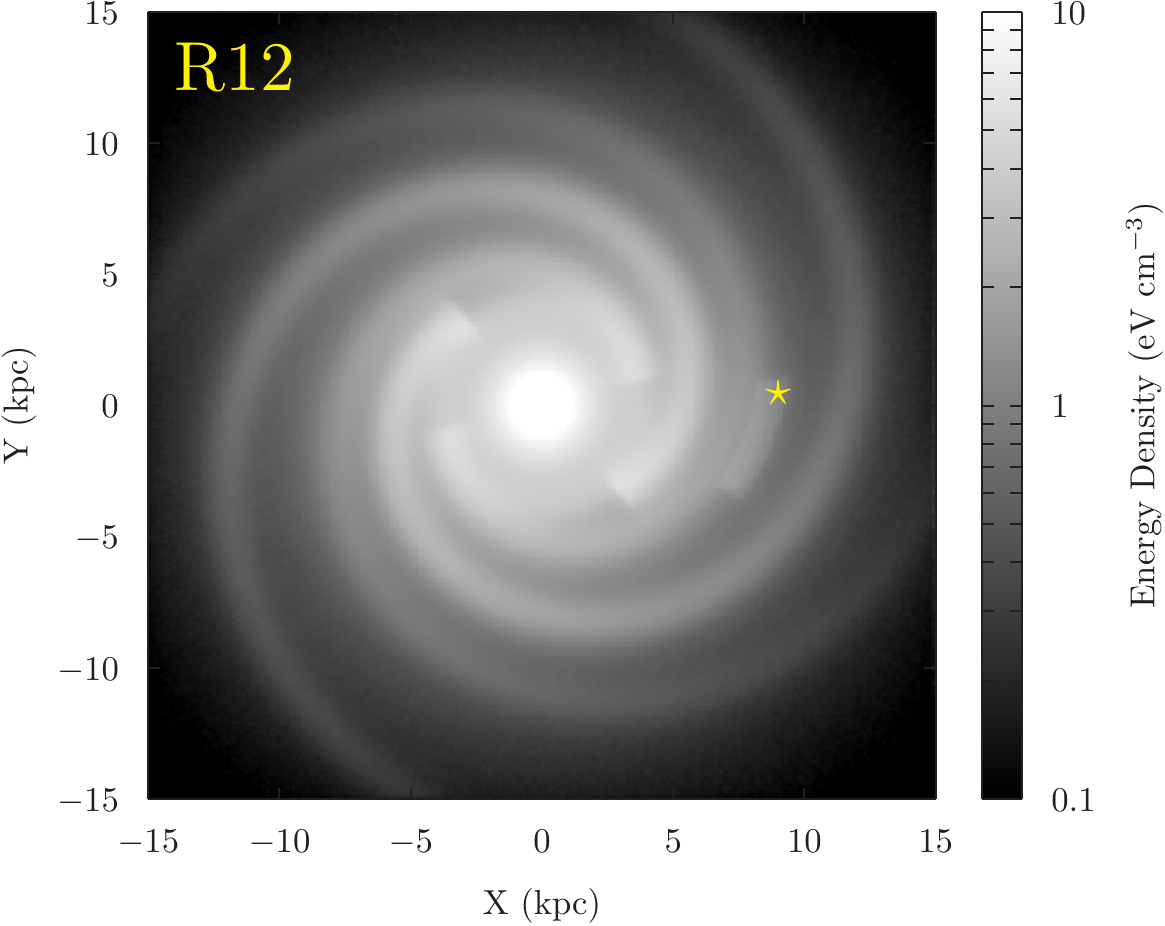}
    \includegraphics[scale=0.8]{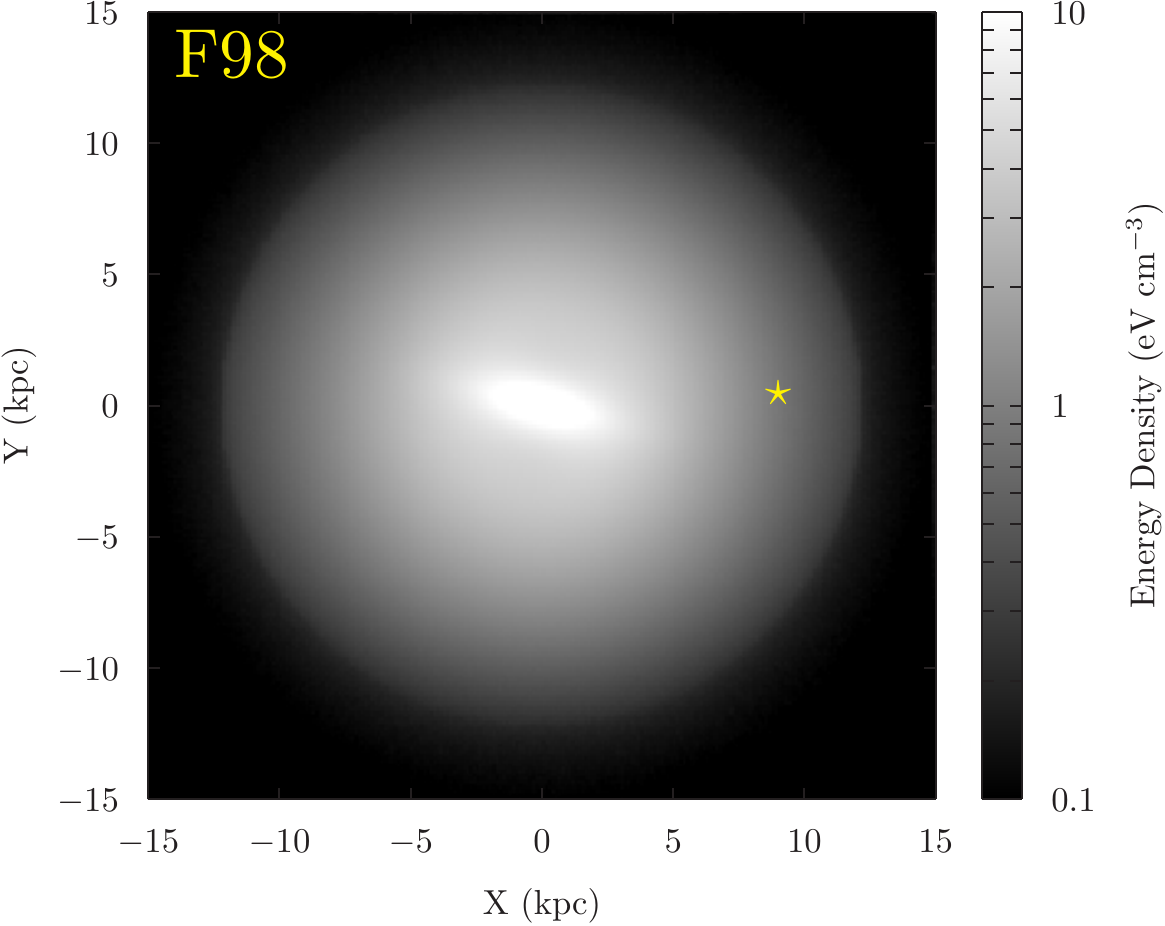}
  }
  \caption{Integrated ISRF energy densities in the Galactic plane for the 
    R12 (left) and F98 (right) models, respectively.
    The yellow star marks the location of the Solar system for each.
    Note that the energy density saturates the scale in and about the GC for
    both models.
  \label{fig:isrfenergydensityplane}}
\end{figure*}

The discrepancies between the R12 and F98 model predictions and local observations are generally minor\footnote{Because the R12/F98 models use densities for the stellar and dust content and finite sized simulation volume segmentation the calculations are not capable of reproducing the fine structure such as seen in the longitude profiles. The general features related to, e.g., the arms (for R12) are however evident.}.
Integrating the intensities over the sky produces all-sky averaged values
for the R12 and F98 models that are close to the data and that are well within the
experimental uncertainties for all wavebands, thus showing the general
consistency for both R12/F98 calculations with the observations.

\begin{figure*}[htb!]
  \includegraphics[width=0.33\textwidth]{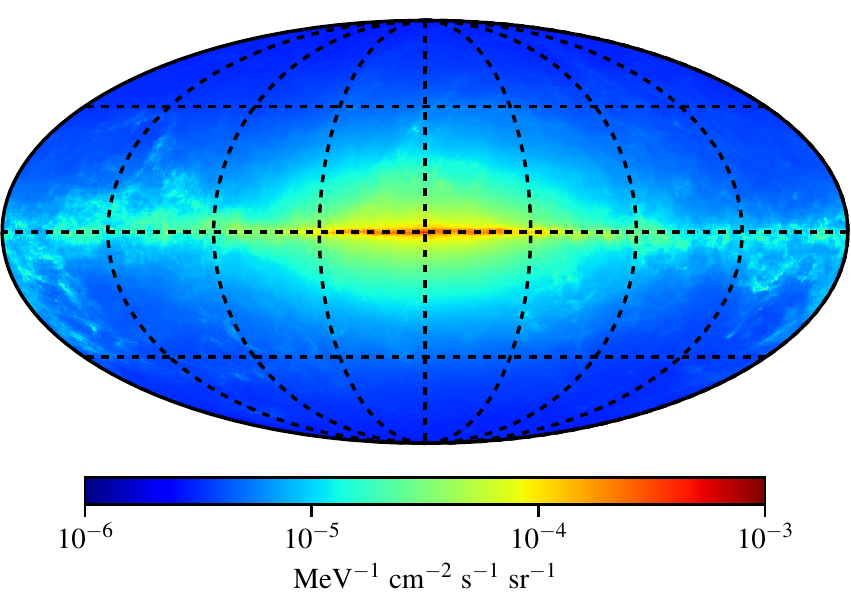}
  \includegraphics[width=0.33\textwidth]{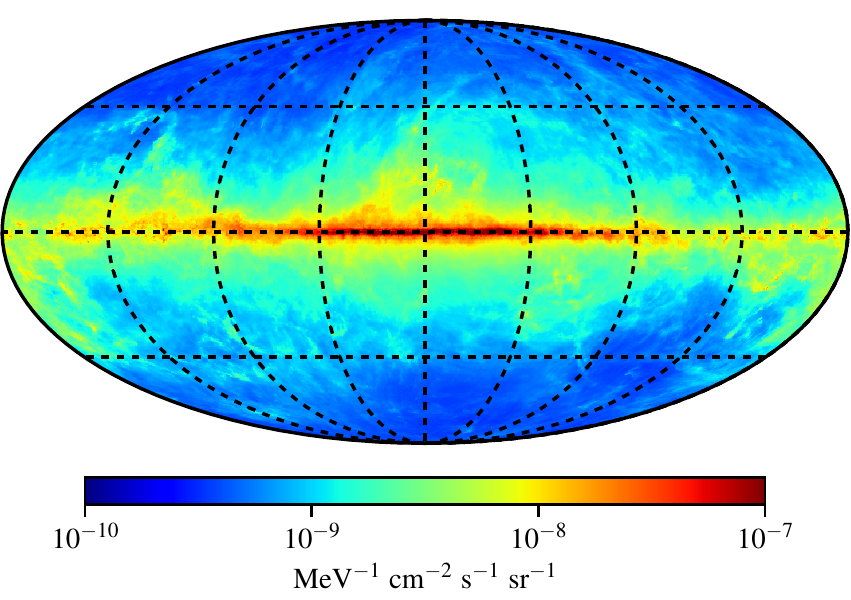}
  \includegraphics[width=0.33\textwidth]{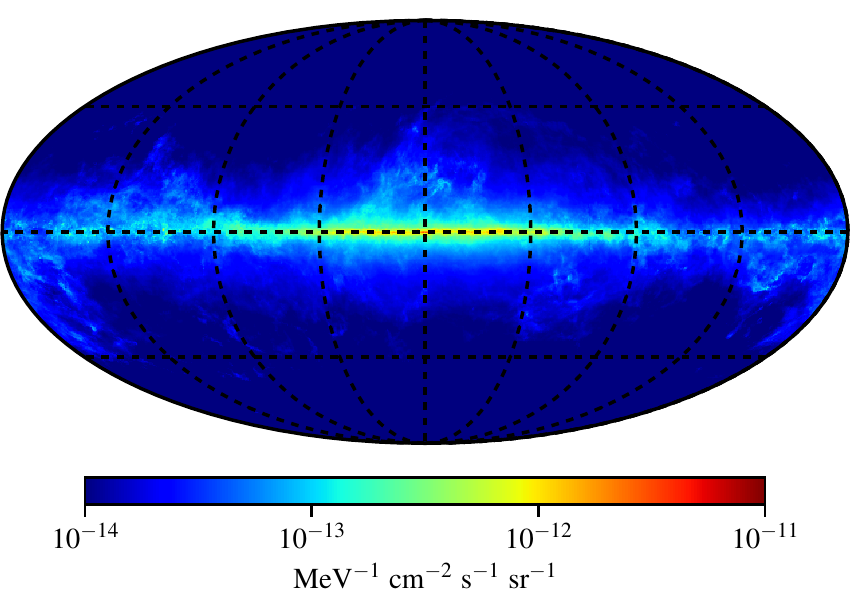}\\

  \includegraphics[width=0.33\textwidth]{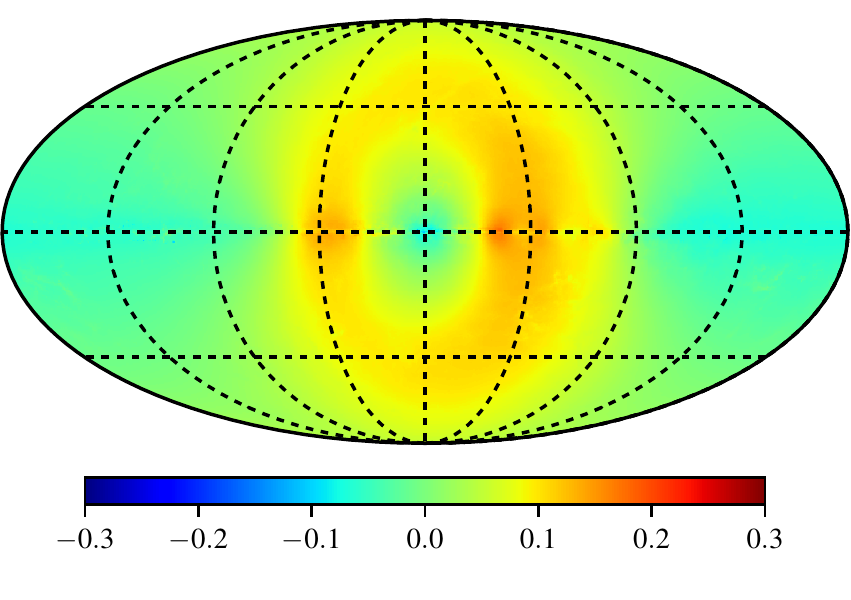}
  \includegraphics[width=0.33\textwidth]{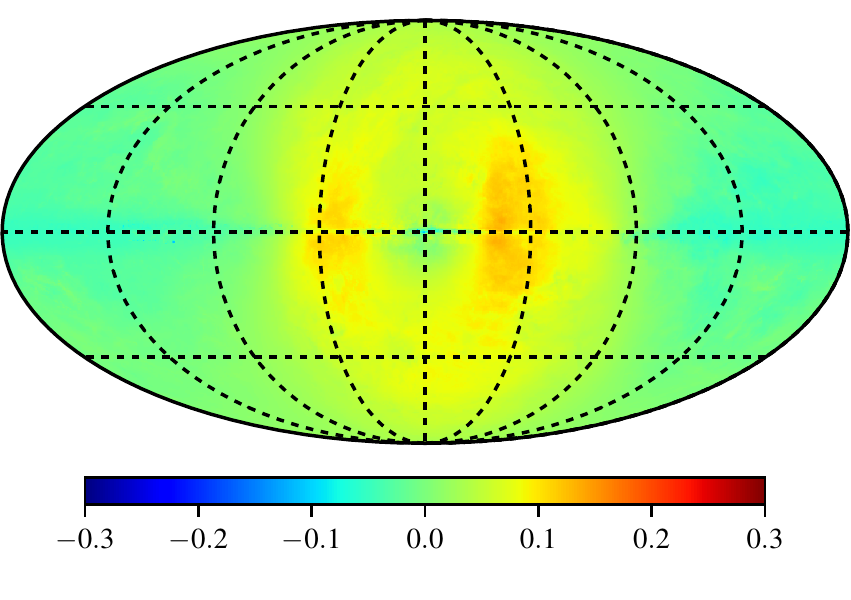}
  \includegraphics[width=0.33\textwidth]{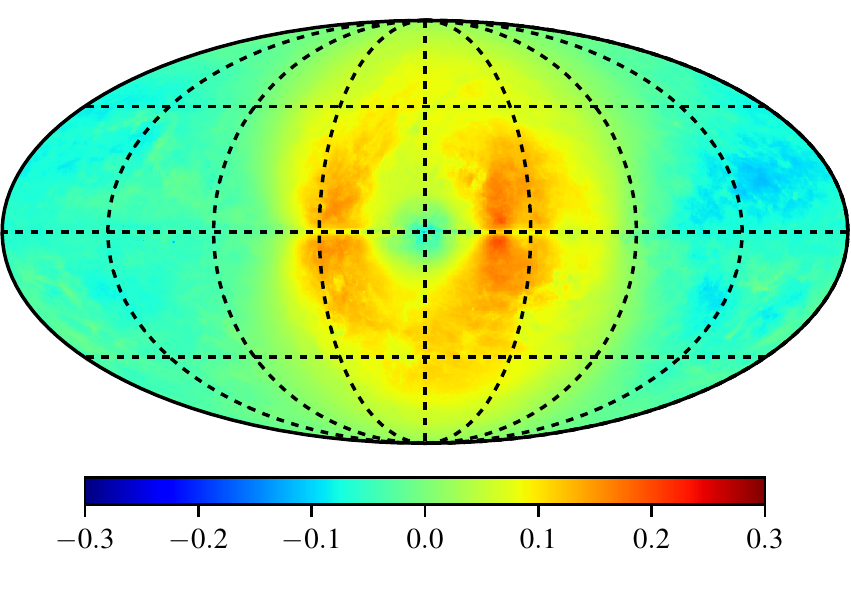}\\
  \includegraphics[width=0.33\textwidth]{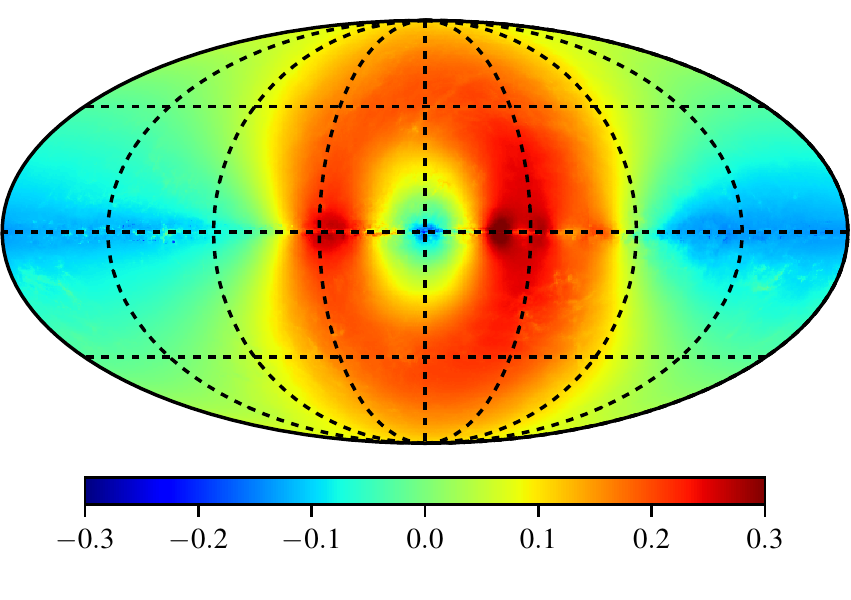}
  \includegraphics[width=0.33\textwidth]{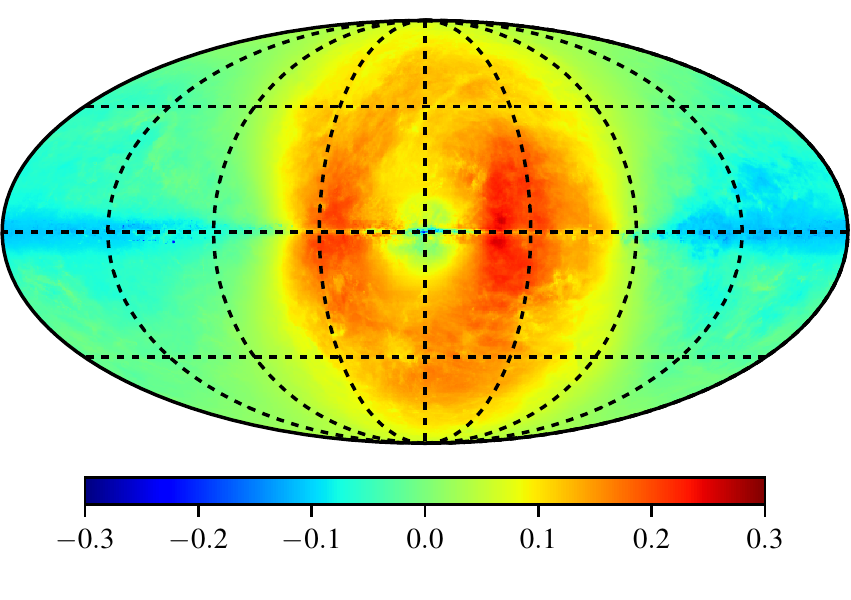}
  \includegraphics[width=0.33\textwidth]{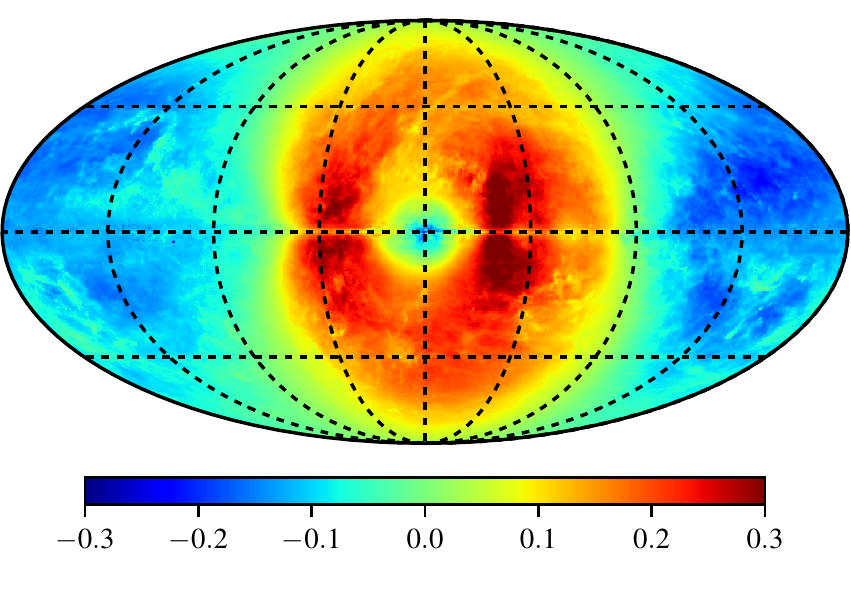}
  \\
  \caption{Top row: total intensity ($\pi^0$-decay, Bremsstrahlung, and IC) at
    10.6~MeV (left), 1.2~GeV (centre), and 79~GeV (right)
    for the SA0--Std reference case.
    Centre and bottom row: fractional difference maps
    for the SA50--Std and SA100--Std model combinations, respectively.
    The maps are in Galactic coordinates with $l,b = 0^\circ,0^\circ$ at the
    centre.
    The longitude meridians and latitude parallels have $45^\circ$ spacing.
    While the fractional residual colour scale is shown for a maximum $\pm30$\%  difference there
    are features for the SA100 CR source density model for the low- and high-energy maps that are outside the scale bounds.
    These are near the direction of the GC and toward $l \sim \pm 30-45^\circ$ where the spiral arm tangents contribute most along the line-of-sight.
    \label{fig:StdSrcfrac}}
\end{figure*}

\begin{figure*}[htb!]
  \subfigure{
    \includegraphics[scale=0.6]{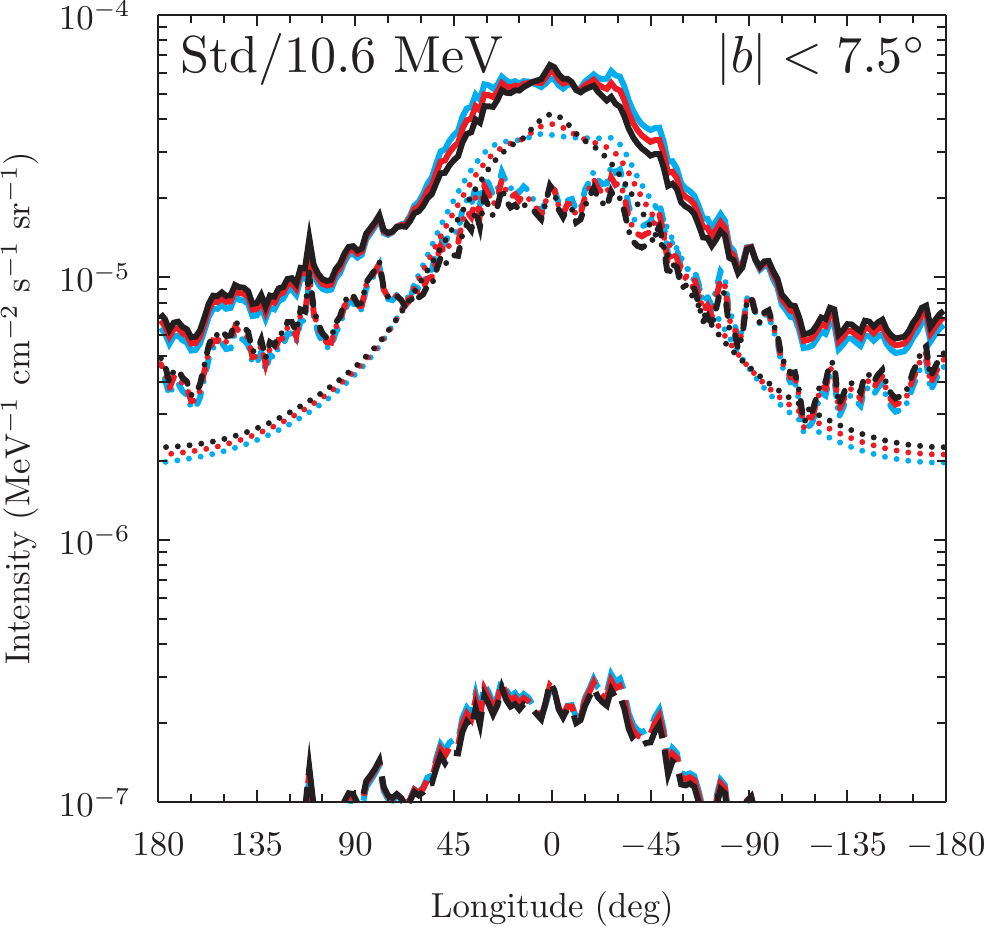}
    \includegraphics[scale=0.6]{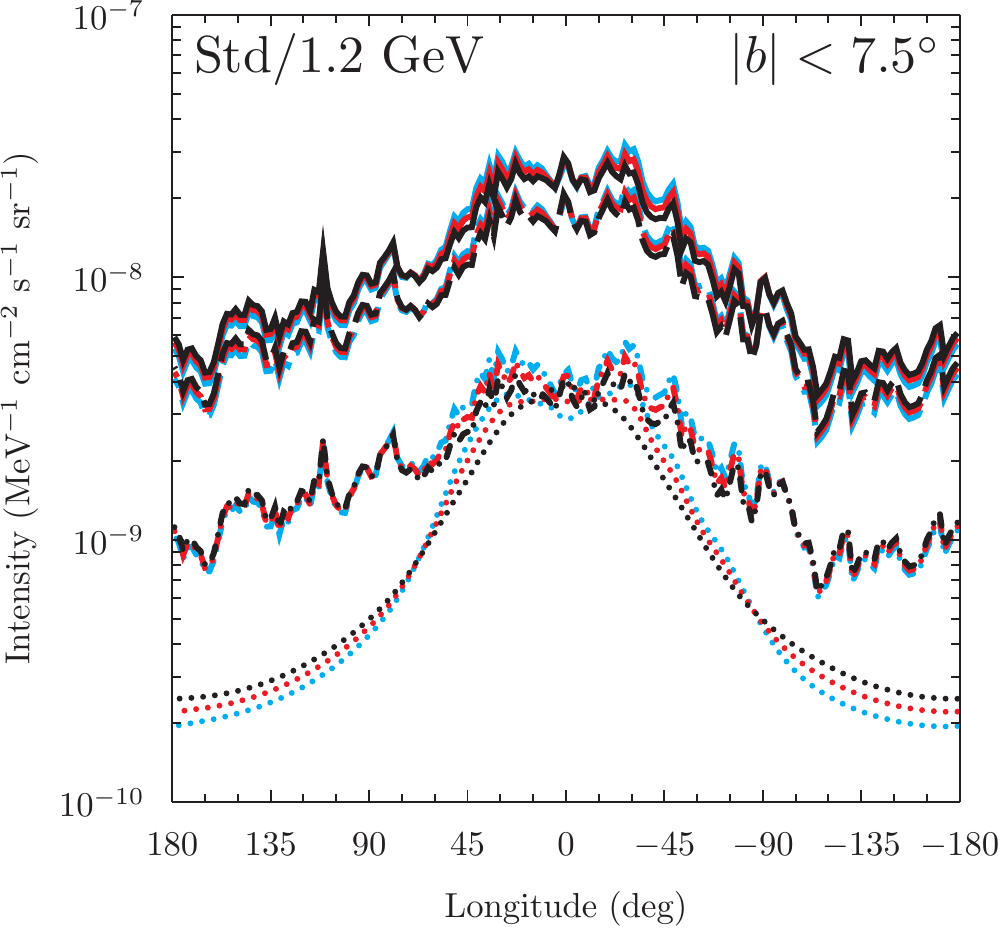}
    \includegraphics[scale=0.6]{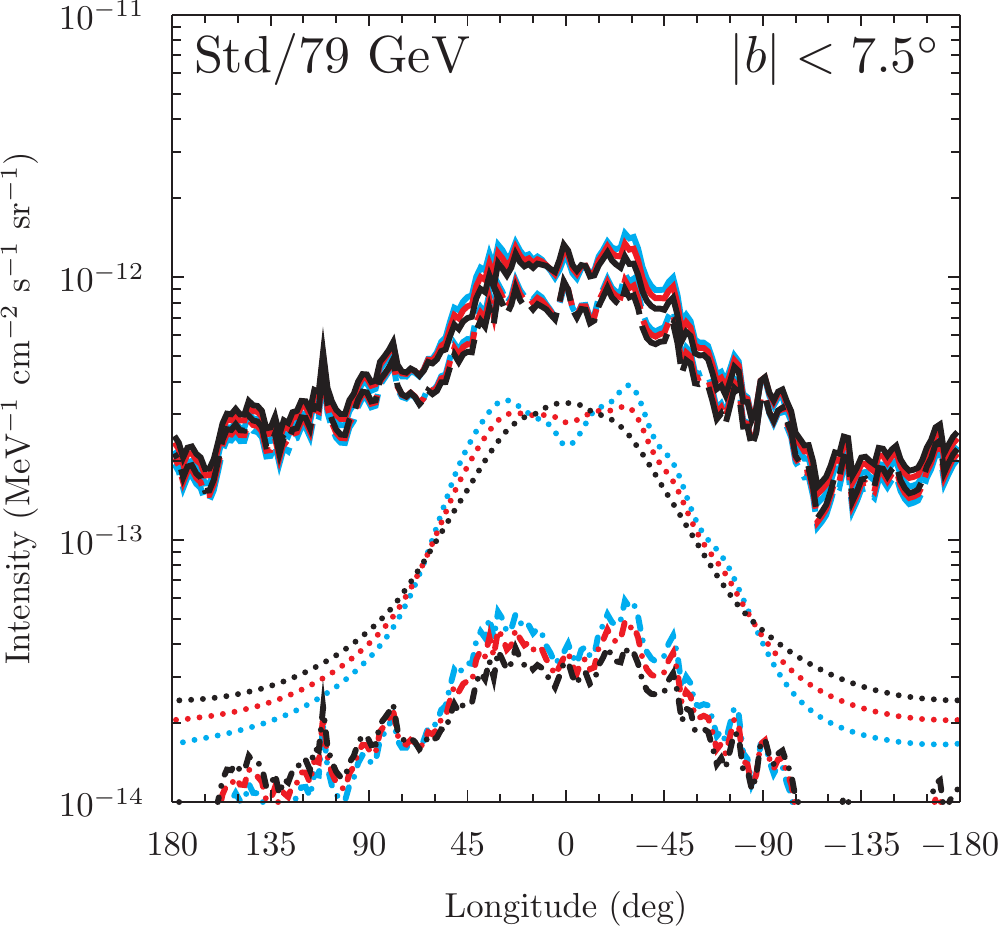}
  }
  \caption{Longitude profiles averaged over $-7.5^\circ \leq b \leq 7.5^\circ$ for the Std ISRF and SA0, SA50, and SA100 CR source density models. Line-styles: solid, total; dashed, $\pi^0$-decay; dash-dot, Bremsstrahlung; dotted, IC. Line colours: black, SA0; red, SA50, cyan, SA100. 
    \label{fig:gammasrcstdlongprofile}}
\end{figure*}

While the predicted intensities locally are very similar those elsewhere
in the Galaxy can vary considerably between the R12 and F98 models.
Figure~\ref{fig:isrfsed} shows 
the spectral energy density (SED) variation with $X$ coordinate in the Galactic 
plane. 
Near the GC the R12 model is considerably more intense at UV/optical to 
near-infrared wavelengths compared to F98.
This comes from several differences: the R12 model has a stellar disc that 
continues toward the GC, while the F98 model has a holed stellar disc; the R12 
bulge luminosity is higher than that for the F98 model; and the hole in the R12
dust disc has a larger radius than that of the F98 model, consequently
the absorption near the GC for the R12 model is lower.
However, the general trend of a shift in the peak of the far-infrared emission to shorter wavelengths from the intense radiation field over the inner Galaxy is present for both R12 and F98 models independent of the dust density there.
The other major difference is the spatial variation of the SED from 
far-UV to $\sim 0.5$~$\mu$m wavelengths that reflects how the early-type
stars are modelled.
The variation of the UV and optical spectral intensity for R12 model 
with $X$ outside the region about the GC is driven predominantly by the 
crossing of the various arms as $X$ increases out to $\sim 12$~kpc.
In particular, for the R12 model the locations shown in the figure sample
inter-arm ($X = 4$~kpc) and in-arm ($X = 8, 12$~kpc) regions.
For the F98 model such variations are absent because the young stars in the arms are treated using spatial averaging.

The variation of the wavelength-integrated SED predicted for
the R12 and F98 models across the Galaxy at the mid-plane is shown in
Fig.~\ref{fig:isrfenergydensityplane}.
The R12 model produces a more structured spatial distribution for the
energy density with features related to the stellar luminosity model clearly
evident.
The variation of the SED shown in Fig.~\ref{fig:isrfsed} (left)
from the crossing of inter-arm/arm regions with $X$-coordinate is also
readily understood where the maxima of emission are clearly visible near
$X \sim 8$ and $\sim 12$~kpc, respectively.
For the F98 model the asymmetric bulge produces a clearly elongated region
of high energy density dominating the inner Galaxy region.
Either side of the minor axis of the bulge are shallow minima in the energy
density distribution, and outside of $\sim 4$~kpc about the GC 
the stellar disc is the only significant contributer.

The spatial variation of the SED outside
of the Galactic plane (not shown) is most sensitive to the details of the
stellar luminosity and dust model for $|Z| \lesssim 1-5$~kpc, with
the effect dependent on position relative to the GC.
Close to the GC the intensities of the respective models differ the most but
become comparable for $|Z|\simeq 5$~kpc, while toward 
the outer Galaxy they are similar for somewhat lower height $|Z| \simeq 1$~kpc.
The fall-off of the intensity with $Z$-coordinate is approximately logarithmic for both models. 
At a distance $\sim 10$~kpc above the plane near the GC the energy density
for either model is similar to that of the CMB, and at a 
distance $\sim 20$~kpc becomes comparable that of the extragalactic background light 
\citep[e.g.][]{2001ARA&A..39..249H,2007A&A...471..439M}.

\subsection{$\gamma$-Rays}
\label{sec:gammacalc}

High-energy interstellar emissions are calculated using \GP\ for the
SA0, SA50, and SA100 source density models (Sec.~\ref{sec:crcalc}), and
the standard 2D \citep[Std,][]{2012ApJ...750....3A} and 3D ISRF models (Sec.~\ref{sec:rtcalcs}).
The SA0 CR source density and Std ISRF (SA0--Std) is used as the reference 
case. 
This combination corresponds to the 2D CR source and ISRF density scenario 
that has been the standard approach for interstellar emission modelling in
the past.

The spatial and CR kinetic energy grid used for the CR propagation model parameter tuning (Sec.~\ref{sec:crcalc}) is also employed here with \gray{s} calculated from 1~MeV to 100~GeV energies using a logarithmic energy grid with 10 bins/decade spacing.
Higher \gray{} energies correspond to CRs with energies $\gtrsim 1$~TeV where the steady-state source injection paradigm employed in this paper is less valid \citep{2001ICRC....5.1964S,2013A&A...555A..48B}.
The ISRF model sampling grid differs from that used for the CR calculations and bi-/tri-linear (2D/3D) interpolation is used to determine the ISRF spectral intensity over the grid used for the \GP{} calculations.
Inelastic collisions by primary CR nuclei with the
interstellar gas yield \gray{s} and other particles.
The secondary CR $e^\pm$s from these interactions are included together with the primary electrons in the interstellar emission calculations because they produce a non-negligible contribution for \gray{}
energies $\lesssim 10$~MeV \citep{2008ApJ...682..400P,2011ApJ...739...29B}.
All calculations of the IC contribution use 
the anisotropic scattering cross section \citep{2000ApJ...528..357M} that
accounts for the full directional intensity distribution for each of the Std, R12, and F98 models.

How the different CR source densities affect the \gray{} intensities 
observed at Earth is examined first using calculations made with the Std ISRF
and the SA0, SA50, and SA100 models.
Changing the CR source density affects both nuclei and leptons and hence 
the summed interstellar emissions from $\pi^0$-decay,
Bremsstrahlung, and IC processes.
Figure~\ref{fig:StdSrcfrac}, top row, shows the total \gray{} intensity for the reference model combination (SA0--Std) for 10.6~MeV, 1.2~GeV, and 79~GeV energies, respectively.
The centre and bottom rows show the fractional residuals, (SA50-SA0)/SA0 and (SA100-SA0)/SA0, for the same energies.
Figure~\ref{fig:gammasrcstdlongprofile} shows the corresponding longitude
profiles separated according to \gray{} production processes ($\pi^0$-decay, Bremsstrahlung, and IC) for each of the CR density models averaged over $|b| \leq 7.5^\circ$ for the same energies.

For the low-energy (10.6~MeV) bin the \gray{} intensity has a negligible $\pi^0$-decay contribution and hence reflects the CR $e^\pm$ densities in the ISM.
For the outer Galactic quadrants the major contribution is by Bremsstrahlung, with Bremsstrahlung and IC approximately 50\% each for $45^\circ \leq |l| \leq 90^\circ$, and IC being dominant for $-45^\circ \leq l \leq 45^\circ$.
The low-energy fractional residuals can be interpreted using this information.
For the quadrants 2 and 3 the depletion of CRs from the smooth disc as the spiral arm injection power fraction is increased produces progressively less Bremsstrahlung because of the correspondingly lower CR $e^\pm$ densities in the outer Galaxy.
Toward the inner Galaxy a doughnut-like feature with excess emission
concentrated near $l \sim \pm 45^\circ$ and extending to high latitudes is evident.
This is caused by the pile-up along the line-of-sight of the IC emissions by CR $e^\pm$ propagating from their spiral arm injection regions, with the more intense
emissions from the inner arms coming from the higher injected luminosities there due to the arm radial profile.
The overall angular width of the excess feature increases with spiral arm injected power fraction and reflects that the lower energy CR $e^\pm$ effectively diffuse about the arms and out of the plane.
In and about the GC the deficit is related to the paucity of CR $e^\pm$ in this region.
The lack of CR sources (e.g., for the SA100 density model there are no CR sources within $\sim 3$~kpc of the GC producing a low CR energy density, see the right panel of Fig.~\ref{fig:crenergydensity}) and quick energy losses mean that there are much lower emissions
from this region.

For the $1.2$~GeV maps the \gray{} intensity is dominated by interactions
of the CR with the interstellar gas (only $\sim 10$\% of the \gray{s} are
produced by the IC process around these energies).
The fractional residuals are more highly structured because of the different CR densities in the ISM produced by the SA50 and SA100 density models.
They are higher in and about the spiral arms and illuminate the gas
that is nearby to these regions differently to the CR densities produced for the SA0 model.
The doughnut-like excess emission, particularly for the SA100 density model,
is present also at these energies.
It is more concentrated toward the plane
than for the lower energies because the gas scale-height is much smaller than that of the ISRF.
Even though the CR nuclei producing \gray{s} with energies $\sim 1$~GeV effectively fill the Galactic volume only relatively nearby ones ($\sim 100-500$~pc) injected in the
arms contribute to produce the excess features around these \gray{} energies.
(The lower energy maps have contributions by CR $e^\pm$ that have propagated far
above the plane, which is the reason for their broader latitude coverage.)
On the other hand the deficit seen at lower energies about the GC is not as deep.
Because the CR nuclei lose energy much slower than the electrons/positrons they
diffuse effectively to fill the inner region, even for the SA100 density model.

The 79~GeV maps generally exhibit the same residual features as the lower 
energies, but the concentration of excess emission is more pronounced about the spiral arm regions.
Because the IC is a greater proportion of the \gray{} emission at these energies ($\sim25-30$\%) the enhancements about the arms are mainly due to the primary CR electrons.
The energy losses are quick and the primary CR electron densities are high close to their injection regions.
Hence the excess emission is much higher for lower latitudes about $l \sim \pm 45^\circ$ than for the lower energies.

At all energies the negative residuals in the outer Galaxy for the SA100 model are deeper than the SA50 case.
This comes from the lack of the smooth CR `disc' component there, the difference between the smooth disc and arm radial scale-lengths, and the effective cut-off in the arm density distributions because of their finite angular extent (Table~\ref{table:r12armparams}).

\begin{figure*}[htb!]
  \includegraphics[width=0.33\textwidth]{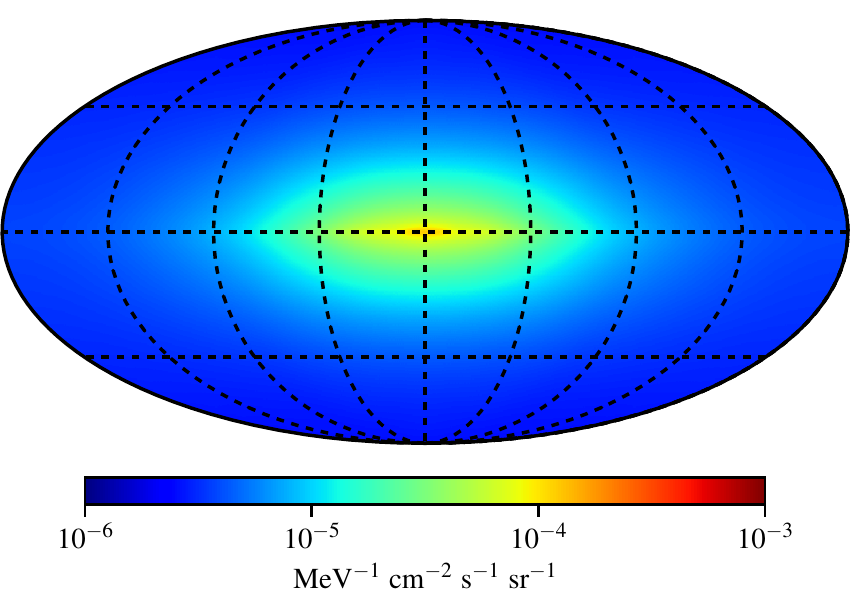}
  \includegraphics[width=0.33\textwidth]{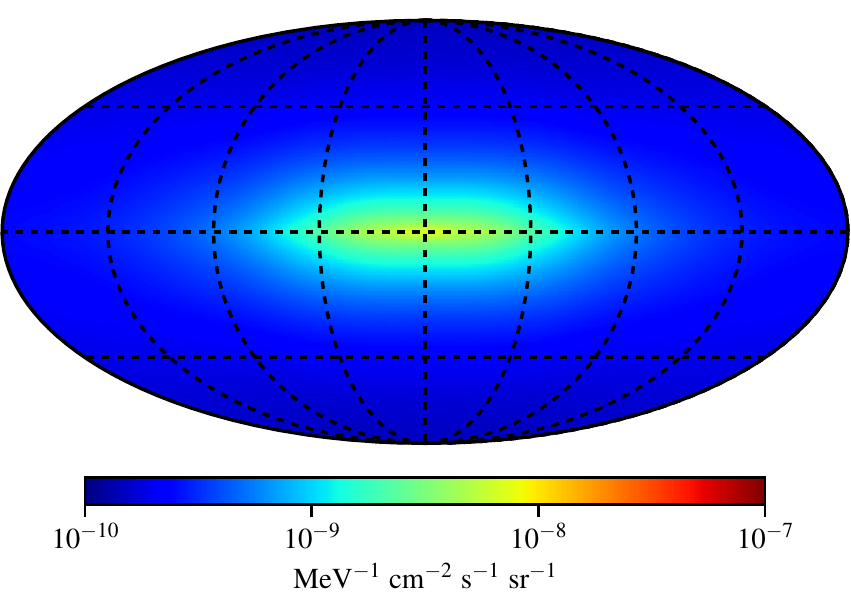}
  \includegraphics[width=0.33\textwidth]{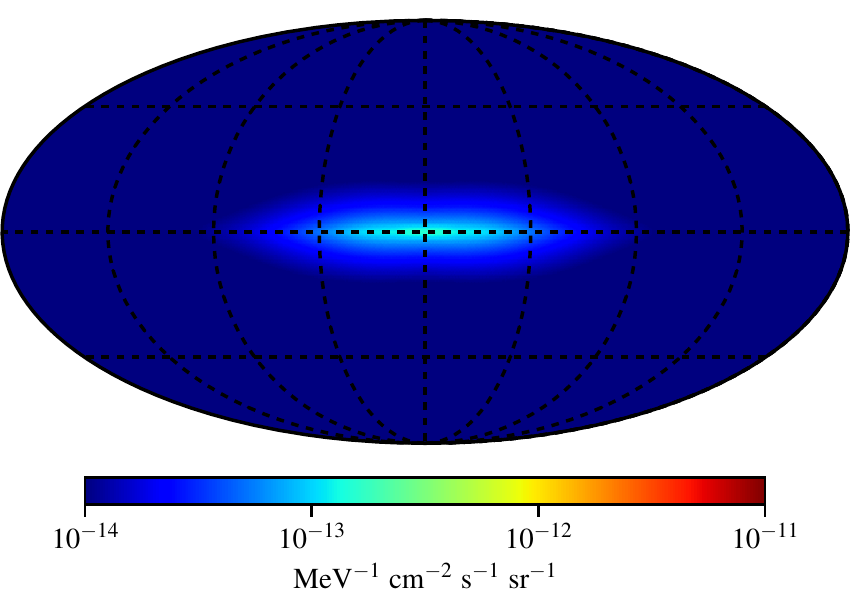}\\
  \includegraphics[width=0.33\textwidth]{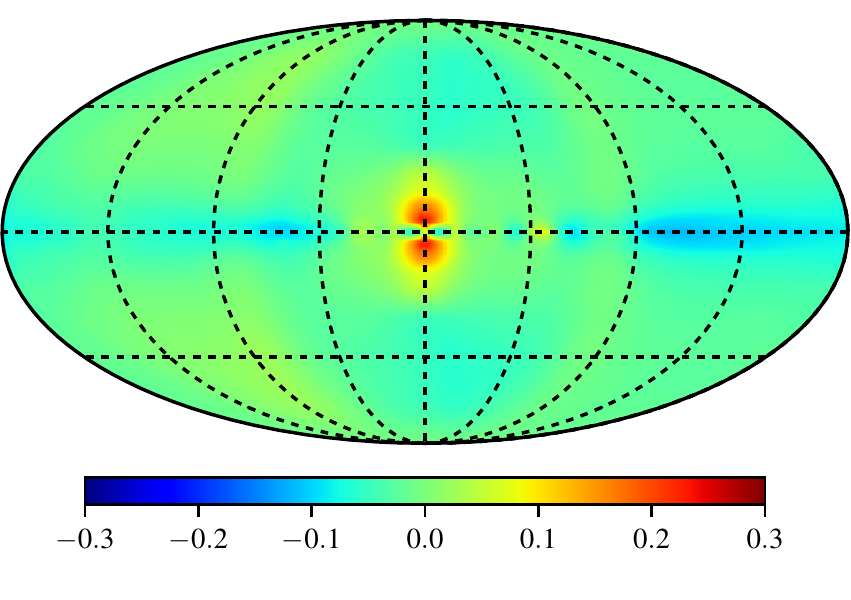}
  \includegraphics[width=0.33\textwidth]{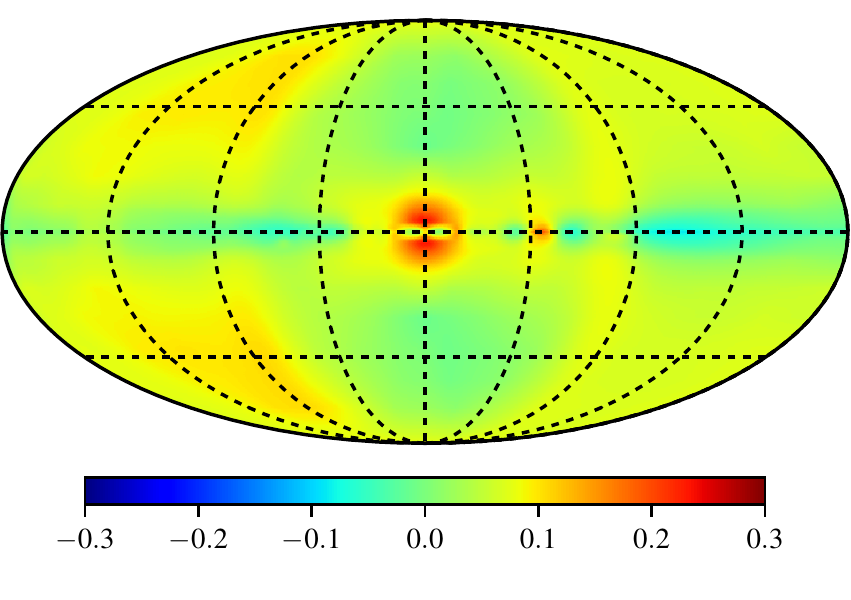}
  \includegraphics[width=0.33\textwidth]{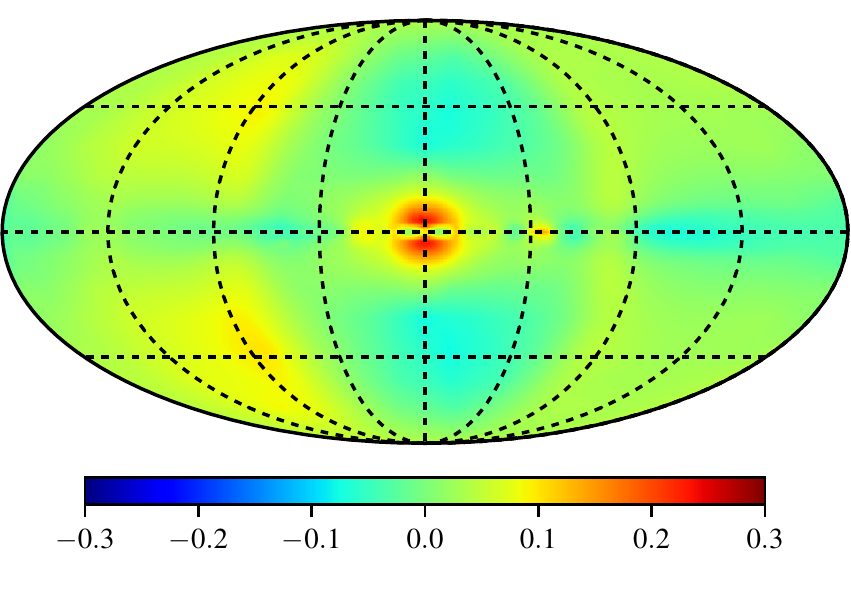}\\
  \includegraphics[width=0.33\textwidth]{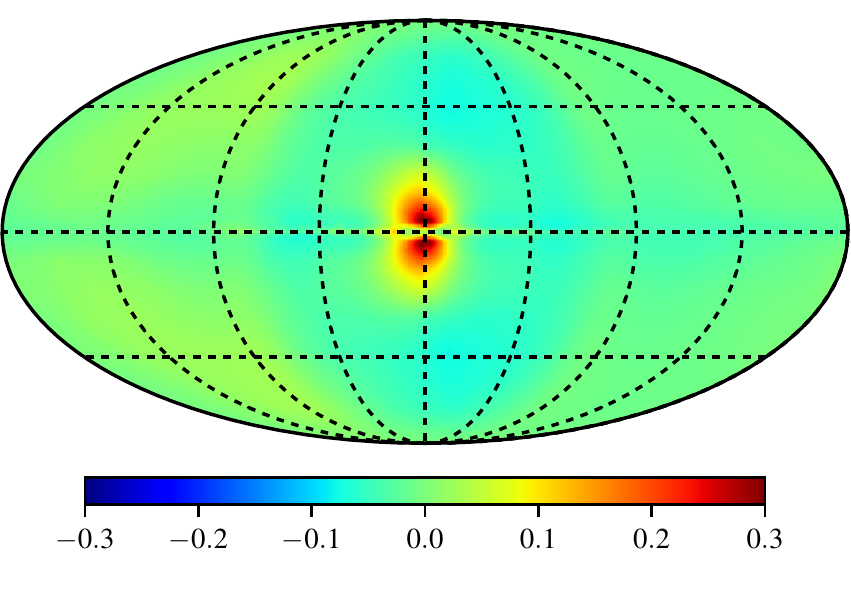}
  \includegraphics[width=0.33\textwidth]{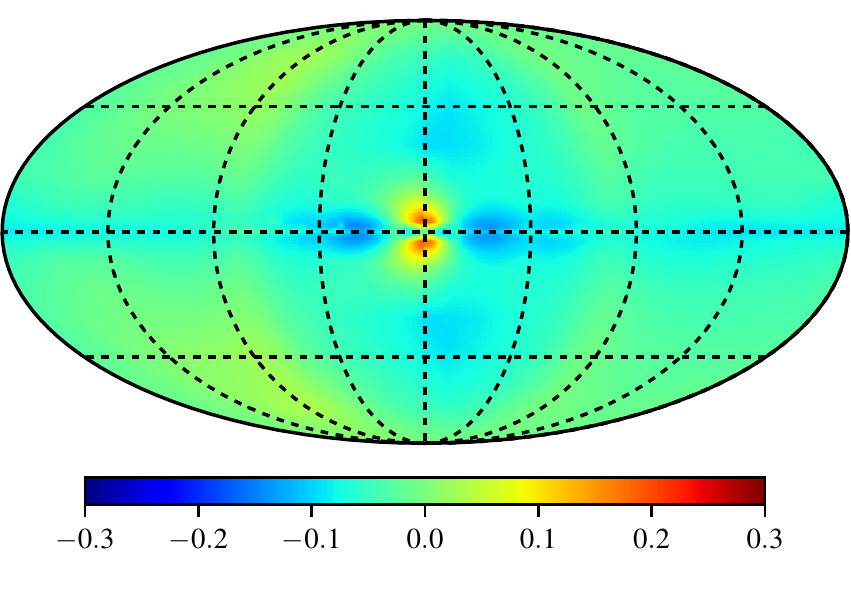}
  \includegraphics[width=0.33\textwidth]{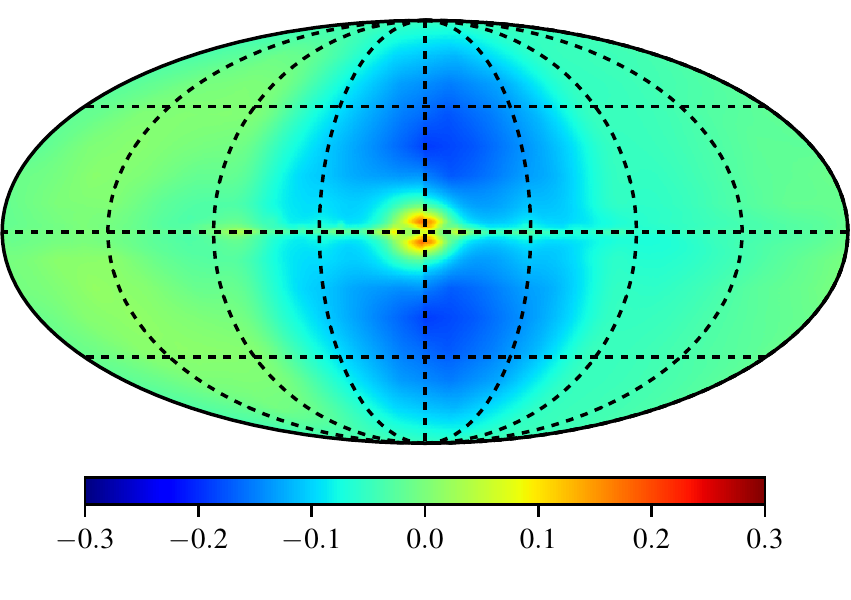}
  \\
  \caption{Top row: IC intensity at 10.6~MeV, 1.2~GeV, and 79~GeV energies (left to right, respectively) for SA0--Std reference case.
    Centre and bottom rows: fractional residuals for the SA0--R12 and SA0--F98 model combinations, respectively.
    The maps are in Galactic coordinates with $l,b = 0^\circ,0^\circ$ at the
    centre.
    The longitude meridians and latitude parallels have $45^\circ$ spacing, as in Fig.~\ref{fig:StdSrcfrac}. 
    \label{fig:StdISRFfrac1}}
\end{figure*}

The effect on the IC component of changing from the Std to R12/F98 ISRF models is illustrated in Fig.~\ref{fig:StdISRFfrac1} where 
the IC intensity for the SA0--Std model combination is shown at 10.6~MeV, 1.2~GeV, and 79~GeV energies (top row, left to right, respectively), together with
the fractional residuals for the SA0--R12 (centre row) and SA0--F98 (bottom row) combinations.
The residuals at all energies exhibit a feature around the GC that 
is associated with the Galactic bulge/bar.
The ISRF spectral intensity over the spatial region dominated by this component
is higher for both the R12 and F98 models compared to the Std.
In addition, 
the R12 and F98 model bulge/bars have larger scale-heights than that of the Std model.
These major differences produce the fractional residuals that show a
spatial morphology with strongly positive `lobes' above and below the GC that are model- and energy-dependent.
The energy dependence reflects the effect of the energy losses on the CR $e^\pm$s toward the inner Galaxy for R12 and F98 models.

Elsewhere the R12 model shows numerous residual features across the sky directly related to
the presence of the spiral arms.
Close to the plane these are due to the 4 main spiral arms (Table~\ref{table:r12armparams}) where hot/cold regions in the residuals come from the
localised enhanced/decreased spectral intensities (Fig.~\ref{fig:isrfenergydensityplane}, left).
At intermediate and high Galactic latitudes the excesses have contributions by both the main and local arms.
For \gray{} energies $\sim 1$~GeV the CR e$^\pm$ producing the IC emission have energies $\sim$~few--10~GeV\footnote{In the Thomson regime the
  \gray{} energy $\sim \gamma_e^2 \epsilon_{\rm ISRF}$ where $\gamma_e$ is the
  electron Lorentz factor and $\epsilon_{\rm ISRF}$ is the target photon energy. Thus a $\sim$~GeV IC \gray{} is produced mainly by $\gtrsim$~few~GeV e$^\pm$ scattering the UV/optical component of the ISRF.} and are distributed throughout the disc and into the halo.
The broadly distributed features for $\sim 1$~GeV come from both nearby and relatively far away in the Galaxy.
Meanwhile, for higher energy \gray{s} the residual maps display features out
of the plane that are due to localised features because
the CR e$^\pm$ producing these emissions are relatively close to their injection sites.

The residuals for the F98 model are less structured than for R12.
The major differences are related to the asymmetric bar, which produces strong
enhancements around the GC and smaller enhancements outside of the plane.
Those close to the GC are related to the higher spectral intensity and spatial
distribution of the bar compared to the Std ISRF model.
Outside the plane the enhancements come from the anisotropic scattering of the
outward moving photons from the inner Galaxy, where the effect is higher for quadrant 2 compared to quadrant 3 because the directional intensity of the bar is stronger there.
About the Galactic plane the F98 model is comparable to, or has a small deficit to the Std ISRF, because the former has a smaller radial scale-length for the
disc populations (the Std ISRF disc radial scale-length is similar to that of the R12 model).
Even though the normalisation of the F98 disc is higher compared to
the Std ISRF the smaller radial scale-length gives a comparatively reduced spectral intensity away from the inner Galaxy.

\begin{figure*}[htb!]
   \includegraphics[width=0.33\textwidth]{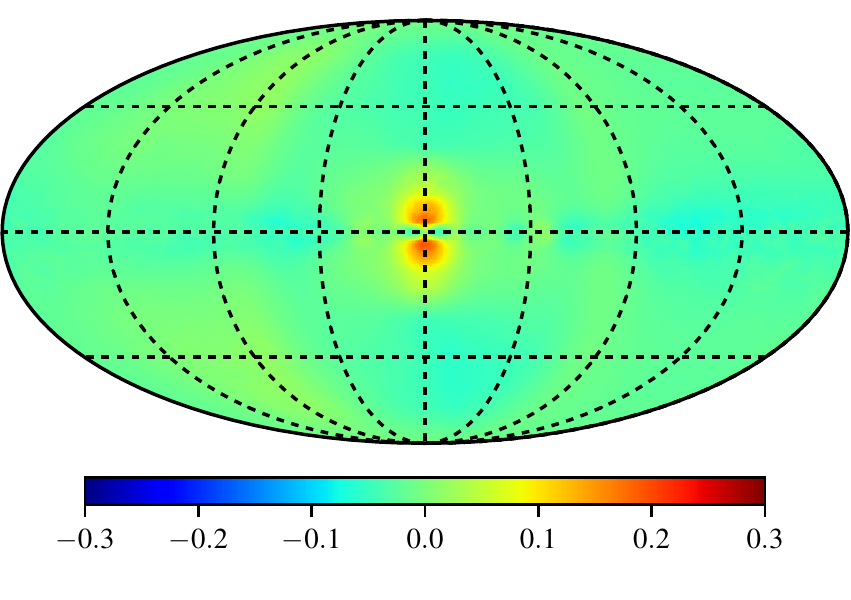}
   \includegraphics[width=0.33\textwidth]{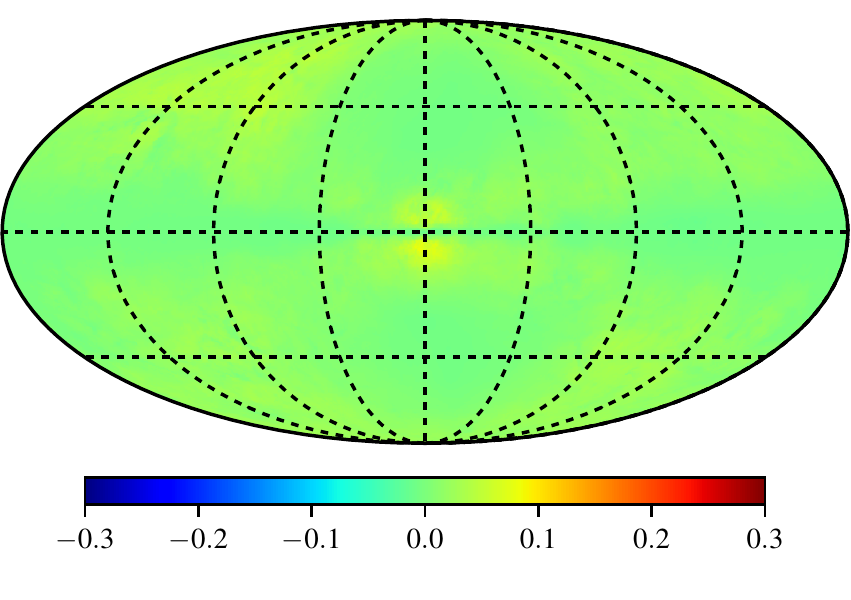}
   \includegraphics[width=0.33\textwidth]{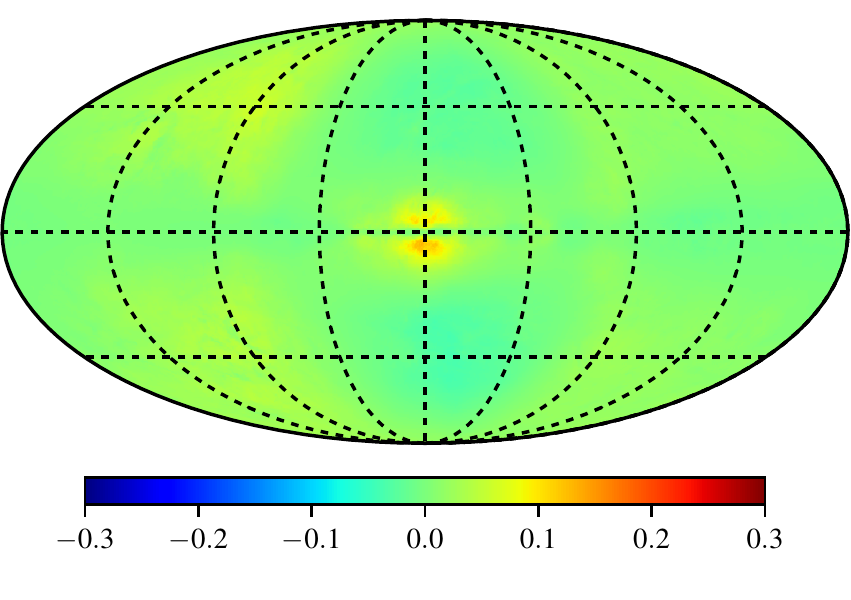}
   \\
   \includegraphics[width=0.33\textwidth]{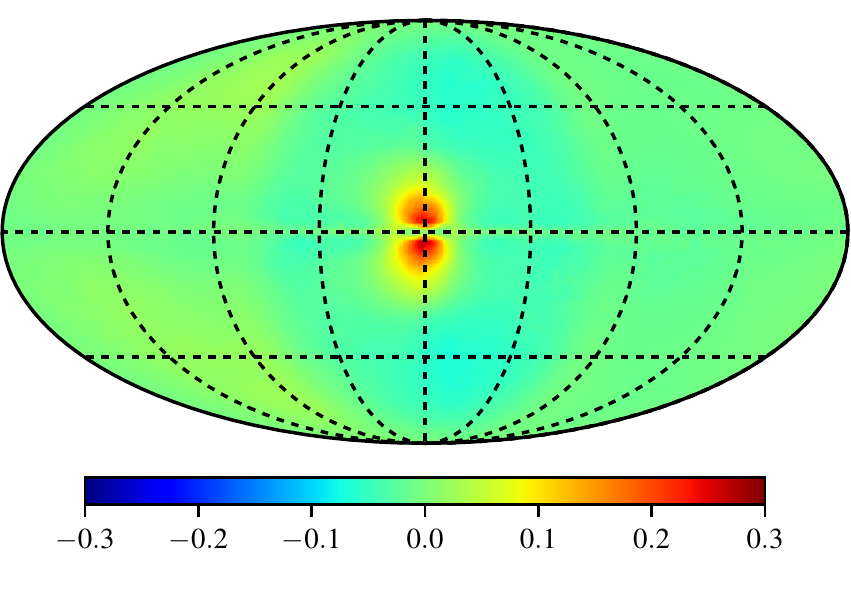}
   \includegraphics[width=0.33\textwidth]{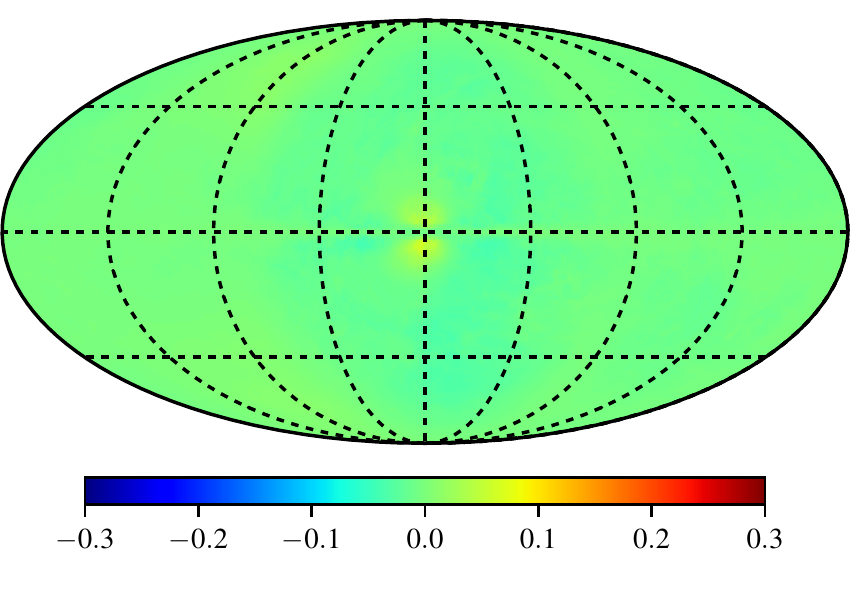}
   \includegraphics[width=0.33\textwidth]{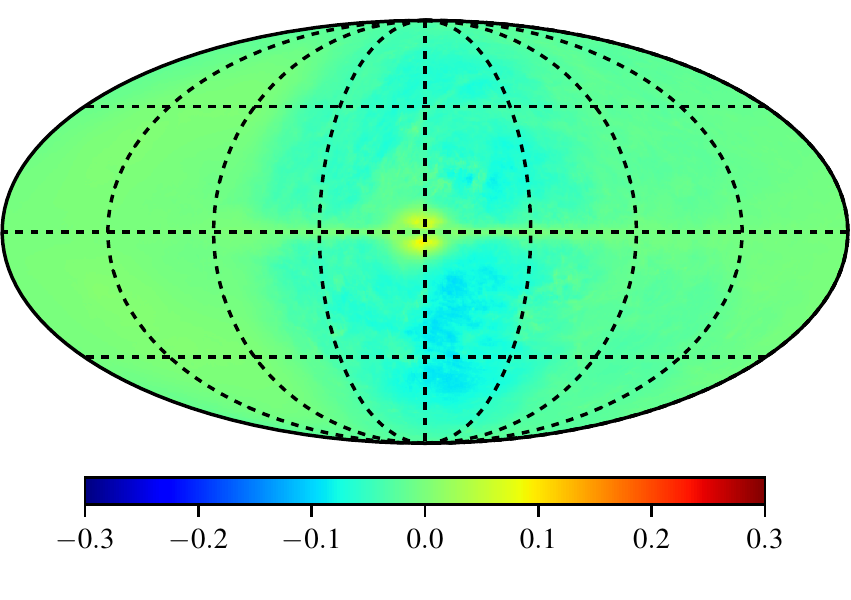}
  \caption{Fractional residuals for the total intensity
    ($\pi^0$-decay, Bremsstrahlung, and IC) at
    10.6~MeV (left), 1.2~GeV (centre), and 79~GeV (right)
    for SA0--Std and the SA0--R12 (top) and SA0--F98 (bottom) model combinations.
    The maps are in Galactic coordinates with $l,b = 0^\circ,0^\circ$ at the
    centre.
    The longitude meridians and latitude parallels have $45^\circ$ spacing, as in Fig.~\ref{fig:StdSrcfrac}.
    \label{fig:StdISRFfrac2}}
\end{figure*}

For the SA0 CR source density model the effect of changing the target density model for the ISRF on the total \gray{} intensity is, however, not as stark as the changes in the IC component maps.
This is illustrated in Fig.~\ref{fig:StdISRFfrac2} where the fractional residuals with the reference case for the SA0--R12 and SA0--F98 model combinations are
shown.
For the lowest energy ($10.6$~MeV) maps the residuals due to the 
higher spectral intensities for the R12/F98 bulge/bar are most strongly seen.
This comes from the primacy of the IC component at the lower energies toward
the inner Galaxy, as discussed earlier.

Combining the SA50 and SA100 CR density models with the 3D ISRF models yields the fractional residuals with respect to the SA0--Std reference that are shown in Fig.~\ref{fig:ISRFSrcfrac}.
The columns from left to right are increasing with energy as for Figs.~\ref{fig:StdSrcfrac} and~\ref{fig:StdISRFfrac1}, respectively.
The rows correspond to the (top to bottom)
SA50--R12, SA100--R12, SA50--F98, and SA100--F98
model combinations and show a rich structure that depends on both the CR source densities and propagation, and the ISRF target density model.

For the R12 ISRF model there is a multiplier effect of the CR densities with the ISRF densities in and about the arm regions.
The emissions are more intense and broadly distributed on the sky than simply altering alone the CR source or ISRF to include spiral structure for either spatial distribution model.
This is the case even around $\sim 1$~GeV where the CRs effectively fill the Galactic volume -- the features from the double arm `enhancement' from the CR source and ISRF density models are evident.
For the low-energy maps (10.6~MeV) the effect of the more luminous bulge in the R12 model is also visible.
The higher spectral intensity of the ISRF in and about the GC region compensates for the reduced CR density to produce an excess, or reduced deficit, compared to the SA0--Std combination.

On the other hand for the F98 ISRF model and SA50/SA100 combinations there is no multiplier effect because distinct spatial components for the spiral arms are only present for the CR source densities.
But the residual maps are more than a simple reflection of the changing CR source densities because the F98 ISRF model also produces structural differences due to its different stellar luminosities and spatial structures compared to the Std model (Fig.~\ref{fig:StdISRFfrac1}).
Toward the inner Galaxy the low-energy maps show more of an enhancement/reduced deficit, compared to the SA50/SA100--R12 cases, because of the bulge luminosity and shape.
Meanwhile, the excesses identified previously associated with CRs injected into the ISM in and about the arm regions are not as intense at higher latitudes because of the reduced IC emissions out of the plane.

\section{Discussion}
\label{discussion}

\begin{figure*}[htb!]
  \hfill
  \includegraphics[width=0.33\textwidth]{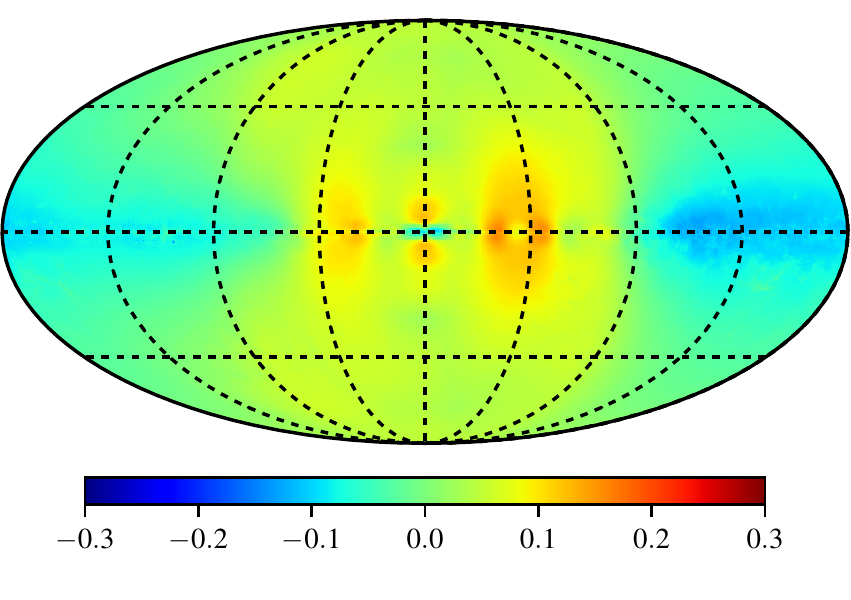}
  \includegraphics[width=0.33\textwidth]{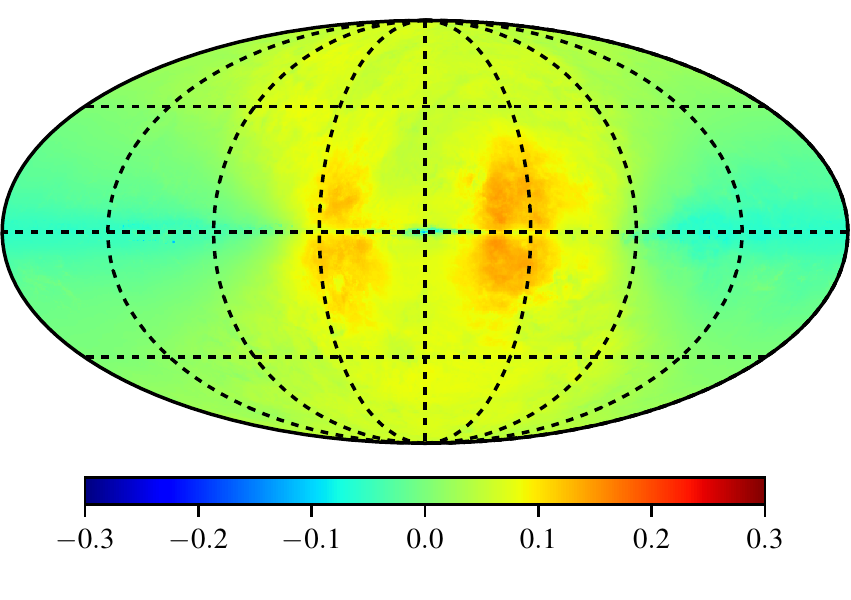}
  \includegraphics[width=0.33\textwidth]{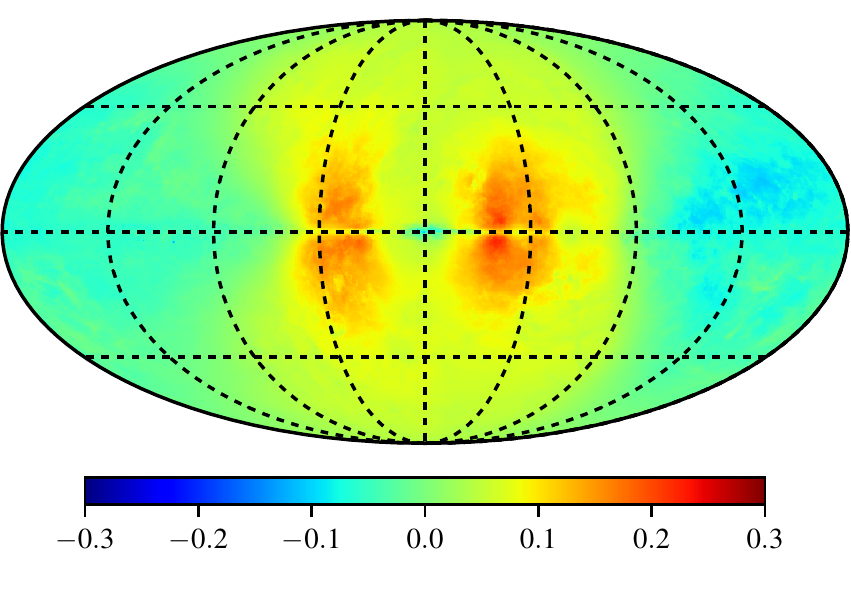}
  \\
  \hfill
  \includegraphics[width=0.33\textwidth]{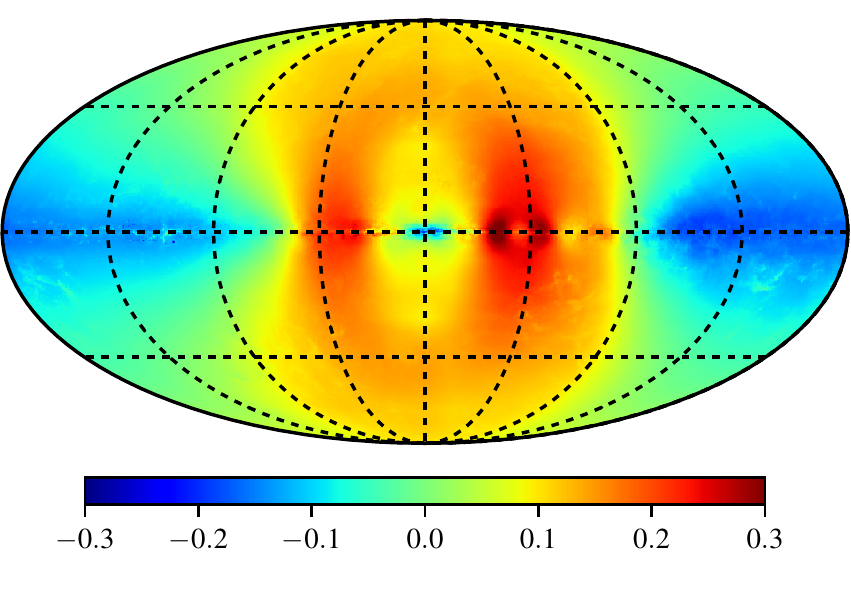}
  \includegraphics[width=0.33\textwidth]{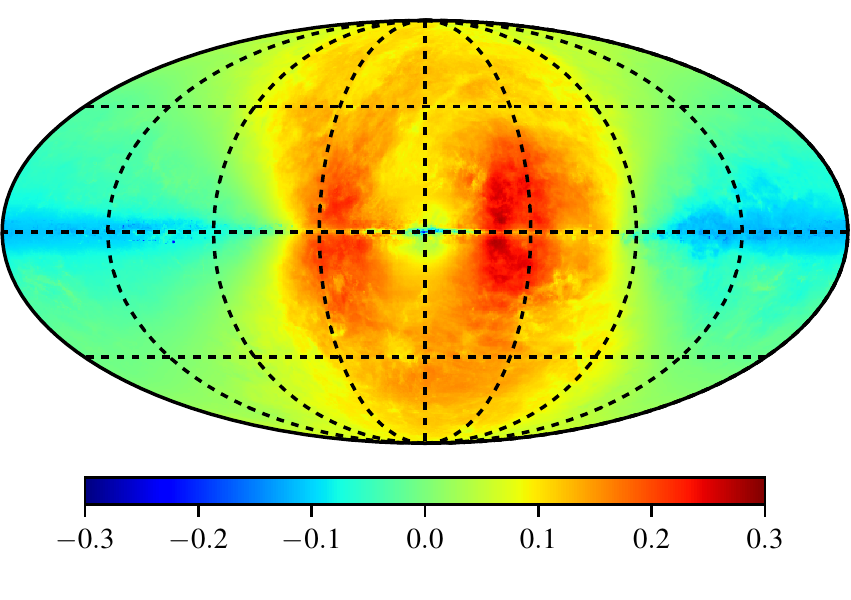}
  \includegraphics[width=0.33\textwidth]{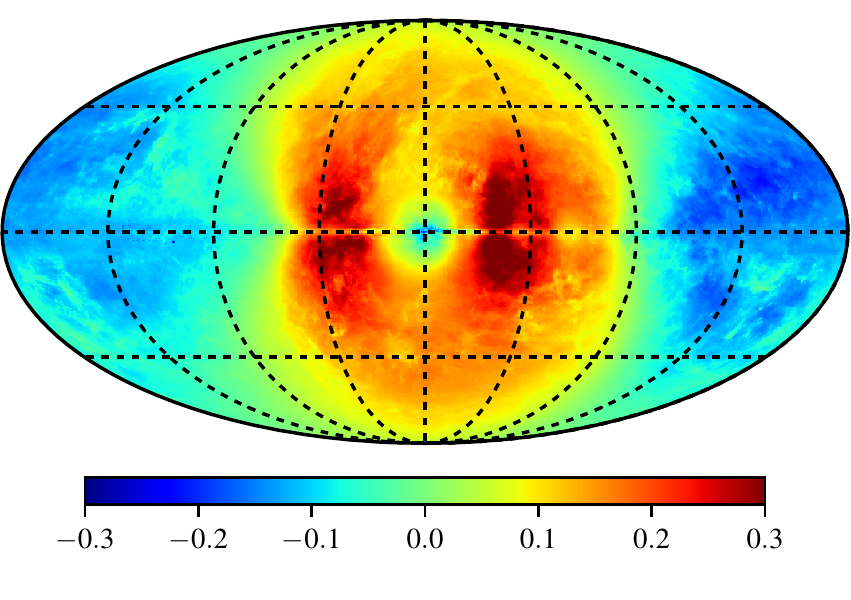}
  \\
  \hfill
  \includegraphics[width=0.33\textwidth]{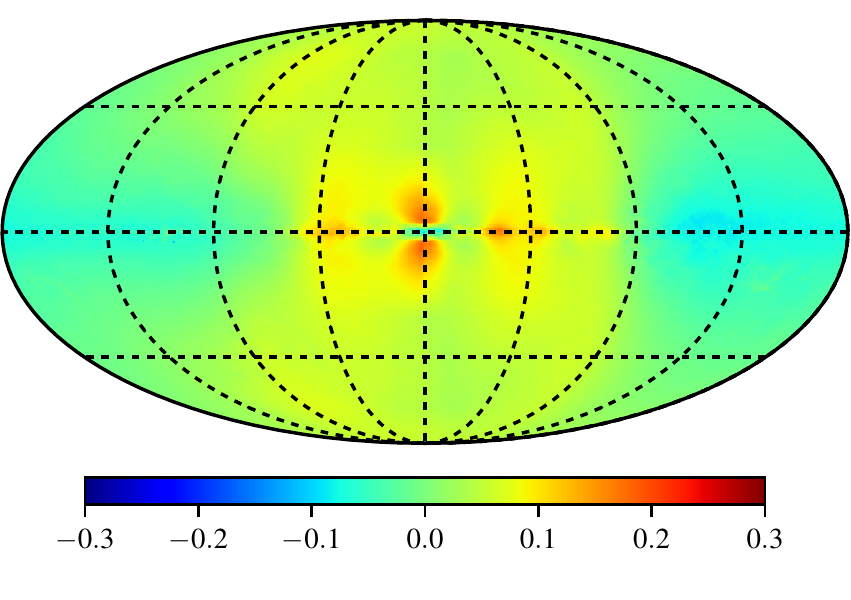}
  \includegraphics[width=0.33\textwidth]{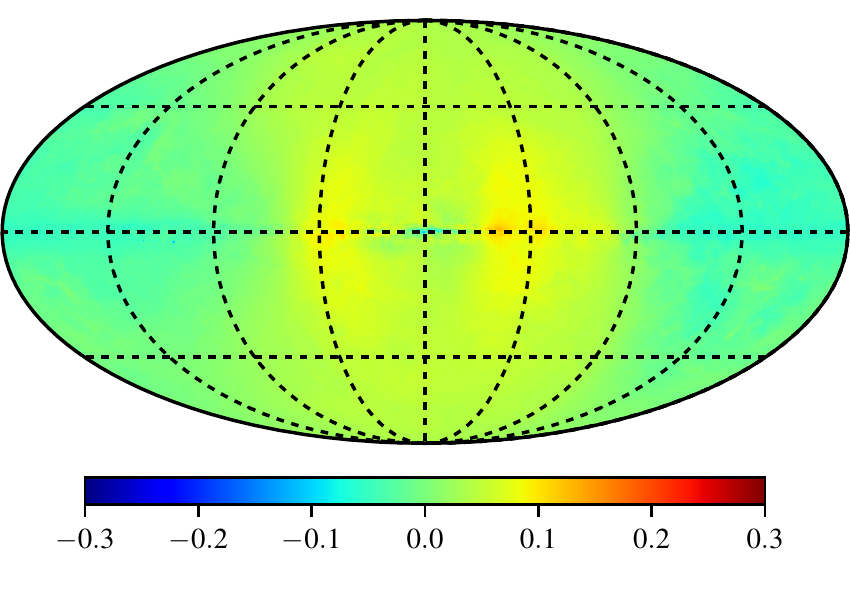}
  \includegraphics[width=0.33\textwidth]{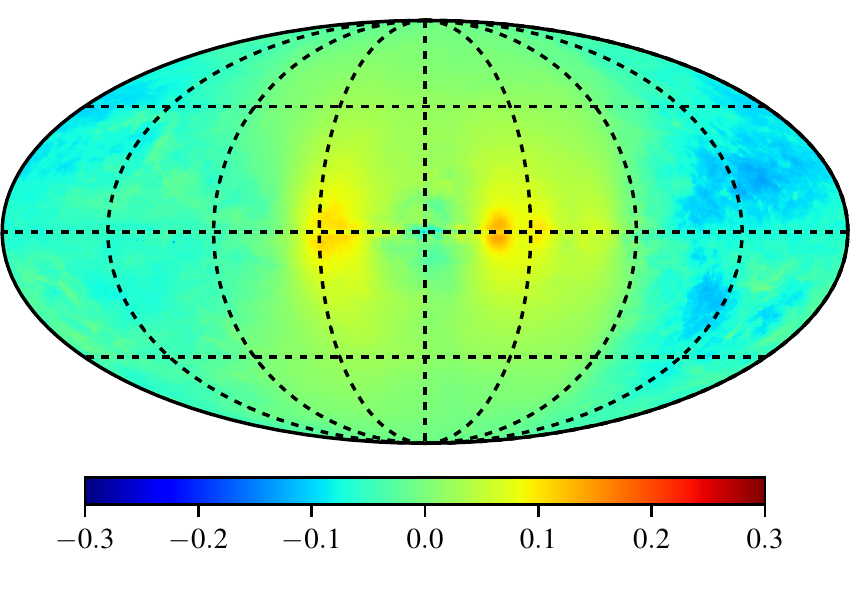}
  \\
  \hfill
  \includegraphics[width=0.33\textwidth]{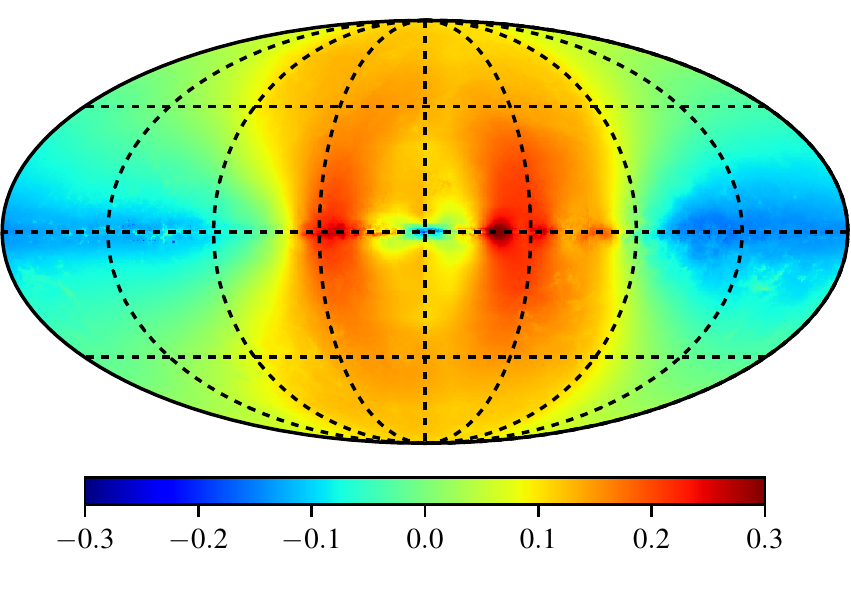}
  \includegraphics[width=0.33\textwidth]{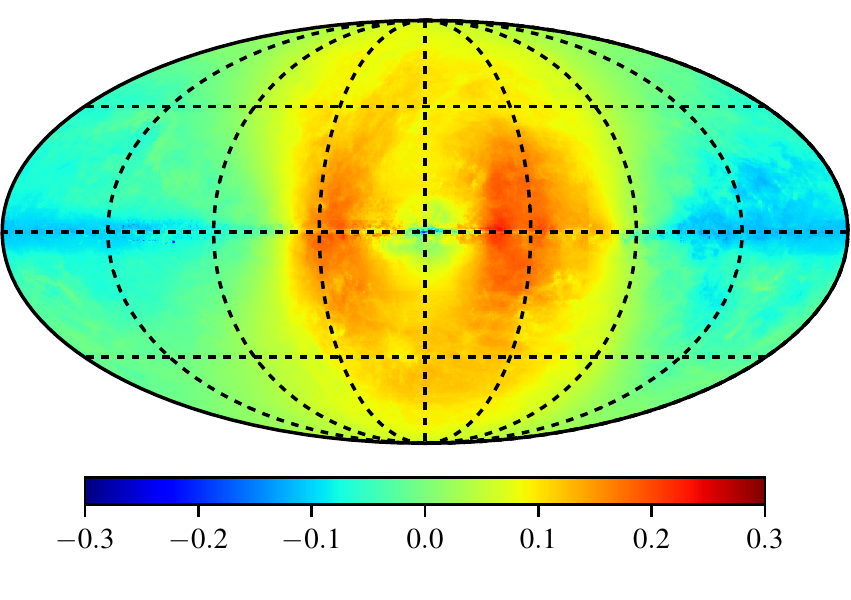}
  \includegraphics[width=0.33\textwidth]{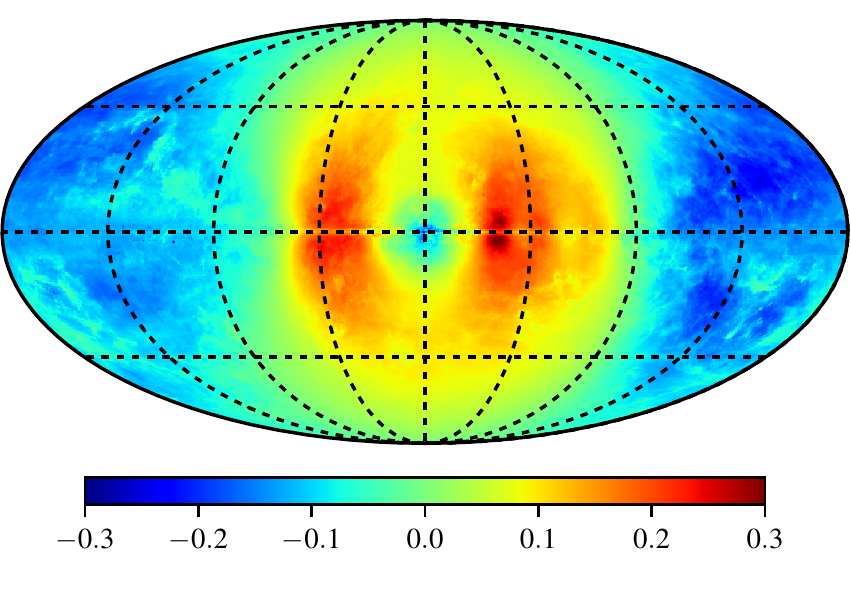}
  \\
  \caption{
    Fractional residual maps for the SA50--R12 (first row), SA100--R12 (second row), SA50--F98 (third row), and SA100--F98 (fourth row) model combinations with the SA0--Std reference case.
    Left to right are the fractional residuals for given model combination
    with increasing energy at 10.6~MeV, 1.2~GeV, and 79~GeV.
    The maps are in Galactic coordinates with $l,b = 0^\circ,0^\circ$ at the
    centre.
    The longitude meridians and latitude parallels have $45^\circ$ spacing.
    \label{fig:ISRFSrcfrac}}
\end{figure*}

The residuals shown in Figs.~\ref{fig:StdSrcfrac} 
and~\ref{fig:ISRFSrcfrac} display many interesting features.
The question is then what resemblance, if any, do they have to large-scale residuals from analysis of real \gray{} data?
A comparison can be made with the results of 
IG16 because the SA0--Std reference model of this paper is very close to the `Pulsars' IEM from that paper\footnote{The SA0--Std IEM of this paper is based on the `Pulsars' IEM with refitting of the propagation model parameters to account for the 3D propagation volume and newer CR data. Taking the fractional residuals between the SA0--Std and `Pulsars' IEM shows differences $\lesssim 5$\% for $-90^\circ \leq l \leq 90^\circ$. The fractional residuals are somewhat higher for the outer Galactic quadrants but for those regions the intensities are generally low compared to the inner quadrants, and the differences between IEMs are mainly from the cylindrical vs. Cartesian geometries.}.
The residuals from the upper panels of Fig.~2 of IG16 are obtained by subtracting the baseline `Pulsars' IEM and a fitted low-intensity isotropic intensity from the \fermilat\ data, which enables direct comparison
with the residuals calculated in this paper.

\begin{figure}
  \subfigure{
    \includegraphics[width=0.5\textwidth]{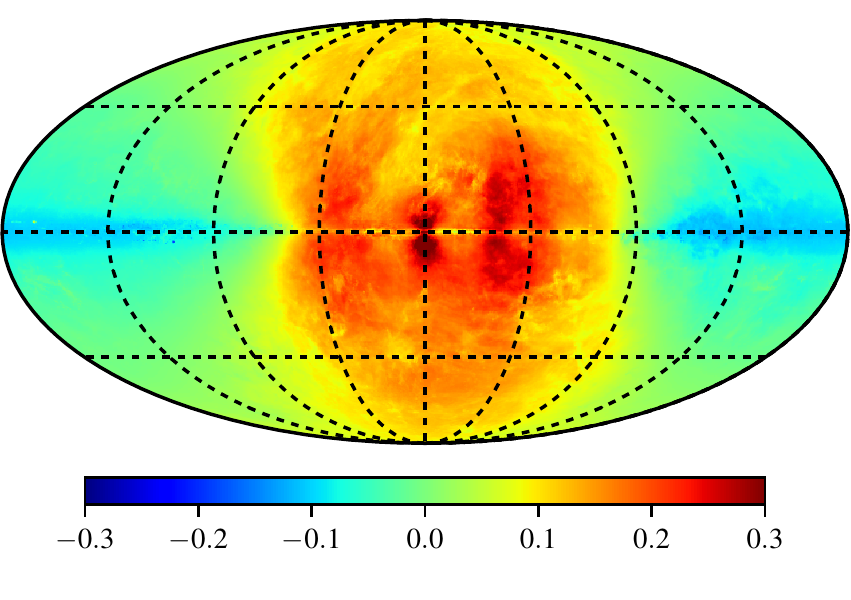}
  }\\
  \subfigure{
    \includegraphics[width=0.5\textwidth]{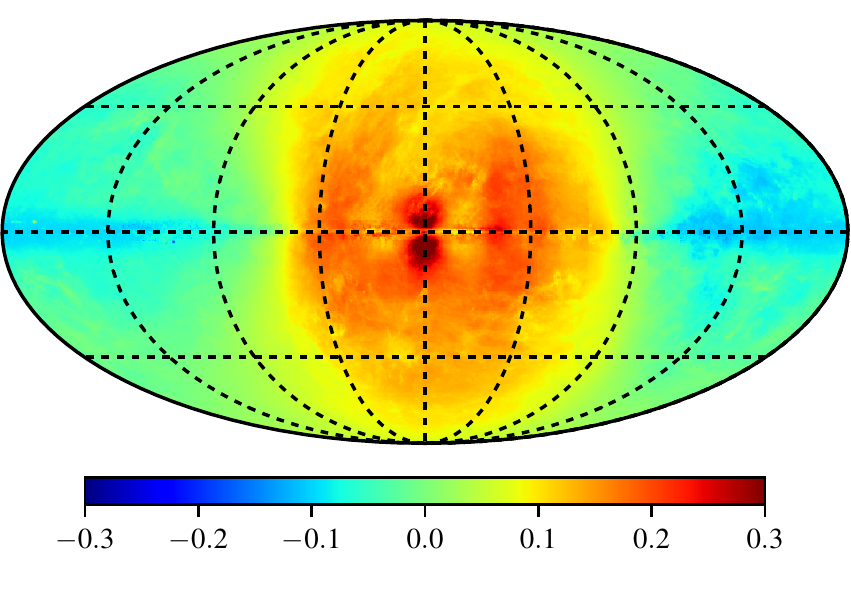}
  }
  \caption{
    Fractional residual map at 1.2~GeV for the SA100/R12B--R12 (top) and SA100/F98B--F98 (bottom) CR source and ISRF density model combinations.
    The map is in Galactic coordinates with $l,b = 0^\circ,0^\circ$ at the
    centre.
    The longitude meridians and latitude parallels have $45^\circ$ spacing.
    \label{fig:CRbulge1}}
\end{figure}

The focus of the discussion will be toward the inner Galactic quadrants 
because it is in this region where the major differences related to the 
3D CR source and ISRF densities are important\footnote{The residuals in the outer Galaxy are where the \gray{} intensities are much lower and differences in the gas column density distribution, CR source density gradient, and halo size, are all plausible explanations to improve the agreement between data and models. See, e.g., \citet{2010ApJ...710..133A}.}.
Generally the IG16 fractional residuals show excesses $\gtrsim 30$\% that extend out of the plane for $-45^\circ < l < 45^\circ$.
These are somewhat higher but similar in spatial distribution to the SA50/SA100 residuals shown in Fig.~\ref{fig:ISRFSrcfrac} where there are peaks near $l \sim 45^\circ$ and $l \sim -30^\circ$ corresponding to the spiral arm enhancements together with a general in-fill up to $|b| \sim 45^\circ$ latitudes.
Because the IEMs calculated in this paper have not been optimised using \gray{} data perfect agreement is not expected.
But it is interesting that for longitudes outside of the 
inner $\pm 45^\circ$ the IG16 positive residuals are mainly clustered near 
the plane, while for longitudes inside they have extensions to high latitudes\footnote{The so-called `Loop-I' feature also contributes about these longitudes with a relatively narrow angular width, but its contribution for $|b| \lesssim 30^\circ$ is relatively small compared to the ISM emissions.}.
The delineation is distinct and aligned with where the spiral arms 
provide major contributions for the SA50/SA100 IEMs, as shown in Fig.~\ref{fig:ISRFSrcfrac}.

However, the SA50/SA100 model combinations have the central `hole' about the GC caused by the lower CR energy density in that region.
The residuals shown in Fig.~2 of IG16 have the region about $|l| \lesssim 20^\circ$ and $|b| \lesssim 50^\circ$ obscured\footnote{The masked regions in the IG16 figures indicate areas on the sky where the \gray{} data was not used in the scaling procedure to develop the fore-/background IEMs employed in that analysis.}, but the level of the residuals under these masks is similar, if not higher, than those over the rest of the $-45^\circ \leq l \leq 45^\circ$ region.
A resolution to obtain comparable residual excesses that in-fill the region about the GC for the SA50/SA100 models is to introduce an extra source density model that provides additional CR power there.
An explanation for such an additional component is that it could be a bulge-related population of CR accelerators that injects a combination of nuclei and leptons\footnote{\citet{2016PhRvD..94f3504C} have considered also a scenario with CRs injected according to a Galactic molecular hydrogen distribution model including a bar and spiral arms.}.
To examine this possibility the SA100--R12 and SA100--F98 model combinations are recalculated using the bulge/bar for each ISRF model as the hypothetical additional CR central source density distribution (Eq.~\ref{r12:bulgedensity} for the SA100--R12 combination, termed ``SA100/R12B'', and Eq.~\ref{f98:baremissivity} for the SA100--F98 combination, termed ``SA100/F98B'', respectively), assuming the same nuclei-to-lepton ratio for the CR bulge/bar as the spiral arms.
Because this is testing a simple `what-if?' scenario no likelihood fitting of \gray{} data is made.
Instead the normalisation of the total injected
CR power for the additional source density component is obtained for each
model combination by adjusting by-eye its value so that residuals around the GC seen in the second/fourth row (centre) of Fig.~\ref{fig:ISRFSrcfrac} are sufficiently filled and at least comparable to the residuals elsewhere for the SA100 models toward the inner Galaxy.

\begin{figure*}[ht]
  \subfigure{
    \includegraphics[scale=0.8]{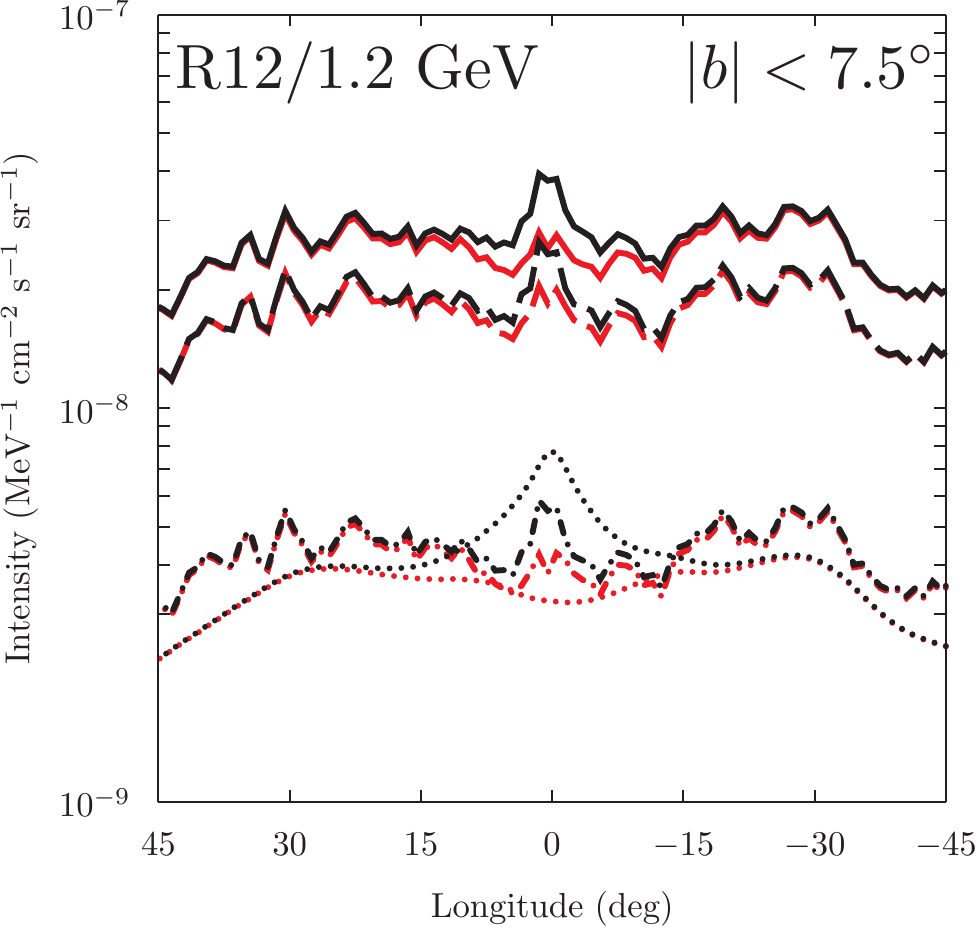}
    \includegraphics[scale=0.8]{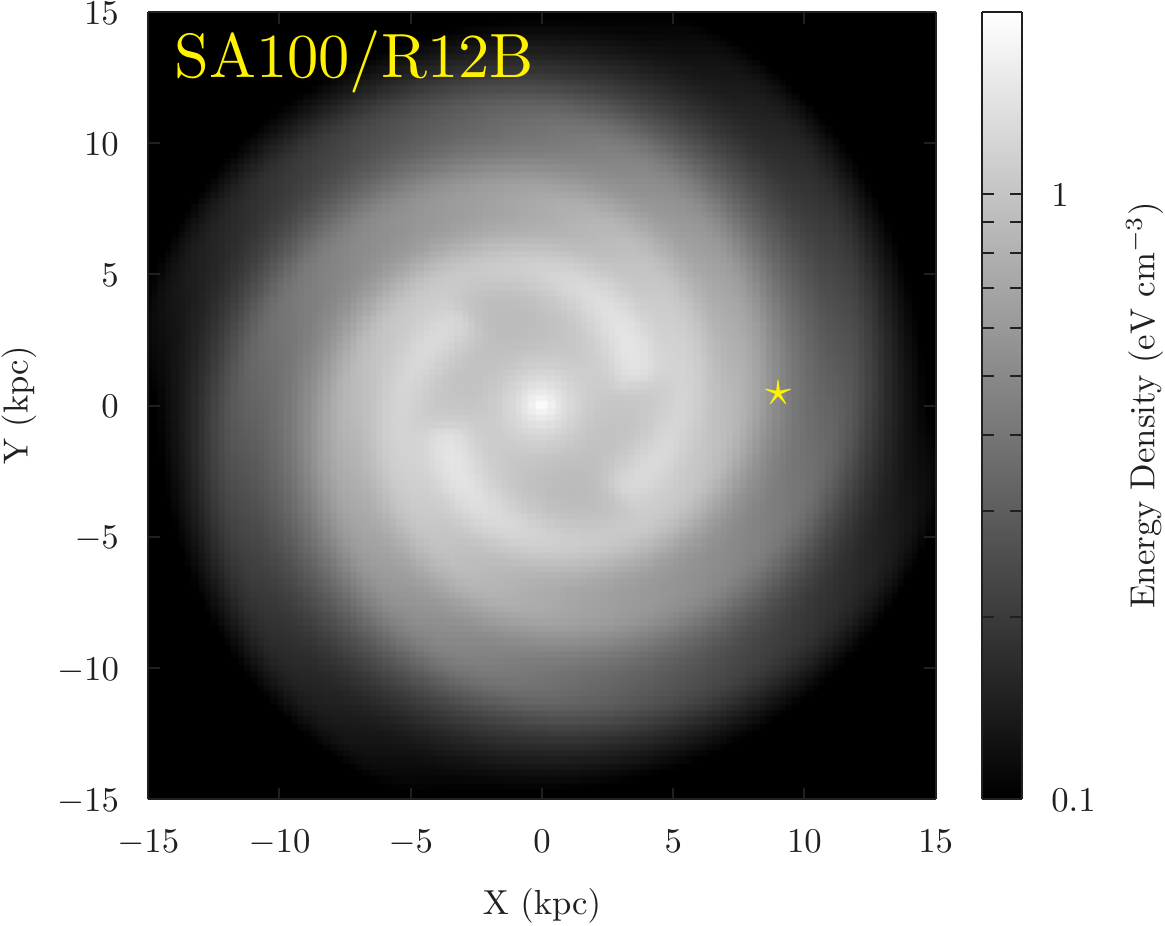}

  }\\
  \subfigure{
    \includegraphics[scale=0.8]{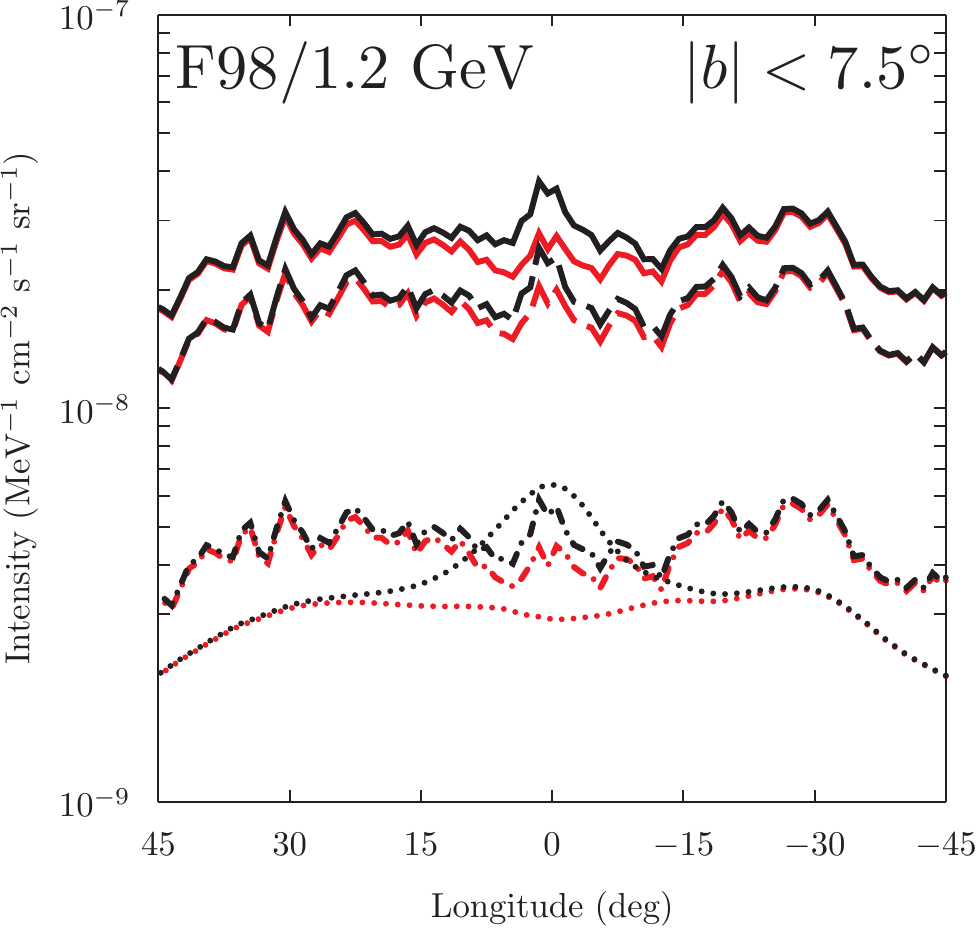}
    \includegraphics[scale=0.8]{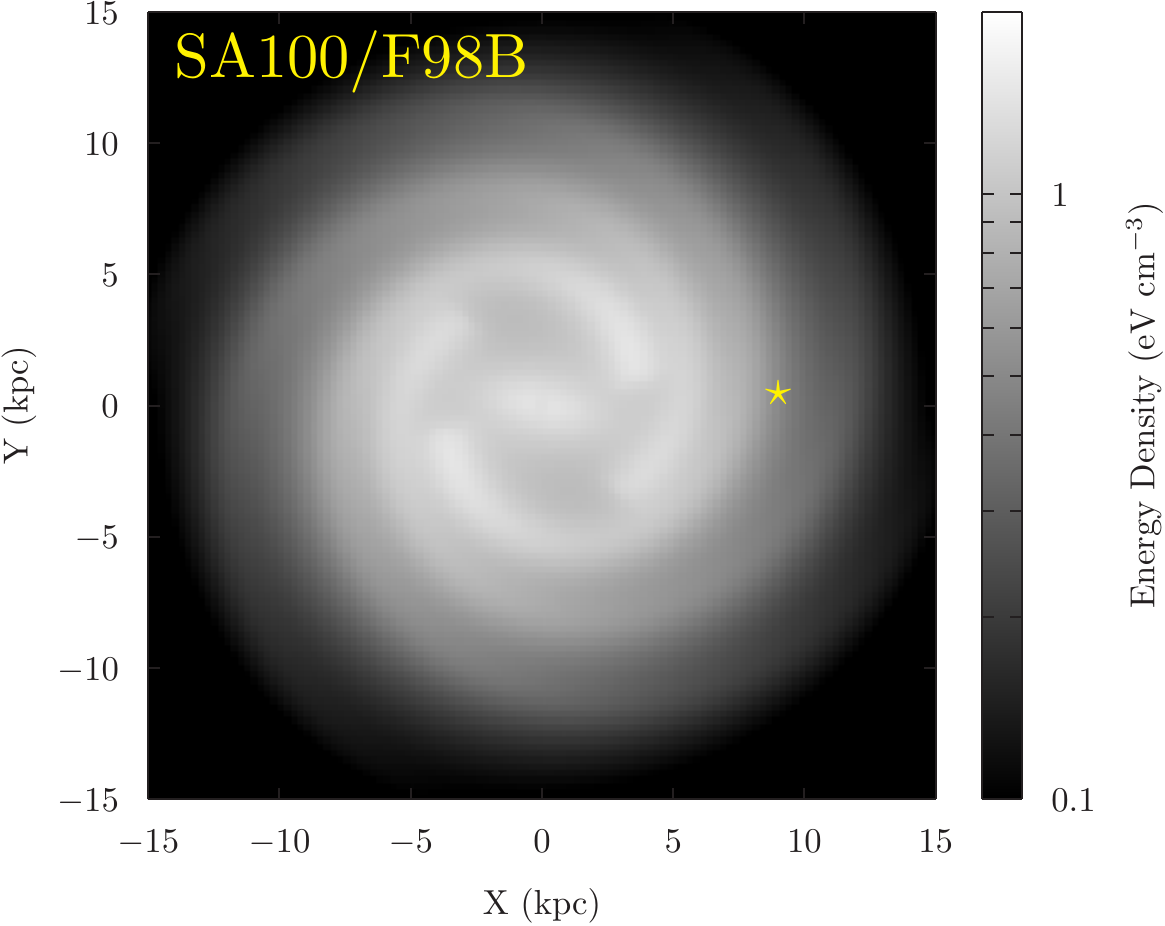}
  }
  \caption{
    Left panel: longitude profile for the intensity at 1.2~GeV averaged over $-7.5^\circ \leq b \leq 7.5^\circ$. Top shows the SA100--R12 combination (red lines) and SA100/R12B--R12 combination (black lines). Bottom shows the SA100-F98 combination (red lines) and SA100/F98B--F98 combination (black lines).
    Line-styles: solid, total; long-dashed, $\pi^0$-decay; dash-dot, Bremsstrahlung; dotted, IC.
    Right panel: spatial distribution of integrated CR energy density at the Galactic plane for the SA100/R12B (top) and SA100/F98B (bottom) CR source density models. The yellow star shows the location of the Solar system.
    \label{fig:CRbulge2}}
\end{figure*}

Figure~\ref{fig:CRbulge1} shows the fractional residuals for energies $\sim 1$~GeV for the SA100/R12B--R12 (top) and SA100/F98B--F98 (bottom) model combinations.
For both modified CR source density models the central `hole' in the residuals is filled to give excesses $\sim 30-40$\%.
The adjustment could be made to give a higher fraction and still be consistent with what is seen in the IG16 residuals, but as described above these calculations are made to serve an illustrative purpose and no additional fine-tuning is made.
Also, note that no refitting of the propagation parameters is performed, or indeed necessary, because the CRs from the bulge/bar addition produce negligible contributions at the Solar system (see Fig.~\ref{fig:CRbulge2} right panels).

The addition of the bulge/bar CR source density model produces residuals that 
show a large excess close to the GC with lobe-like structures above/below
the GC, similar to those seen at lower energies (Figs.~\ref{fig:StdISRFfrac2} and~\ref{fig:ISRFSrcfrac}).
The interactions of both the nuclei and leptons however contribute at $\sim1$~GeV energies, as can be seen in 
the left panel of Fig.~\ref{fig:CRbulge2} that show the longitude profiles of the intensity at 1.2~GeV averaged over $-7.5^\circ \leq b \leq 7.5^\circ$ for
the SA100/R12B--R12 (top) and SA100/F98B--F98 (bottom) combinations.
The increased injection of the nuclei and leptons creates more $\pi^0$-decay and Bremsstrahlung with the gas in the inner Galaxy.
But the strong enhancement for the IC emission is due to the combined effect 
of the higher CR and ISRF intensities in this region -- similar to the 
density effect enhancing the IC emissions from the arms for the SA50/SA100--R12 model combinations.
There is also asymmetry of the IC emission profile from the SA100/F98B-F98 combination about the GC because of the angular offsets with respect to the Sun-GC line of the bar spatial distributions for the CR and ISRF densities (Figs.~\ref{fig:isrfenergydensityplane} and~\ref{fig:CRbulge2}).

The results for the SA100/R12B and SA100/F98B provide a framework for understanding the results obtained by IG16 related to re-scaling of the 2D baseline IEMs to produce their final fore-/background IEMs, and the strong contribution by IC emission in and about the GC region.
Figure~15 of IG16 compares the variation with Galactocentric radius of the \GP{}-predicted fluxes $>1$~GeV for the adjusted components ($\pi^0$-decay and IC) of the baseline IEMs and the fitted results over the $15^\circ \times 15^\circ$ region about the GC.
The fit results for the `Pulsars' IEM, which uses the SA0 CR source density model of this paper, for the range $3.5 \lesssim R \lesssim 8$~kpc give an upscaling approximately the same ($\sim 20-30$\%) for the $\pi^0$-decay and IC emission, while for $R \lesssim 1.5$~kpc the IC emission is upscaled by larger factors.

Comparing the energy density distributions for the SA0 and SA50/100 models (Fig.~\ref{fig:crenergydensity}) those with spiral arms have higher CR intensities for $3\lesssim R \lesssim 7$~kpc by $\sim 20$\%.
Because the propagation smoothes the CRs injected by the arms into a quasi-axisymmetric ring it is perhaps no accident that the fits using the 2D models by IG16 find a broader (in $R$) distribution of CR intensities are needed to reproduce the data.
The interpretation then is that at least some non-negligible fraction of CR sources can be located in the spiral arms.

The SA100 and additional CR bulge/bar source density is one piece of the puzzle that can be used also to understand the origin of the IC emission
in and about the GC found by IG16.
The SA0 CR source density with its normalisation condition at the Solar circle is insufficiently peaked toward the GC to provide the injected CR power that would approximate that by the additional R12B/F98B bulge/bar components introduced in this paper.
The other piece of the puzzle is the intensity of the ISRF in and about the GC from the bulge/bar for the R12/F98 models.
The Std ISRF has a lower intensity over the inner Galaxy compared to either of the ISRF models used in this paper.
The dual effect of the higher CR and ISRF intensities associated with these centrally peaked distributions is the critical combination required to understand how the dominant IC \gray{} emissions from the central few kpc about the GC can be produced. 
Comparing the SA0--Std IC emission averaged over $-7.5^\circ \leq l \leq 7.5^\circ$ in Fig.~\ref{fig:gammasrcstdlongprofile} (centre) to that obtained by averaging over the same longitude range for the SA100/R12B--R12 and SA100/F98B--F98 combinations (Fig.~\ref{fig:CRbulge2} left panels), the increase for the additional bulge/bar models is a factor $\sim 2-3$ higher.
IG16 find that fitting the IC for the same region $>1$~GeV requires an increase by a factor $\sim 4$ compared to their baseline `Pulsars' model (the same as the SA0--Std combination).
This is fairly close to the enhancements that are calculated with the additional CR bulge/bar model combinations considered in this paper.

Extracting the physical properties of a putative CR bulge/bar and ISRF density across the inner Galaxy will require further work.
Because the additional CR bulge/bar population is a negligible contribution at the Solar system local CR data cannot provide useful information on its characteristics and the viable techniques will rely on electromagnetic signatures.
For the current illustrative calculations the 
corresponding injected CR powers are $1.84\times 10^{39}$ erg~s$^{-1}$ (SA100/R12B) and $1.94\times10^{39}$ erg~s$^{-1}$ (SA100/F98B), which are a factor $\sim 25$ smaller than the injected power for the SA100 source density model (the injection spectrum parameters are taken to be the same as the spiral arm components, as given in Table~\ref{tab:CRparameters} for the SA100 density model).
Increasing the input CR power for either the R12B/F98B additional component produces correspondingly higher residuals, e.g., a 50\% increase for the R12B component gives positive residuals that are instead $\sim 40-50$\% about the GC, which is comparable to the IG16 residuals there.
But these numbers assume the spatial parameters for the R12/F98 bulge bar distributions used in this paper and that the CR bulge/bar injects nuclei and leptons with the same ratio as the spiral arms.

Optimisations to the models for the CR source and ISRF density distributions can be made.
Determining a more realistic 
mixture of CR sources attributed to spiral arms and a smoothly distributed disc-component than used in this paper (50 or 100\%) is one such improvement.
This paper has assumed that the CR spiral arm density and disc components have the same averaged spectral characteristics, and that the spiral arms all inject the same CR power.
However, most likely the arms are not as equally balanced: the R12 ISRF model has higher (stellar) luminosities for arms 2 and 2$^\prime$ (Fig.~\ref{fig:compluminosity}) coming from the hot, young stars that are the progenitors for the typical CR sources like pulsars and supernova remnants.
The spatial distributions of the arms may also not be as symmetric as assumed.
The fractional residuals in Fig.~2 of IG16 are higher near $l\sim 45^\circ$ than $l\sim -30^\circ$, somewhat opposite the trend particularly for the combinations using the R12 ISRF model shown in Figs.~\ref{fig:ISRFSrcfrac} and~\ref{fig:CRbulge1}.
To remedy this may require modifications of the spatial parameters and weighting the injected CR powers for individual arms differently than done here.
While these aspects of the CRs injected in the spiral arms have not been investigated in this paper because of the already considerable number of models and parameters, they are an additional element that should be addressed by future work.
Possible distributions that could also be used are the NE2001
model of \citet{2002astro.ph..7156C} and \citet{2004ASPC..317..211C}, or that of \citet{2001ApJ...556..181D}.
The NE2001 model has been used to investigate
3D CR distributions in the ISM by \citet{2015APh....64...18W}, but so far not
to calculate IEMs with 3D ISM target density distributions.
Studies of other nearby spiral galaxies can also guide the construction of suitable source density distributions for the bulge/bar and arms \citep[e.g.,][]{2016MNRAS.456.2848H,2016MNRAS.459.3130A}.
In addition the production of an optimised ISRF model is a key ingredient that will require further work: better determination of arm model parameters including distributing dust in them as well as stars, and whether or not stellar/dust arms are co-aligned \citep[e.g.,][]{2014ApJS..215....1V}, the description of the non-axisymmetric structures across the inner Galaxy \citep[e.g.,][]{2005A&A...439..107L,2005MNRAS.358.1309B,2011ApJ...733...27G,2016PASA...33...25Z}, warps and flaring in the stellar disc \citep[e.g.][]{2014A&A...567A.106L}, and possibly other details, are essential.

Lower energy \gray{} data ($\lesssim50$~MeV) can be used to test the predictions made for the CR lepton source and ISRF density models developed in this paper, because of the minimal emissions by $\pi^0$-decay production in this energy regime.
\citet{2008ApJ...679.1315B} and \citet{2011ApJ...739...29B} analysed
INTEGRAL/SPI data using 2D \GP{}-generated models and find that the interstellar emission from the Galactic ridge toward the inner Galaxy finding that it likely has an IC origin but needing some combination of higher electron and/or ISRF intensities to be consistent with the data. 
Bearing in mind the sensitivity of the data, re-analysis with models as developed in this paper can enable better determination of the CR and ISRF densities necessary.
For the $\sim 1-30$~MeV energy range COMPTEL on {\it CGRO} has provided the best data to date but the resolution and sensitivity are limited.
Major changes to the background rejection algorithms for the \fermilat\ have significantly enhanced its sensitivity below 100~MeV energies for the so-called Pass-8 data\footnote{See https://fermi.gsfc.nasa.gov/ssc/data/analysis/documentation/Pass8\_usage.html and https://fermi.gsfc.nasa.gov/ssc/data/analysis/LAT\_caveats.html.} opening the way to using these data down to the COMPTEL energy range with the possibility of even some small overlap in coverage.
Also, proposed medium energy \gray{} instruments such as e-ASTROGAM \citep{2016SPIE.9905E..2NT,2016arXiv161102232D} and AMEGO\footnote{https://asd.gsfc.nasa.gov/amego/index.html} will significantly improve the data quality in this energy range.

The emissions from CRs losing energy nearby to their sources at higher energies also offers opportunities to test the predictions made by models such as have been considered in this paper.
For leptons the rapid energy losses $\gtrsim 100$~GeV energies in and about the enhanced ISRF energy densities of the spiral arms produces signatures that can be related to the average injected CR power for different arms.
This is true also toward the inner Galaxy where the hypothetical CR bulge/bar population can be similarly tested because of its localisation.
Analysis of \fermilat\ data can accomplish this up to energies $\sim 100$s of GeV energies, which is sufficient for modelling and testing the CR injection in the steady-state picture that is assumed in this work.
Higher energies probe the regime where the CR injection by individual sources becomes important, but these are beyond the scope of the calculations made in this paper.

Finally, it should be noted that the lobe-like excesses and ``pinching'' toward the GC shown in Fig.~\ref{fig:CRbulge1} have similarity to the so-called ``EEE'' residual component of the IEM developed to 
support the generation of the Third Fermi Point-Source Catalogue \citep[see Figs.~10 and~11 of][]{2016ApJS..223...26A}.
It is suggested by \citet{2016ApJS..223...26A} that these are the end-points toward the GC of the so-called ``Fermi Bubbles'' \citep[e.g.,][]{2010ApJ...724.1044S} whose origin has been widely speculated.
\citet{2016ApJS..223...26A} employed a 2D axisymmetric \GP{}-calculated IEM using a slightly different CR source density distribution and the Std ISRF as its basis, and the long-established method of dividing the interstellar emission components into Galactocentric annuli and fitting to the \gray{} data for their respective emissivities \citep[e.g.,][]{1988A&A...207....1S,1996A&A...308L..21S}.
However, \citet{2016ApJS..223...26A} did this only for the gas-related emission components ($\pi^0$-decay and Bremsstrahlung) and used an all-sky \GP-predicted IC intensity map with re-fitting only for its spectral characteristics.
This does not have the flexibility of the method of IG16 for accounting for mismodelling of the IC emission. (Note that IG16 found no evidence for a residual like the ``EEE'' component.)
The residuals shown in Fig.~\ref{fig:CRbulge1} (for 1.2~GeV) are very similar to those at 3.4~GeV and 22~GeV, which are the energies that the ``EEE'' residual is shown for in the lower centre and right panels of Fig.~10 of \citet{2016ApJS..223...26A}.
Their shape about the GC is somewhat similar to that of the ``EEE'' component there, and the difference between the averaged IC intensities for either the SA100/R12B--R12 or SA100/F98B--F98 combination and reference model within the $5^\circ \times 5^\circ$ region about the GC are $\sim 5\times10^{-10}$ and $\sim 7\times 10^{-12}$ MeV$^{-1}$ cm$^{-2}$ s$^{-1}$ at those energies, respectively.
These are within $\sim 30$\% those for the ``EEE'' component for the same region, suggesting that a plausible explanation for its origin there is a mismodelling of the CR source and ISRF densities by the 2D IEMs combined with the inflexibility of the fits by \citet{2016ApJS..223...26A} for treating the IC emission spatial distribution on the sky.
Fitting for the IC intensity in Galactocentric annuli as done by IG16 would allow the 2D approach to better account for mismodelling of this component, which can enable these simplified IEMs to maintain some utility for supporting generation of point-source catalogues and analyses of \gray{} data in localised regions where the physical interpretation of the large-scale interstellar emission is not the major objective.

\section{Summary}
\label{summary}

New calculations of the high-energy interstellar emissions produced by CRs interacting with the ISM of the Galaxy have been made with the latest release of the \GP{} code.
These have used 3D spatial models for the CR source and ISRF densities. 
The 3D ISRF models are also new calculations with the \frankie\ code that include spatial distributions to account for non-axisymmetric features of the stellar distribution in the Milky Way, such as the spiral arms and bar.
Compared with 2D Galactocentric axisymmetric models those including 3D structure show features in all-sky \gray{} maps that can be attributed to specific spatial elements, such as the spiral arms.
These features depend on the CR source and ISRF density models, with the models that have localised increases in both CR and ISRF intensities producing enhanced IC emissions.
This is evident for all model combinations involving the SA50/SA100 CR source density models, the R12 ISRF model, and the additional `CR bulge/bar' SA100/R12B and SA100/F98B combinations.

The calculations presented in this paper can be used to better interpret analyses of \gray{} data.
They have been used here to provide a plausible physical interpretation for results obtained by the \fermilat\ team in their analysis of \gray{} data toward the inner Galaxy.
In particular, the up-scaling outside of the inner Galaxy needed for the interstellar emission predicted by the 2D-based IEMs that they employed can be understood as mismodelling of the CR source density there: the spiral arms produce a higher CR energy density in the ISM compared to a smooth disc component.
The spatial distribution of CR energy density in the ISM for the spiral arms can be fit somewhat successfully even with an axisymmetric model because the propagation smoothes the CR source density from these localised injection regions into a quasi-axisymmetric distribution for Galactocentric $R\sim 3-7$~kpc.
The strong IC emission found by the \fermilat\ team toward the GC comes from the combined effect of a population of CR accelerators localised in and about the inner Galaxy injecting CRs that interact with the intense ISRF in this region.
The spatial characteristics of this additional CR `bulge/bar' source density are modelled using the same distributions employed for the respective calculations of the 3D ISRF models in this paper.

This work has demonstrated the need for detailed modelling of the distribution of CR sources and the ISRF taking into account the 3D structure of the ISM.  The residual structure in the calculated maps resemble features that have been previously interpreted as possible signs of new physics. However, further work is needed to optimise the models and more carefully tune them to the available data.

\acknowledgements
T.~A.~P. would like to thank Simona Murgia and Thomas Robitaille for useful discussions.
GALPROP development is partially funded via NASA grants NNX13AC47G and NNX17AB48G.

Some of the results in this paper have been derived using the 
HEALPix~\citep{2005ApJ...622..759G} package.

\bibliographystyle{aasjournal}
\bibliography{isrf}

\begin{appendix}

  \section{\GP\ v56 features}
  \label{app:gpv55}

  The improvements to \GP\ made include significant modifications to its architecture as well as numerous technical improvements and inclusion of additional physics code.
  An overview of the major changes are given here, while specific examples of command line configuration/build/installation/execution are provided at the \GP\ website (http://galprop.stanford.edu).
  Forums and a bugzilla are also available at the website.

  Architecturally, previous versions of \GP\ are monolithic with the code and configuration required to detect external libraries and build the \GP\ library and executable contained in the source distribution.
  With the new release this changes to have a single external dependency on the \galtoolslib\ library.
  Because of re-use considerations -- the \frankie, \galgas\ \citep{Johannesson:2015qqi}, and \gardian\ \citep{2012ApJ...750....3A} packages also developed by members of the \GP-team use many common elements and similar configuration and build procedures -- core functionality across all code bases is abstracted into this library.
  \galtoolslib\ includes utility code for parameter parsing (e.g., reading the {\it galdef} configuration file of a \GP\ run), specifying spatial distributions (e.g., for CR source densities), libraries for the representation of results (e.g., skymaps with HEALPix), core physics routines for the nuclear reaction network and energy losses, and other commonly reused code.
  Its required external library dependencies are: CFitsIO\footnote{https://heasarc.gsfc.nasa.gov/fitsio/fitsio.html} and CCfits\footnote{https://heasarc.gsfc.nasa.gov/fitsio/ccfits/}, CLHep\footnote{http://proj-clhep.web.cern.ch/proj-clhep/}, the Gnu Scientific Library\footnote{https://www.gnu.org/software/gsl/}, HEALPix\footnote{http://healpix.jpl.nasa.gov/}, and Xerces-C\footnote{https://xerces.apache.org/xerces-c/}.
  Optional external library dependences are: libastro from XEphem\footnote{http://www.clearskyinstitute.com/xephem/}, WCSLIB\footnote{http://www.atnf.csiro.au/people/mcalabre/WCS/}, OpenCL\footnote{https://www.khronos.org/opencl/}, and CppUnit\footnote{https://sourceforge.net/projects/cppunit/}.
  WCSLIB and libastro are used both for reading and writing data in different map projections.
  OpenCL can be used to distribute calculations across CPUs and compute accelerators for \frankie\ and \galgas\ but is not needed for \GP\ runs.
  CppUnit is a framework that is used by the various packages developed by the \GP\ team for unit testing.
  
  The configuration and build process previously employed the Gnu autoconf\footnote{https://www.gnu.org/software/autoconf/autoconf.html} and automake\footnote{https://www.gnu.org/software/automake/} tools.
  These are deprecated with the new release and now the configuration/build/installation uses CMake\footnote{https://cmake.org/}. 
  The minimum requirement for building \galtoolslib\ and \GP\ is CMake 3.0.
  The base language support requires a C/C++ compiler that implements the
  C++11 standard, and a Fortran 77/90 compiler.
  Minimum requirement compiler versions tested for this release include gcc/g++ 4.9 and recent Intel compilers.
  Almost any recent Fortran compiler is sufficient.
  
  Supported build targets are Linux and OSX, with other Unix-like variants possible.
  Likely if the target system has the minimum language support noted above and allows for the installation of necessary external libraries then \galtoolslib\ can be installed.
  Following successful installation of \galtoolslib, the straightforward build and installation of the separate \GP\ library and executable is enabled by pointing the command line configuration invocation to the \galtoolslib\ location.
  This configures the paths using scripts generated by the \galtoolslib\ installation.
  The standard ``make \&\& make install'' following the configuration will build and install the libraries and executable in the specified directories.
  To ensure end-user sanity it is highly recommended to employ consistency with the configuration/build/installation process for \galtoolslib\ and \GP. 
  That is, use the same compilers and configuration for all steps.

  Technical and physics improvements have focussed on better memory layout and computational speed and making \GP\ more flexible as a general code for calculating CR propagation and interstellar emissions from galaxies, instead of just for the Milky Way.
   Many of the internal structures and loops have been reorganised to take advantage of data and code caches on modern CPUs.
  Parallelisation has been mainly at the loop level with OpenMP for earlier versions\footnote{http://www.openmp.org/}, but now more advanced constructs are taken advantage of including vectorisation facilities of the OpenMP 4 specification where implemented by compiler vendors.
  New solvers for the diffusive transport equations, in particular, take advantage of the vectorisation to dramatically increase the speed of the 3D mode so that the solutions take only a small fraction of time compared to earlier versions.

  More options are now available for the distributions of the interstellar gas and CR sources. This is done via the {\it galstruct} library included in \galtoolslib, which reads XML files describing the distributions of the gas and CR sources.  
  The {\it galstruct} library is easily extended with new modules and functionality, plus it includes many pre-defined distributions.
  With this new mechanism it is possible to include many different source classes in a single \GP\ run.
  The current release includes two source classes, one for backwards compatibility so prior run configuration files still work without modification, and a new class using the {\it galstruct} library for the spatial properties and injection spectra.
  The new framework allows easy incorporation of multiple spectral models.
  It currently includes a multiple broken power-law and a smoothly joined multiple broken power-law.
  The power-laws can in both cases be a function of rigidity, kinetic energy, momentum or total energy through a user selected parameter.
  Each isotope can also have a separate injection spectrum.
  Relative normalisation of the isotopes has also been improved and can now be specified in terms of either a single point in energy or an integrated band in energy.
  Using the integrated bands has helped stability when modifying \GP\ parameters in maximum-likelihood fits because it reduces degeneracy of the spectral parameters of different isotopes.
  
  The system of the propagation equations has been generalised to allow for spatial variations in the diffusion coefficient.  
  Scaling of the diffusion coefficient and the Alfv\'en speed with the strength of the Galactic magnetic field is used for modelling of the interstellar emissions by \citet{2015ApJ...799...86A} (their model C). 

  The skymap integrator has been re-written using a variable step size integrator that is both faster and more accurate.  
  The location of the observer can also now be arbitrary in $(X,Y,Z)$. 
  The integrator also allows for the absorption of \gray{s} on the ISRF, which is implemented following \citet{2006ApJ...640L.155M}. It is a user-selectable option to include this or not. If selected the $e^\pm$ pairs resulting from the absorption are included as an additional source of secondary leptons. The individual lepton production cross section is taken from \citet{1997A&A...325..866B}. The position dependent pair source function assumes an isotropic \gray{} and ISRF photon distribution, which is not formally correct, but significantly reduces the computational cost and does not introduce significant error.

  The output IC maps are generalised to allow for an arbitrary spatial splitting, in particular, their splitting can be set to match the splitting of the gas-related emission.
  This allows for more flexible template fitting as used, e.g., in the analysis of the emission from the inner Galaxy by IG16.

  The 3D treatment of the magnetic field now includes regular and random components that can be specified differently in the disk and the plane. 
  Temperature and polarisation of radio and microwave synchrotron emission can be calculated. Synchrotron $I$, $Q$ and $U$ Stokes parameters 
  are calculated and output as HEALPix maps. Radio absorption and free-free emission are included, more details can be found in 
  \citet{2011A&A...534A..54S}, \citet{2013MNRAS.436.2127O}, and \citet{Orlando:2015bsa}

  CR antiprotons by CR nuclei interactions with interstellar gas and are, therefore, called secondary. The same interactions produce charged and neutral mesons that decay to secondary $e^\pm$ and \gray{s}.
  \citet{2015ApJ...803...54K} analysed $\bar p$ production in $pp$-, \mbox{$pA$-,} 
  and $AA$-interactions using EPOS-LHC and QGSJET-II-04.
  The $\bar p$ yields
  of the two MC generators agree reasonably well with each other
  and the available experimental data. Therefore, the results 
  of these generators can be used to predict reliably the $\bar p$ 
  yield outside the energy range covered by fixed target accelerator data, 
  $E_{\bar p}\approx 10-$100\,GeV.
  The $\bar p$-yield differ 
  by a factor of few from yields of parameterisations based on the fixed target data commonly used in astrophysics, and that are also the basis of the $\bar p$ calculations made with earlier \GP\ versions.
  The new $\bar p$ yield calculations are a user-selected option with this release of \GP.
  
\end{appendix}

\end{document}